\documentclass[prd,nofootinbib,twocolumn,superscriptaddress,preprintnumbers,balancelastpage,longbibliography]{revtex4-1}

\usepackage{graphicx}
\usepackage{color}
\usepackage{hepunits}

\usepackage[english]{babel}

\newcommand{\es}[2] {\begin{equation} \label{#1} \begin{split} #2 \end{split} \end{equation}}

\newcommand{\be}{\begin{equation}}
\newcommand{\ee}{\end{equation}}

\begin{document}
\title{Detecting Axion Dark Matter with Radio Lines \\
from Neutron Star Populations
 }

\author{Benjamin R. Safdi}
\email{bsafdi@umich.edu}
\affiliation{Leinweber Center for Theoretical Physics, Department of Physics, University of Michigan, Ann Arbor, MI 48109 U.S.A.}

\author{Zhiquan Sun}
\email{zqsun@umich.edu}
\affiliation{Leinweber Center for Theoretical Physics, Department of Physics, University of Michigan, Ann Arbor, MI 48109 U.S.A.}

\author{Alexander Y. Chen}
\email{alexc@astro.princeton.edu}
\affiliation{Department of Astrophysical Sciences, Princeton University, Princeton, NJ 08544, USA}

\preprint{LCTP-18-22}

\date{\today}

\begin{abstract}
It has been suggested that radio telescopes may be sensitive to axion dark matter that resonantly converts to radio photons in the magnetospheres surrounding neutron stars (NSs).  In this work, we closely examine this possibility by calculating the radiated power from and projected sensitivity to axion dark matter conversion in ensembles of NSs within astrophysical systems like galaxies and globular clusters.  We use population synthesis and evolution models to describe the spatial distributions of NSs within these systems and the distributions of NS properties.  
Focusing on three specific targets for illustration, the Galactic Center of the Milky Way, the globular cluster M54 in the Sagittarius dwarf galaxy, and the Andromeda galaxy, we show that narrow-band radio observations with telescopes such as the Green Bank Telescope and the future Square Kilometer Array may be able to probe the quantum chromodynamics axion over roughly two orders of magnitude in mass, starting at a fraction of a $\mu$eV.
\end{abstract}

\maketitle

\section{Introduction}
It was recently shown that axion dark matter (DM) may be detectable through radio telescope observations of narrow lines from axion DM conversion to photons in neutron star (NS) magnetospheres~\cite{Hook:2018iia,Pshirkov:2007st,Huang:2018lxq}.  The axion DM falls into the gravitational potential of the NS, where it is accelerated to semi-relativistic speeds.  The axion-photon conversion takes place in the strong magnetic fields in the NS magnetosphere, though the conversion is stymied by the fact that the axion mass prevents efficient mixing with the massless photon.  However, as was shown in~\cite{Hook:2018iia,Pshirkov:2007st,Huang:2018lxq}, the plasma in the magnetosphere gives the photon an effective mass, which increases with decreasing radius from the NS surface.  There exists a radius, which we refer to as the conversion radius, where the axion mass equals the effective photon mass, and in the vicinity of this radius resonant conversion takes place between the axion DM and electromagnetic radiation.  

In the present work, we explore the sensitivity to axion DM from radio observations of systems that contain large ensembles of NSs.  For example, we examine observations towards the Galactic Center of the Milky Way, nearby dwarf galaxies and globular clusters, such as M54, and nearby galaxies such as M31.  Calculating the predicted axion-induced radio flux in these systems requires combining spatial models for the distribution of NSs, models for the distribution of DM, and models for the distributions of NS properties.  We address these models within this work, pointing out those with the largest quantifiable systematic uncertainties, and demonstrate that radio observations with current and near-term telescopes of the targets listed above have the potential to cover vast regions of so-far uncharted axion DM parameter space. 

The possibility of detecting axion DM through radio observations of systems of NSs is important considering that the quantum chromodynamics (QCD) axion is one of the best-motivated DM candidates at present.  The QCD axion was originally proposed as a new particle to solve the strong {\it CP} problem~\cite{Peccei:1977hh,Peccei:1977ur,Weinberg:1977ma,Wilczek:1977pj}, though it was shortly thereafter realized that this particle could also naturally explain the observed DM of the Universe within a cosmological framework~\cite{Preskill:1982cy,Abbott:1982af,Dine:1982ah}.  In addition to coupling to QCD, the axion $a$ also couples to electromagnetism through the operator ${\mathcal L} =  g_{a\gamma\gamma} a {\bf E} \cdot {\bf B}$, where ${\bf E}$ (${\bf B}$) is the electric (magnetic) field and $g_{a\gamma\gamma}$ is the coupling constant.
For the QCD axion, $g_{a \gamma \gamma} = g \alpha_\text{EM} / (2 \pi f_a)$, with $\alpha_\text{EM}$ the fine-structure constant, $f_a$ the axion decay constant, and $g$ a number order unity that depends on the UV completion of the model; for the DFSZ~\cite{Dine:1981rt,Zhitnitsky:1980tq} (KZVZ~\cite{Kim:1979if,Shifman:1979if}) models, $g \approx 0.75$ (-1.92).
  In the presence of a static, external magnetic field, this operator causes an initial axion field to rotate into an electromagnetic wave of the same frequency polarized along the direction of the magnetic field~\cite{Sikivie:1983ip,Raffelt:1987im}.

  The axion decay constant, which normalizes the axion's coupling to QCD, also determines the QCD axion's mass: $m_a \approx 1.3 \times 10^{-5} \, \, \text{eV} \, (10^{12} \, \, \text{GeV} / f_a)$.  Axion-photon conversion in NSs produces a narrow line at this frequency, which is in the radio band for $f_a \sim 10^{12}$ GeV~\cite{Hook:2018iia,Pshirkov:2007st,Huang:2018lxq}.  However, in string compactifications it is common to find a plethora of light psuedo-scalars~\cite{Svrcek:2006yi} called axion-like particles (ALPs), which may couple to QED but not to QCD.  The ALP masses are thus not necessarily related to their electromagnetic couplings, unlike for the QCD axion.  The primary goal of this work is to reach sensitivity to the QCD axion, though we will see that in the process large regions of ALP parameter space may also be covered.   

To calculate the axion-induced signal from ensembles of NSs, we must use NS population and evolution models to predict the distributions of NSs and their properties in galaxies.
We use population models such as~\cite{FaucherGiguere:2005ny,2010MNRAS.401.2675P}, which have been tuned to observable pulsar data, to describe the distributions of initial NS magnetic field strengths and spin periods. 
We then evolve the NSs to the present day using spin-down and magnetic field decay models~\cite{FaucherGiguere:2005ny,Spitkovsky:2006np,2009A&A...496..207P,2010MNRAS.401.2675P,2010MNRAS.407.1090B,Philippov:2013aha,Johnston:2017wgm}, though there are uncertainties on these processes which we describe.   

{To date we have detected more than 2000 rotation-powered pulsars, but this only makes up a very small fraction of the total NS population. The active isolated pulsars tend to be have an age $\tau \lesssim 100\,\mathrm{Myr}$, whereas a large population of undetectable ``dead'' pulsars should be lying dormant, undetectable, with a small number of them recycled by accretion from a companion star, spun up and detectable as millisecond pulsars.}
The majority of NSs will not produce any pulsed emission in the radio band.  This is because as the pulsar spins down, at some point it will not be able to sustain the voltage required for pair creation, {which is believed to produce the required plasma for coherent radio emission.} This happens around the so-called pulsar ``death line'' (see {\it e.g.}~\cite{Zhang:2000rd}, though also the recent~\cite{Tan:2018rhg} for a pulsar that appears beyond the death line).

Following previous works~\cite{Hook:2018iia,Pshirkov:2007st,Huang:2018lxq}, we model the magnetospheres of the active pulsars by the analytic model developed by  Goldreich and Julian (GJ)~\cite{1969ApJ...157..869G}.
However, both
analytical models ({\it e.g.}~\cite{1985MNRAS.213P..43K}) and numerical
simulations ({\it e.g.}~\cite{Spitkovsky:2002wg}) have shown that without pair production,
the magnetosphere of a NS falls back to a solution with finite
extent, called the electrosphere solution, that differs qualitatively from the GJ model. Apart from a ``torus'' of plasma around the equator and a ``dome'' above each
pole, the solution is very similar to a vacuum spinning dipole.
For oblique rotators the spin-down rate is well described by the vacuum approximation, and for aligned rotators the system is stable and does not spin down anymore. A small amount of plasma can still leak from the tip of the equatorial torus due to the diochotron instability, which is the non-neutral plasma analog of the Kelvin-Helmholtz instability \cite{2002A&A...387..520P}.
As this is a very robust configuration for the NS magnetosphere, we believe it well describes the fate of all pulsars after
they cross the death line~\cite{Cerutti:2016ttn}.  We use numerical simulations of the electrosphere solution to describe the magnetospheres of the dead NSs.

As we show, ensembles of active pulsars and dead NSs in systems like the center of the Milky Way produce narrow and bright radio lines in the presence of axion DM, which may be detected through dedicated observations with current and near-term telescopes, such as 100-m class telescopes like the Effelsberg telescope and the Green Bank Telescope (GBT) and radio interferometer arrays such as the Square Kilometer Array (SKA).  This new indirect path to axion DM detection is complementary to traditional axion DM direct detection.  For example, the Axion Dark Matter eXperiment (ADMX), which is a resonant cavity experiment, has already constrained a narrow region of axion DM parameter space with axion masses around $10^{-6}$ eV, and ongoing as well as future runs of ADMX may be able to cover nearly a decade of possible axion masses, from $\sim$$10^{-6}$ eV to $\sim$$10^{-5}$ eV~\cite{Shokair:2014rna,Rosenberg:2015kxa,Du:2018uak}.  At higher frequencies, planned experiments such as HAYSTAC~\cite{Brubaker:2016ktl,Kenany:2016tta,Brubaker:2017rna} and MADMAX~\cite{Majorovits:2016yvk} may extend the reach in axion mass to $\sim$$10^{-4}$ eV, while at lower masses experiments including ABRACADABRA~\cite{Kahn:2016aff,Foster:2017hbq,Henning:2018ogd,Ouellet:2018beu}, DM-Radio~\cite{Chaudhuri:2014dla,Silva-Feaver:2016qhh}, and CASPEr~\cite{Budker:2013hfa} may potentially reach all the way down to $\sim$$10^{-9}$ eV or below.  The indirect radio searches proposed in this work may reach sensitivity to the QCD axion from masses in the range $\text{few} \times 10^{-7}$ eV to $\text{few} \times 10^{-5}$ eV, depending on the target, telescope, and systematics such as the DM density profiles and the distributions of NS properties.  

An indirect detection discovery with a radio telescope at a specific frequency may be verified with one of the direct detection efforts described above, as for most of these experiments the most difficult aspect is scanning over all of the possible axion masses.  Resonant cavity experiments such as ADMX, for example, must scan over $\sim$$10^6$ independent frequencies, by stopping and changing the resonant frequency of the cavity each time, to cover a decade of possible axion mass.  If the frequency of the axion is known in advance, this would thus cut down the amount of time needed to verify the signal by a factor $\sim$$10^6$ compared to a blind axion search.  Similarly, a laboratory axion signal would give a concrete prediction for the frequency that the radio lines discussed in this work should appear.   

The remainder of this paper proceeds as follows.  First, in Sec.~\ref{sec: mags} we review the calculation of the radiated power in the GJ model from axion-DM conversion to radio photons in the NS magnetosphere, and we use numerical simulations to discuss how the radiated power changes in the electrosphere model.  In Sec.~\ref{sec: population-model} we develop population models to describe the distributions of NS magnetic fields and spin periods.  In Secs.~\ref{sec: NS-DM},~\ref{sec: M54}, and~\ref{sec: M31} we present models for the expected spatial distributions of NSs and DM in the Milky Way, the globular cluster M54, and M31, respectively.  Then in Sec.~\ref{sec: radio} we combine the formalisms developed to predict the sensitivity to axion DM from radio observations with current and future telescopes, such as GBT and SKA.  We conclude in Sec.~\ref{sec: discussion}.   

\section{Axion-photon conversion in neutron-star magnetosheres}
\label{sec: mags}

In this section we discuss the axion-induced radio flux in the electrosphere model.  First, we review the GJ-model calculation from~\cite{Hook:2018iia}.  Then, we generalize that framework to apply to arbitrary plasma distributions, and we use these results to numerically calculate the radio flux in the electrosphere model that applies to inactive, ``dead" NSs. 

\subsection{Resonant conversion in the Golreich-Julian model}
\label{sec: res-GJ}

Axion DM of mass $m_a$ and axion-photon coupling $g_{a \gamma \gamma}$ may convert to radio photons as the DM falls into the potential well of the NS.  Following~\cite{Hook:2018iia}, the radiated power per solid angle for resonant conversion in the GJ model is given by
\es{power}{
{d P^\text{GJ} \over d \Omega} = &4.5 \times 10^{8} \, \, \text{W} \left( {g_{a \gamma \gamma} \over 10^{-12} \, \, \text{GeV}^{-1} } \right)^2   \\
 &\left( {224 \, \, \text{km} \over r_c(\Omega)} \right)^4 \left( {B(\Omega) \over 5 \times 10^{13} \, \, \text{G}} \right)^2 \left( {1 \, \, \text{GHz} \over m_a} \right)  \\ &\left( {\rho_\infty \over 0.3 \, \, \text{GeV} / \text{cm}^3} \right) \left( {M_{\rm NS} \over 1 \, \, M_\odot} \right) \left( {200 \, \, \text{km}/\text{s} \over v_0} \right)\,,
}
where the conversion radius $r_c(\Omega)$ and the magnetic field $B(\Omega)$ depend on the angular coordinates $\Omega$.  We will discuss the calculation of the conversion radius within the GJ model shortly.  Above, $\rho_\infty$ is the ambient DM density, not accounting for gravitational focusing from the NS itself, and $v_0$ is a parameter specifying the velocity dispersion of the DM distribution, which will be defined more precisely later.  The NS is assumed to have a mass $M_\text{NS} \approx 1$ $M_\odot$ and a radius $r_0 \approx 10$ km.

  For an aligned NS, where the magnetic dipole axis is aligned with the rotation axis, 
\es{}{
B(\theta) = {B_0 \over 2} \left( 3 \cos^2\theta + 1\right)^{1/2} \,,
}
where $\theta$ is the polar angle from the rotation and magnetic axis, and $B_0$ is the magnetic field at the poles.  We focus on the aligned NS throughout this section for simplicity.  Note that $B(\theta)$ is the magnetic field at the NS surface; since the field is a dipole configuration, the magnetic field is assumed to fall of with radius like $B(r) \sim B(r_0) (r_0 / r)^3$.  In this case, no physical quantities depend on the azimuthal angle $\phi$ by symmetry.

In the GJ model, the plasma charge density is given by 
\es{GJ_charge}{
n_c & = \frac{2 \mathbf{\Omega} \cdot \mathbf{B}}{e} + \, \, \text{(relativistic corrections)} \,,
}
where ${\bf \Omega}$ is the angular velocity vector for the NS and ${\bf B}$ is the magnetic field vector.  This implies that when $\mathbf{\Omega} \cdot \mathbf{B} < 0$, the plasma is negatively charged, while when the product is positive the plasma is positively charged.  In the GJ model, the plasma is completely charge separated, so that the number density of electrons (ions) in the negatively (positively) charged regions is given by $-n_c$ ($n_c$). 
If the NS is inactive, then there are no positrons and the plasma simply consists of charge-separated electrons and ions.  The difference between the electron and ion dominated regions is important because the plasma frequency is given by $\omega_\text{pl} \approx \sqrt{4 \pi \alpha n_e / m_c}$, where $m_c$ is the mass of the charge carrier.  The conversion radius is solved for by setting $m_a = \omega_\text{pl}$, which implies that the conversion radius $r_c^p$ in the ion-dominated regions is smaller by a factor $\sim$$(m_e / m_p)^{1/3} \approx 0.08$ compared to the conversion radius $r_c^{e^-}$ in the electron-dominated regions, where $m_e$ ($m_p$) is the electron (proton) mass.  

To solve for the conversion radius, first note that combining~\eqref{GJ_charge} with the definition of the plasma frequency and the dipole assumption for the magnetic field yields a plasma-density profile that scales with radius like $\omega_\text{pl} \propto r^{-3/2}$.  As expected, the plasma frequency increases monotonically towards the NS surface.
In the electron-dominated region, the conversion radius is given by~\cite{Hook:2018iia}
\es{rc}{
& r_c^{e^-}(\theta)  = 224 \, \, \text{km} \times \big|3 \cos^2 \theta - 1 \big|^{1/3} \times \\
 & \left( {r_0 \over 10 \, \, \text{km}} \right) \times \left[  {B_0 \over 10^{14} \, \, {\rm G} } \ { 1\, \, \text{sec} \over P} \left( {1 \, \, \text{GHz} \over m_a} \right)^2 \right]^{1/3} \,,
} 
where $P = 2 \pi / \Omega$ is the rotation period of the NS. 

Note that resonant conversion only takes place if $r_c > r_0$.  
When the resonant conversion criterion is satisfied, then the conversion takes place over a length $L_\text{GJ} = \sqrt{2 \pi r_c v_c / (3 m_a)}$, where $v_c$ is the DM velocity at the conversion radius.  The radiated power in~\eqref{power} may be understood from the fact that the axion-photon conversion probability scales like $p_{a \gamma}^\infty \sim g_{a \gamma \gamma}^2 B(r_c)^2 L_\text{GJ}^2 / v_c$ and $dP/d\Omega \sim p_{a \gamma}^\infty \rho_\text{DM}^\infty v_c^2 r_c^2 / v_0$ (see~\cite{Hook:2018iia} for details).  The velocity $v_c$ is well approximated by $\sqrt{2GM_{\rm NS}/r_c}$, since the DM is assumed to start off non-relativistic far away from the NS.
Note that when the resonant conversion criterion is not satisfied, radiated flux may still originate from non-resonant conversion, though we have checked explicitly that this is not important in practice for the systems we consider.

\subsection{Resonant conversion in the electrosphere model}
\label{sec: res-electrosphere}

\begin{figure}[htb]
\includegraphics[width = 0.49\textwidth]{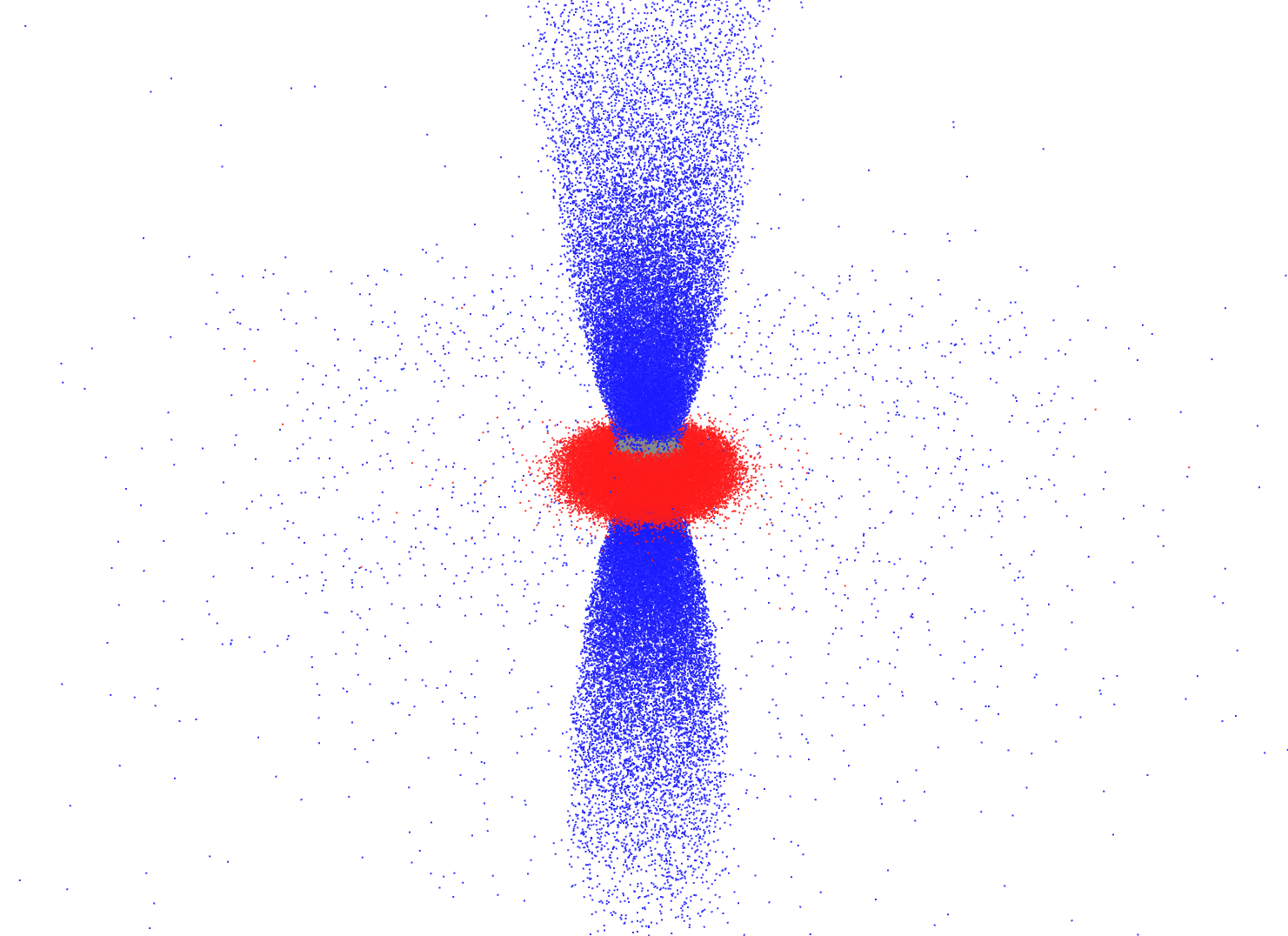}
\vspace{-0.7cm}
\caption{{An illustration of the electrosphere from our simulations. The blue particles are electrons and red are ions. Electrons concentrate around the two poles to form two ``domes'' while the ions stay around the equator to form a ``torus.''  In this example, the magnetic axis is parallel to the rotation axis; if the two are anti-parallel, then the electrons and ions are swapped.}}
\label{fig:electrosphere}
\end{figure}

In this subsection we numerically calculate the radiated power in the electrosphere model for the NS magnetosphere. 
We simulate the electrosphere using the Particle-in-Cell code \emph{Aperture} first described in \cite{2014ApJ...795L..22C}. We solve the time-dependent plasma dynamics with Maxwell equations on a spherical grid with logarithmic spacing in $r$. We assume axisymmetry, and the magnetic axis of the star is taken to be aligned with the rotation axis. Although we have a 2D grid, all 3 components of ${\bf E}$ and ${\bf B}$ fields are evolved, and particle motion is integrated in full 3D. A schematic diagram of the charge distribution is given in Fig. \ref{fig:electrosphere}. The figure shows the tracked particles from one of our runs, clearly demonstrating that the magnetosphere settles down to a ``dome-and-torus'' structure. This is qualitatively very similar to the result obtained in \cite{Spitkovsky:2002wg}. We take the electron and ion density profiles from the simulation for our subsequent analysis.

In the electrosphere model, the magnetic field is still distributed like a dipole, but the plasma no longer follows the relation in~\eqref{GJ_charge}.  However, given the numerical plasma density profile, we may, for each $\theta$, solve for the conversion radius $r_c$.  Around this conversion radius, the plasma will take a profile 
\es{omega_n}{
\omega_\text{pl}(r) \approx m_a \left({ r_c \over r}\right)^n \,,
}
for some index $n$, which equals $3/2$ in the GJ model.  We fit the numerical plasma profile to the functional form above in the vicinity of $r_c$ to find the best-fit $n$.  Note that this is only an approximation, since the plasma does not have a simple power-law form in general, but we have verified explicitly through numerical simulations along the lines of those in~\cite{Hook:2018iia} that this approximation works well for the electrosphere for the purpose of calculating the radiated power.

For a plasma with a radial dependence as in~\eqref{omega_n}, the axion-photon conversion probability still scales like $p_{a \gamma}^\infty \sim g_{a \gamma \gamma}^2 B(r_c)^2 L_n^2 / v_c$, for some length $L_n$, but that length now differs from the GJ value.  In particular, it is straightforward to generalize the formalism described in~\cite{Hook:2018iia} to account for a general index $n$, and this leads to the result $L_n = L_\text{GJ} \sqrt{1.5 / n}$, where $L_\text{GJ}$ is the GJ conversion length.  This, in turn, implies that 
\es{power-n}{
{d P \over d \Omega} = {dP^\text{GJ} \over d \Omega} \left(3/2 \over n \right)\,,
}
where ${dP^\text{GJ} / d \Omega} $ is given in~\eqref{power}.  In the electrosphere model we find $n \sim 3/2$ at small $r$, though $n$ increases quickly at large $r$ since the electrosphere model has a smaller spatial extent than the GJ model.

\begin{figure*}[htb]
\includegraphics[width = 0.49\textwidth]{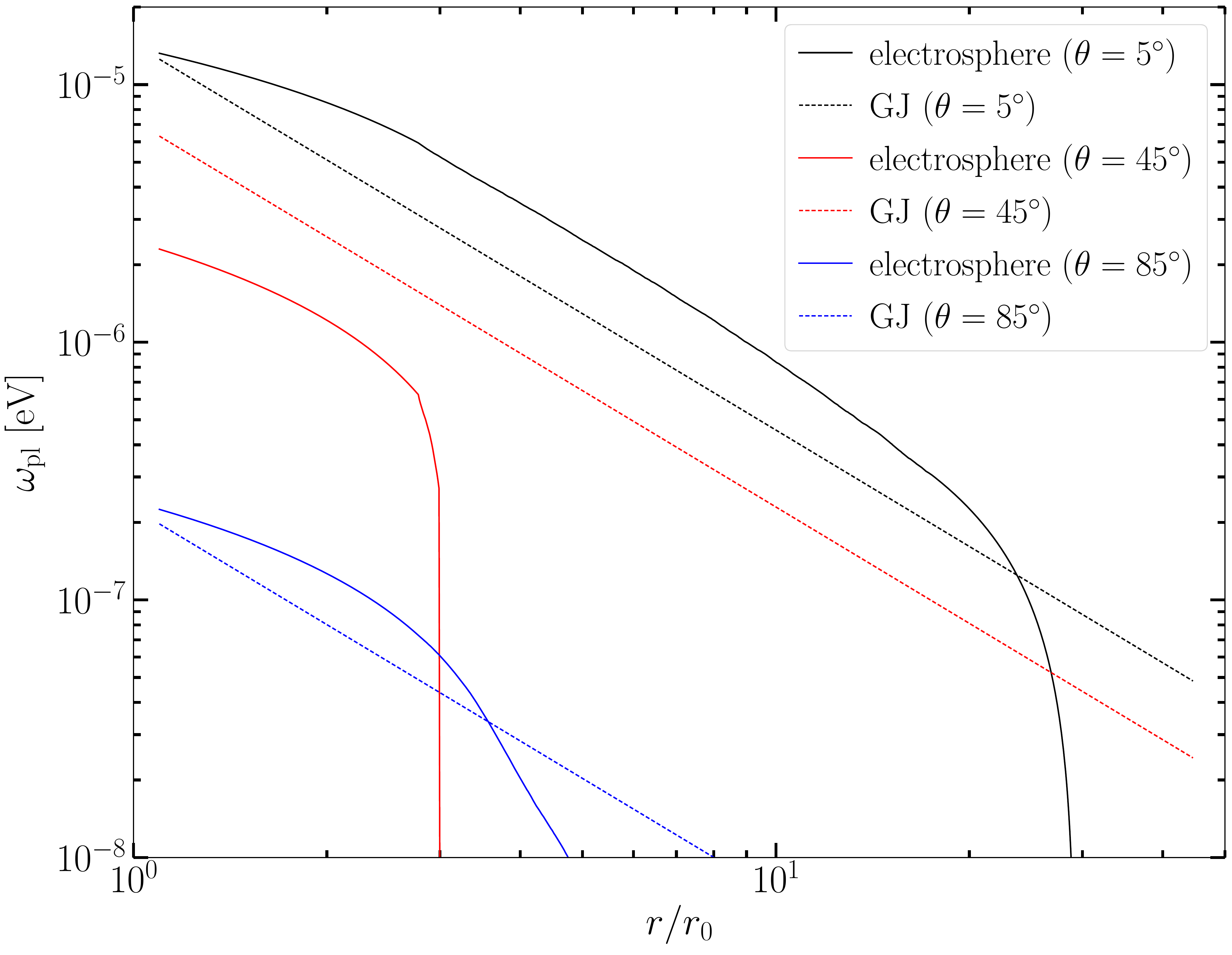} 
\includegraphics[width = 0.49\textwidth]{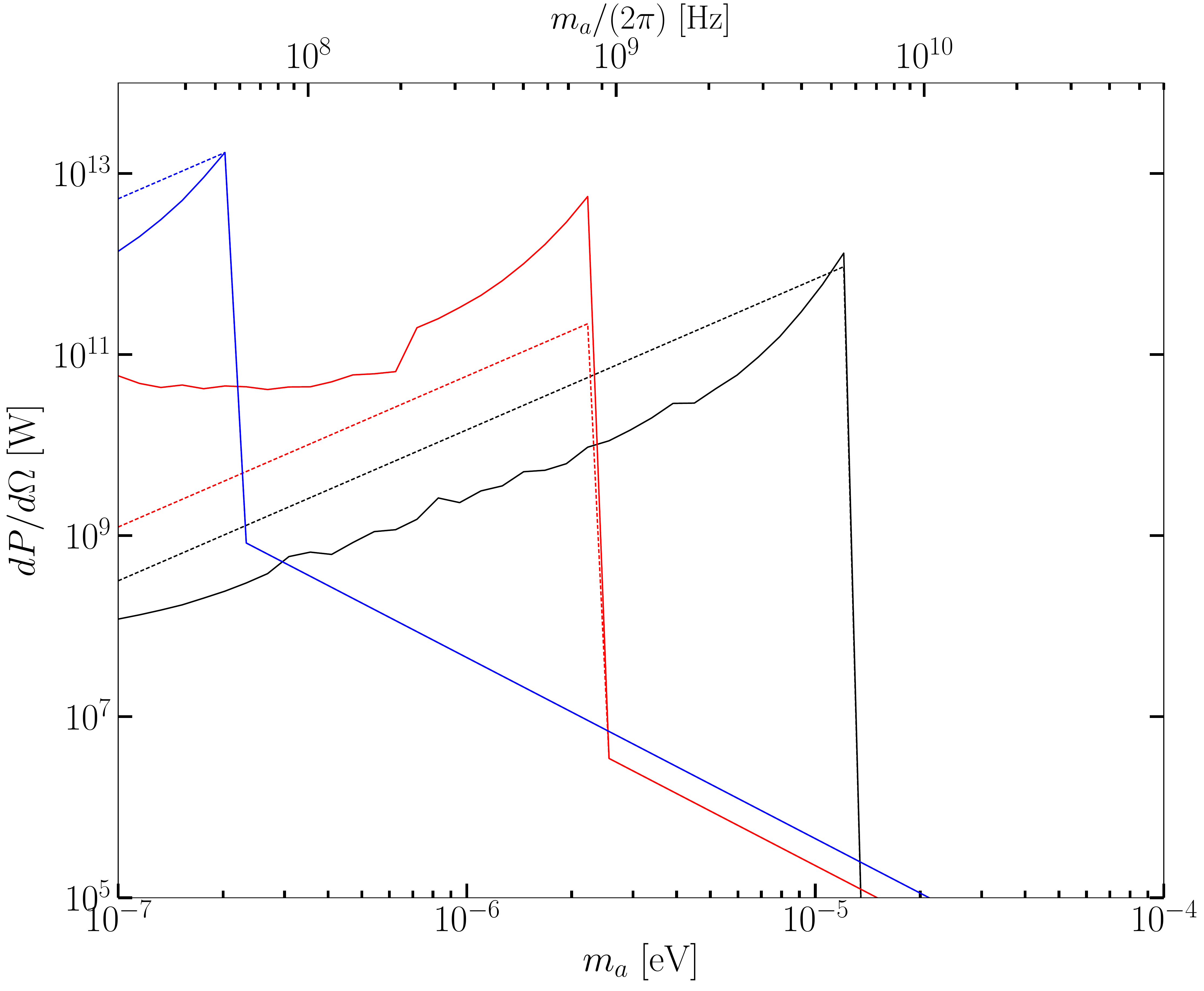}
\caption{(Left) A comparison between the plasma frequency profiles in the electrosphere model and the GJ model for three different polar angles $\theta$, which indicate the alignment angle between the NS and Earth.  For illustration, these figures assume NS properties similar to the nearby isolated NS J0806.4-4123 ($P \approx 11$ s, $B \approx 2.5 \times 10^{13}$ G~\cite{Kaplan:2009ce}).  (Right) The radiated power for the example NS illustrated in the left panel and assuming $g_{a \gamma \gamma} = 10^{-12}$ GeV$^{-1}$. This estimate includes both resonant and non-resonant conversion, which is responsible for the low-power tails at high masses.  
}
\label{fig: plasma}
\vspace{-0.5cm}
\end{figure*}

In the left panel of Fig.~\ref{fig: plasma} we plot the plasma frequency in the magnetosphere in the electrosphere model as a function of radius $r$ for three different angles ($\theta = 5^\circ, 45^\circ, 85^\circ$), and we compare these profiles to those in the GJ model.
For this example, we have taken the NS to have the parameters of the isolated NS J0806.4-4123; namely, we took $P \approx 11$ s, $B \approx 2.5 \times 10^{13}$ G, $M_\text{NS} \approx 1 M_\odot$, and $r_0 \approx 10$ km~\cite{Kaplan:2009ce}.  Changing between different NS models simply rescales the $y$-axis.
While the GJ model and electrosphere plasma frequency profiles are similar for $r \sim r_0$, it is important to note that at larger $r$, the electrosphere plasma profile falls faster than the GJ model.  This results in an increased flux at low masses, for some angles $\theta$, in the electrosphere model, as compared to the GJ model.  This is illustrated in the right panel of Fig.~\ref{fig: plasma}, which shows the power $dP/d\Omega$ as a function of mass for this NS and for the different angles $\theta$ illustrated in the left panel.  Note that we assume $g_{a \gamma \gamma} = 10^{-12}$ GeV$^{-1}$ for illustration.  The sharp cutoffs at large $m_a$ in these curves comes from the transition where the conversion radius fall below $r_0$ and resonant conversion no longer takes place.  At higher frequencies, non-resonant conversion still takes place, and an estimate for the non-resonant power is included in the figures.  However, the non-resonant-induced power is orders of magnitude below the resonant power, making the resonant power the most important contribution across the relevant frequency range. 

On a final note, the discussions and illustrations above have assumed that the NS rotation axis is parallel to the magnetization axis.  If there is an offset between these two, the flux as observed on Earth can acquire periodic time dependence~\cite{Hook:2018iia}.  We will not consider this possibility in this work, since we are mostly interested in the ensemble of signals over populations of NSs, for which the time dependence is washed out.  Moreover, older NSs will have less time dependence, as their alignment angles decay over time.  However, another possibility that needs to be considered is that the rotation axis is anti-aligned with the magnetization axis, as this should happen roughly half of the time.  In this case, the physics is exactly as discussed above, except that the positively and negatively charged regions are swapped according to {\it e.g.}~\eqref{GJ_charge}.  When constructing NS population models, as discussed below, we assign each NS a 50\% chance of having its magnetic field aligned parallel or anti-parallel to the rotation axis.

\section{Neutron Star Population model}
\label{sec: population-model}

In this section we construct a population model for NSs.  We build upon the significant effort that has gone into constructing population models for active pulsars (see, {\it e.g.},~\cite{FaucherGiguere:2005ny,2010MNRAS.401.2675P}), tuned to the available radio and $X$-ray datasets that probe these populations.  However, little attention has been paid to the properties of the  dead NSs, which have crossed below the pulsar death line and are no longer easily detectable by radio or even X-ray emission.  These dead NSs, though, greatly outnumber the pulsar population.  As we will show, depending on how their magnetic fields evolve over time, the dead NSs may be the dominant contributors to the axion-induced radio signal.
We construct two different models for the NS populations, which we label as model 1 and model 2, in order to begin to address the level of systematic uncertainty mismodeling the NS population has on the axion-induced radio flux.  The first model does not incorporate magnetic field decay while the second does.  

The initial conditions for our NS models follow closely Refs.~\cite{FaucherGiguere:2005ny} and~\cite{2010MNRAS.401.2675P}.  Our NS model 1 uses the results of~\cite{FaucherGiguere:2005ny}, while our NS model 2 uses~\cite{2010MNRAS.401.2675P}.  
The key difference between these two models is that model 2 allows the NS magnetic fields to decay over time.  Both analyses yield similar populations of active pulsars, as we will show, but there are important differences in the populations of dead NSs.  Both models, however, share a similar framework.  A population of NSs is produced with an initial distribution of magnetic fields, spin periods, alignment angles, and ages.  The NSs are evolved until the present day.  Following Refs.~\cite{FaucherGiguere:2005ny,2010MNRAS.401.2675P}, we then distribute the NSs within the Galactic disk and perform simulated radio observations consistent with the main surveys that make up the ATNF pulsar catalog~\cite{2005AJ....129.1993M}, and we confirm that the properties of the simulated pulsar catalogs match those of the real ATNF catalog.
 Below, we summarize the key features of these models, show they there are able to reproduce the observed properties of pulsars in the ATNF pulsar catalog, and discuss how the properties of the dead NSs differ between the models.  

\subsection{Neutron star birth properties}

In this subsection, we summarize the distributions of NS initial conditions taken in models 1 and 2.  
Since pulsar lifetimes tend to be much less than the age of a galaxy like the Milky Way, the precise form of the NS birth rate is important for properly calculating the ratio of active to dead NSs.  However, for definiteness we assume that the NS birth rate has been constant over the age of the galaxy, which is $\sim$$13 \times 10^9$ yrs for the Milky Way.  All NS masses and radii are fixed to $M_\text{NS} = 1 M_\odot$ and $r_0 = 10$ km, respectively, as small deviations from these values do not significantly affect the axion-induced radio flux.

\begin{table}[htb]
  \centering
  \caption{Initial distribution parameters for the magnetic fields and periods of NS models 1 and 2.  Note that model 2 includes magnetic field decay, while model 1 does not.  These parameters are taken from Refs.~\cite{FaucherGiguere:2005ny} and~\cite{2010MNRAS.401.2675P} for models 1 and 2, respectively.}
  \label{tab: IC}
  \begin{tabular}{l|c|c|c|c|c|c}
     & $\langle \log (B_0/G) \rangle$ & $\sigma_{\log (B_0/G)}$ & $\langle P_0 \rangle$ [s] & $\sigma_{{P_0}}$ [s]  \\
    \hline
    \hline
    Model 1 & 12.95 & 0.55 & 0.3 & 0.15 \\
    \hline
    Model 2 & 13.25 & 0.6 & 0.25 & 0.1\\
    \hline
  \end{tabular}
\end{table}

 The initial magnetic field distribution of the NSs is taken to follow a log-normal distribution with central value $\langle \log B_0 \rangle $ and standard deviation $\sigma_{\log B_0}$, so that
\es{}{
p(B_0) = {1 \over \sqrt{2 \pi} \sigma_{\log B_0}} e^{-{(\log B_0 - \langle \log B_0 \rangle)^2 \over 2 \sigma_{\log B_0}^2} } 
}
represents the probability to find initial field value $B_0$ at the magnetic pole.\footnote{In this section, all magnetic field values will refer to the values at the poles, and $B_0$ will be reserved for the initial field value at birth.  Explicitly, $B(t)$ will give the magnetic field value at the pole a time $t$ after birth.}  The best-fit values for $B_0$ and $\sigma_{\log B_0}$ as found in Refs.~\cite{FaucherGiguere:2005ny} and~\cite{2010MNRAS.401.2675P} for models 1 and 2, respectively, are given in Tab.~\ref{tab: IC}.  Note that~\cite{FaucherGiguere:2005ny} quotes magnetic fields in terms of those at the midplane, which are half as large as the fields at the poles.  In Tab.~\ref{tab: IC}, all field values have been converted to those at the poles.  Also note that the central field value $\langle B_0 \rangle$ is larger for model 2 than model 1 because model 2 includes field decay.

While Ref.~\cite{2010MNRAS.401.2675P} included magnetic field decay in their analysis, it is important to note that the form of their field decay differs from the form we consider in this work.  This is because~\cite{2010MNRAS.401.2675P} concentrated on time scales much shorter than those considered in this work and thus did not consider possible late-time effects that could cause NS magnetic fields to decay on time scale  $\sim$$10^{10}$ years.  Still, we find that on the $1$--$100$ Myr timescale, our decay model is a reasonable match to that taken in~\cite{2010MNRAS.401.2675P}, which justifies our use of their initial conditions.

The distribution of initial spin periods $P_0$ is taken to follow a normal distribution with central value $\langle P_0 \rangle$ and standard deviation parameter $\sigma_{P_0}$.  Note that only positive values of $P_0$ are considered.
The best-fit values for $\langle P_0 \rangle$ and $\sigma_{P_0}$ for models 1 and 2 are given in Tab.~\ref{tab: IC}.

The initial misalignment angle $\alpha_0$ between the NS rotation and magnetic axis is taken to be randomly distributed, following the simple geometric distribution $p(\alpha_0) = \sin(\alpha_0)$.  This initial alignment angle distribution, however, will turn out to play only a minor role in determining the properties of the final NS population, as we will discuss.    

\subsection{Neutron star evolution}

The NS spindown for active pulsars is a complicated dynamical question that involves a careful consideration of both the dipole radiation and the plasma effects (see, {\it e.g.},~\cite{FaucherGiguere:2005ny,Spitkovsky:2006np,Philippov:2013aha,Johnston:2017wgm}).   If the pulsar is inactive, then it is assumed to spin down through dipole radiation alone.  In this case, the period and alignment angle evolve through the differential equations~\cite{1970ApL.....5...21M,Philippov:2013aha}
\es{dipole_eom}{
P(t) P'(t) = {2 \over 3} {P_0^2 \over \tau_0} \sin^2 \alpha(t) \\ 
{d \over dt} \log \sin \alpha(t) = - {2 \over 3} {\cos^2 \alpha_0 \over \tau_0} \,,
}
where\footnote{Note that we use SI units, and not the more conventional Gaussian units, to describe the pulsar magnetic moment.} 
\es{}{
\tau_0 = {I \over \pi \mu^2 f_0^2} \approx 10^4 \left( { 10^{12} \, \, \text{G} \over B } \right)^2 \left( {P_0 \over 0.01} \right)^2 \, \, \text{yr} \,.
}
Above, we have taken the canonical value $I = 10^{38}$ kg m$^2$ for the moment of inertia and related the magnetic moment $\mu$ to the magnetic field at the pole $B$ by $\mu = 2 \pi B r_0^3$.  The alignment angle approaches zero exponentially, 
\es{alpha_evolve}{
\sin \alpha(t) = \sin \alpha_0 e^{- {2 t \cos^2 \alpha_0 \over 3 \tau_0} } \,,
}
while the pulsar period evolves according to 
\es{dipole-spindown}{
P(t) = P_0 \sqrt{1 + \tan^2 \alpha_0 \left( 1 - e^{- {4 t \cos^2 \alpha_0 \over 3 \tau_0}} \right) } \,.
}
Note, however, that since the period evolution stops as $\alpha(t) \to 0$, as may be seen in~\eqref{dipole_eom}, it is difficult in this case for the final pulsar period to be significantly larger than the initial period $P_0$.  

It is thought that active pulsars lose energy not only by dipole radiation but also by plasma effects, which cause the spin-down luminosity to remain non-zero as $\alpha(t) \to 0$, yielding significantly longer final pulsar periods.  For example,~\cite{Spitkovsky:2006np} found that in numerical simulations of the pulsar magnetosphere in the ``force-free" limit, the spin-down luminosity may be described by 
\es{}{
L_\text{pulsar} \approx {\mu^2 \Omega^4 \over 4 \pi} \left(1 + \sin^2 \alpha \right) \,,
}
which differs qualitatively from the vacuum formula
\es{}{
L_\text{vac} = {2 \over 3} {\mu^2 \Omega^4 \over 4 \pi} \sin^2 \alpha \,,
}
used to derive~\eqref{dipole-spindown}, in the limit $\alpha \to 0$.  In particular, the plasma-filled pulsar continues to spin down even when $\alpha \to 0$, while this is not the case for the NS in vacuum that spins down only from dipole radiation. This is important because if the NS continues to spin down even as $\alpha \to 0$, then much longer periods are obtainable.  We model this effect following~\cite{FaucherGiguere:2005ny} by modifying the period spin-down equation for active pulsars from~\eqref{dipole_eom} to 
\es{dipole_eom}{
P(t) P'(t) = {2 \over 3} {P_0^2 \over \tau_0} \,, \\ 
}
which, assuming a constant magnetic field, yields the solution
\es{period_simp}{
P(t) = P_0 \sqrt{1 + {4 t \over 3 \tau_0} } \,.
}

For our NS model 1, we model the period evolution using~\eqref{period_simp} until the pulsars reach the death line and become inactive. 
 For the pulsar death line, we use the approximation that active pulsars have $B/P^2$ greater than~\cite{1992A&A...254..198B} 
\es{}{
{B \over P^2} = 0.34 \times 10^{12} \, \, \text{G s}^{-2} \,,
}
where again we emphasize that $B$ refers to the magnetic field at the pole.  While the pulsars are active, the alignment angles should also evolve according to~\eqref{alpha_evolve}.  The result is that by the time the pulsars reach the death line, most of the alignment angles are close to aligned.  After the pulsars reach the death line, we continue to evolve them until today using the pure dipole spin-down equations.  However, we find that this post-death evolution makes little difference, given~\eqref{dipole-spindown} and that $\tan^2\alpha_0$ after the death line is close to zero for most NSs.  

Model 2 proceeds in an analogous manner, except now the magnetic field evolves over time.  In general, magnetic fields in NSs decay over time through one of three main channels: Ohmic dissipation, ambipolar diffusion, and Hall drift~\cite{1992ApJ...395..250G}. Ohmic dissipation is simply the effect of small but finite conductivity inside the NS, which slowly dissipates magnetic field through Ohmic heating whenever there is nonzero current. Ambipolar diffusion is the net drift motion of electrons and protons with respect to the neutron fluid. The motion is driven by magnetic stress and creates local drag between the charged particle species and the neutrons, as well as breaking chemical equilibrium between the species and driving weak interactions, dissipating magnetic energy. Hall drift does not dissipate energy directly, but it can tangle the magnetic field and enhance Ohmic dissipation especially in the NS crust.

The question of how the magnetic fields in NSs evolve is far from settled, and answering this question involves performing detailed simulations (see, {\it e.g.},~\cite{Cumming:2004mf,Aguilera:2007xk,2014MNRAS.438.1618G,Beloborodov:2016mmx,Bransgrove:2017jzs}).
{In this work, we take as a starting point a simple parametric model to roughly capture the physics of magnetic field decay described above, combining some recent theory as well as simulation results.}

If a NS is born with a high magnetic field ($B\gtrsim 10^{16}\,\mathrm{G}$) and high temperature ($T_\mathrm{core}\gtrsim 10^9\,\mathrm{K}$), the dominant process to dissipate magnetic energy in its core is ambipolar diffusion, as the rate of this process scales as $B^2$ \cite{1992ApJ...395..250G}:
\begin{equation}
t_\mathrm{ambip} \sim 3\times 10^9\frac{T_8^2L_5^2}{B_{12}^2}\,\mathrm{yr} \,,
\label{ambip}
\end{equation}
where $T_8 \equiv T_\text{core} / 10^8 \, \, \text{K}$, $L_5 = L / 10^5 \, \, \text{cm}$ with $L$ the thickness of the crust, and $B_{12} = B / 10^{12} \, \, \text{G}$ with $B$ the magnetic field at the pole.
According to \cite{Beloborodov:2016mmx}, the temperature of the NS core rapidly cools to $T_\text{core} \sim 10^9\,\mathrm{K}$ in $\sim 1\,\mathrm{yr}$ even if it is born with a much higher temperature. Furthermore, the decay of the magnetic field through ambipolar diffusion in the core will heat the star, bringing the temperature to a plateau for $\sim 1\, \,\text{kyr}$. Therefore we assume that all NSs that are born with high magnetic fields evolve according to $t_\mathrm{ambip}$, until the timescale becomes comparable to the dissipation timescale of the field in the NS crust.  We take $T_\text{core} \approx 10^8 ( {10^6 \, \, \text{yr} / t})^{1/6}$ K, where $t$ is time since the NS birth~\cite{1983bhwd.book.....S}.  This cooling law arises from assuming that the NS cools according to Urca neutrino emission~\cite{1983bhwd.book.....S}.

{When ambipolar diffusion is no longer efficient, the evolution of the NS magnetic field is dominated by the drift of magnetic flux tubes from the core to the crust, which are then dissipated in the crust through a combination of Hall drift and Ohmic heating \cite{2006MNRAS.365..339J}. Detailed numerical simulations of this process were carried out recently by \cite{Bransgrove:2017jzs}, which found that the time scale for this process is $\tilde{t}_\mathrm{ohm}\sim 150\,\mathrm{Myr}$ for young NSs with higher crust temperatures, and can be as large as $\sim$$1.8\,\mathrm{Gyr}$ when the crust has cooled down to $T_\text{crust} \sim 10^6\,\mathrm{K}$. The actual decay timescale at low temperatures depends on the level of impurities in the crust $Q_\mathrm{imp}$:
\begin{equation}
t_\mathrm{ohm} \sim \frac{1.8\,\mathrm{Gyr}}{Q_\mathrm{imp}}.
\label{impur}
\end{equation}
Note that NSs with $Q_\mathrm{imp} \lesssim 0.13$ will decay on the Hubble timescale, which means their magnetic field will be frozen out for the purpose of our work.  
Estimates for $Q_\mathrm{imp}$ in the literature span the range from $10^{-3}$~\cite{1977ApJ...215..302F} to $10$~\cite{doi:10.1046/j.1365-8711.2001.03990.x}.
In our simplified model, we simply take the low-temperature impurity-dominated timescale in~\eqref{impur} and consider a log-flat distribution for $Q_\mathrm{imp}$ over the range $[10^{-3}, 10]$.  That is, for each NS in our simulation, we randomly assign a value $Q_\mathrm{imp}$ from this distribution.  

Explicitly, we solve for the $B$-field evolution using the equation
 \es{Bevol}{
 {d B \over dt} = -B \left( {1 \over \tau_\text{ohm}} + \left({B \over B_0} \right)^2 {1 \over \tau_\text{ambip}} \right) \,,
 }
 where $\tau_\text{ambip}$ is the decay rate for ambipolar diffusion~\eqref{ambip}, evaluated with $B = B_0$, and $\tau_\text{ohm}$ is the timescale for Hall drift and Ohmic heating with impurities~\eqref{impur}.  We fix the crust thickness at $L = 10^5$ cm. In Fig.~\ref{fig: B_evol}, we show examples of the B-field evolution in our decay model, for a range of initial magnetic field values.  Note that in constructing these curves we have averaged of a population of NSs with randomly assigned impurities as discusses above.

\begin{figure}[htb]
\includegraphics[width = 0.49\textwidth]{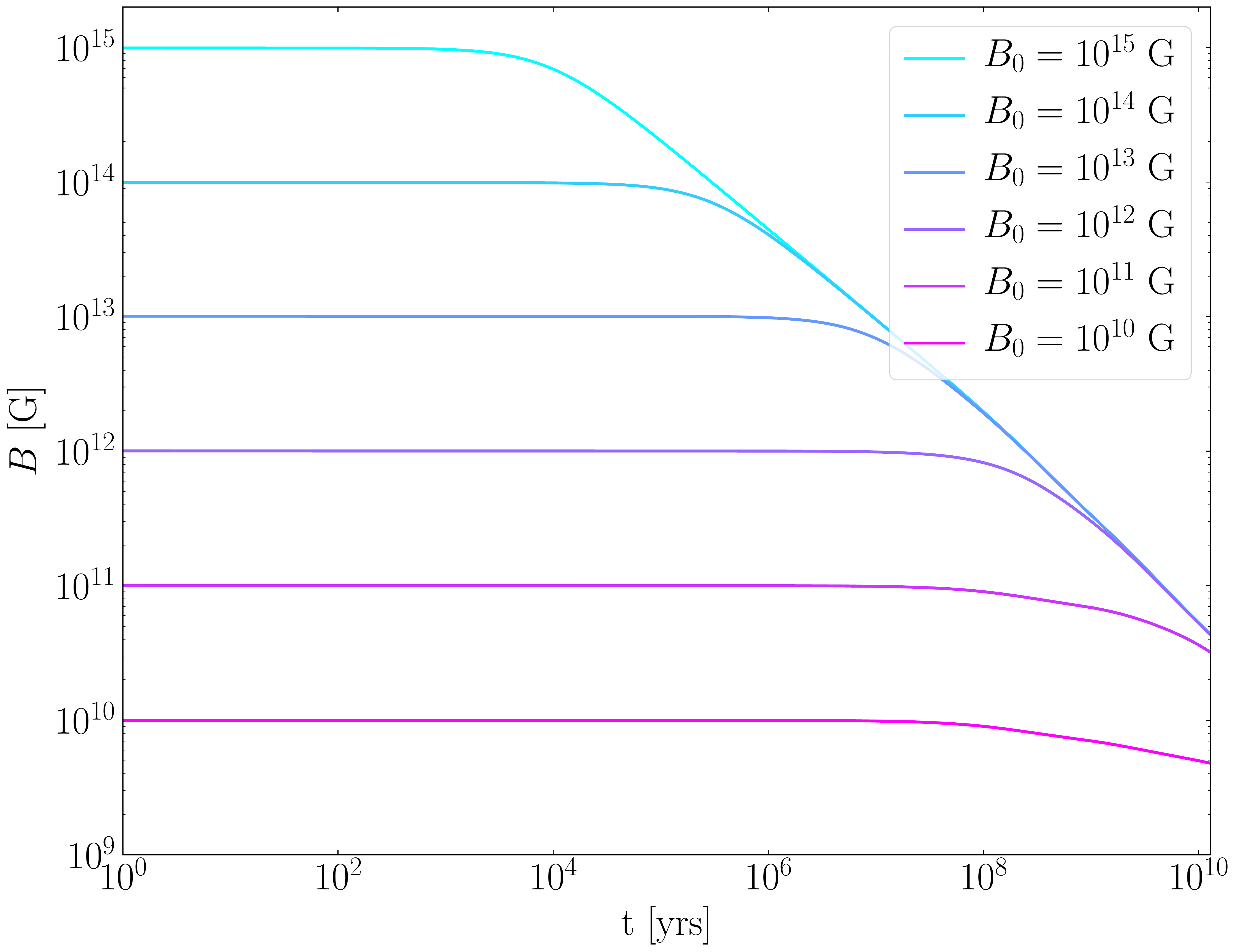}
 \vspace{-0.7cm}
\caption{Magnetic field evolution in NS model 2, averaged over a population of NSs with randomly assigned impurities from a log-flat distribution over $[10^{-3}, 10]$.  Larger initial magnetic fields cause the magnetic field to decay sooner as a result of ambipolar diffusion.  At late times, almost all magnetic fields are at $\sim$$10^{11}$ G or below in this model, which will lead to the result that active pulsars typically dominant the axion-induced radio flux in NS model 2.
}
\label{fig: B_evol}
\end{figure}    

\subsection{Population simulations and comparison to ATNF pulsars}

Following the procedure above, we generate a large ensemble of NSs for both models 1 and 2, keeping track of the active pulsars and the dead NSs.  For the active pulsars, it is useful to cross-check the simulations by comparing them to the measured properties in the ATNF catalog.  Of course, since the works~\cite{FaucherGiguere:2005ny,2010MNRAS.401.2675P} tuned their models to the measured pulsar properties in the $P$-$\dot P$ plane, it is unsurprising that we find good agreement between the simulated pulsar measurements and the observed pulsar properties.  Still, this is a useful check to both validate the models and also understand how the measured pulsar properties differ from those of the pulsar population as a whole and from the populations of dead NSs.

We simulate ATNF pulsar catalogs from the simulated NS populations following a simplified version of the procedure outlined in~\cite{FaucherGiguere:2005ny}.  First, we distribute the simulated pulsars across the Galactic disk using the spatial distribution described in Sec.~\ref{sec: NS-DM}.  Note that, for simplicity, we do not account for the spiral arms structure.  For each NS, we model the intrinsic radio luminosity $L$ at $1.4$ GHz using the best-fit empirical model from~\cite{FaucherGiguere:2005ny}, for which 
\es{}{
\log L = \log \left[ L_0 P^{\epsilon_P} \left( {\dot P \over 10^{-15}} \right)^{\epsilon_{\dot P}} \right] + L_\text{corr} \,.
}
Here, $L_\text{corr}$ is randomly drawn for each NS from a Gaussian distribution with standard deviation $\sigma_{L_\text{corr}}$.  Ref.~\cite{FaucherGiguere:2005ny} found best-fit parameters $L_0 = 0.18$ mJy kpc$^2$, $\epsilon_p = -1.5$, $\epsilon_{\dot P} = 0.5$, and $\sigma_{L_\text{corr}} = 0.8$, and we use these parameters in our simulations.  We note that the period derivative $\dot P$ may be inferred using~\eqref{dipole_eom}, which implies that 
\es{B_to_PdotP}{
B \approx 6.4 \times 10^{19} \sqrt{ {P \dot P} \over \text{s} } \text{  G} \,.
}

After simulating the luminosity of each pulsar, we may calculate the flux observed at Earth.  We then simulate observations with the Parkes~\cite{Manchester:2001fp} and Swinburne~\cite{2001MNRAS.326..358E} Multibeam surveys at 1.4 GHz, which make up most of the ATNF catalog.  The observational parameters for these surveys, such as latitude and longitude ranges, integration times, sampling intervals, bandwidths, and signal-to-noise thresholds, are summarized in~\cite{FaucherGiguere:2005ny}.  We also follow~\cite{FaucherGiguere:2005ny} in accounting for the fraction of pulsars that are beamed towards Earth and in accounting for the degradation to the minimum detectable spectral flux density by the dispersion measure.  To calculate the dispersion measure, we use the free electron parameterization with position along the Galactic disk presented in~\cite{Cordes:2002wz}. 

\begin{figure*}[htb]
\includegraphics[width = 0.49\textwidth]{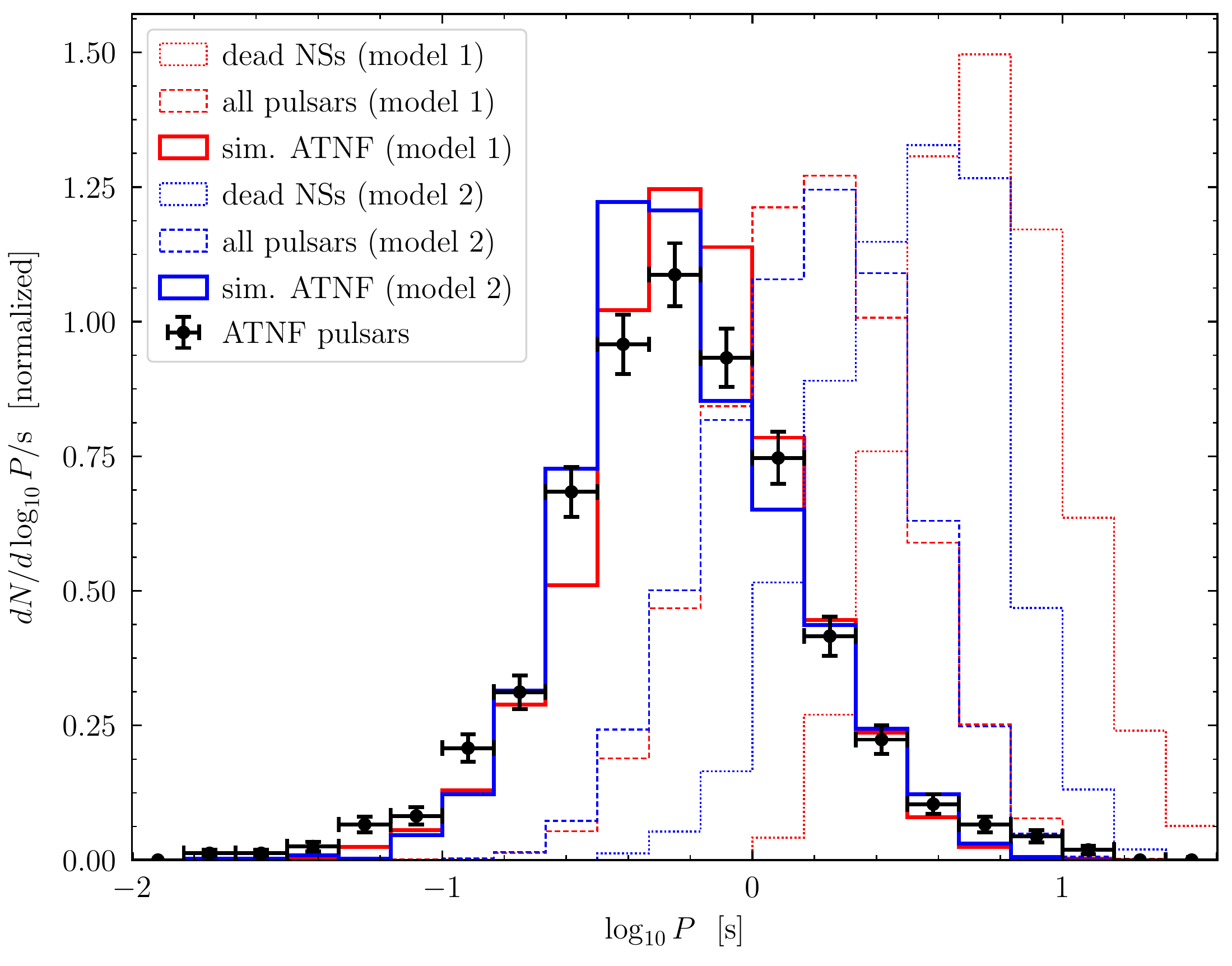} \includegraphics[width = 0.49\textwidth]{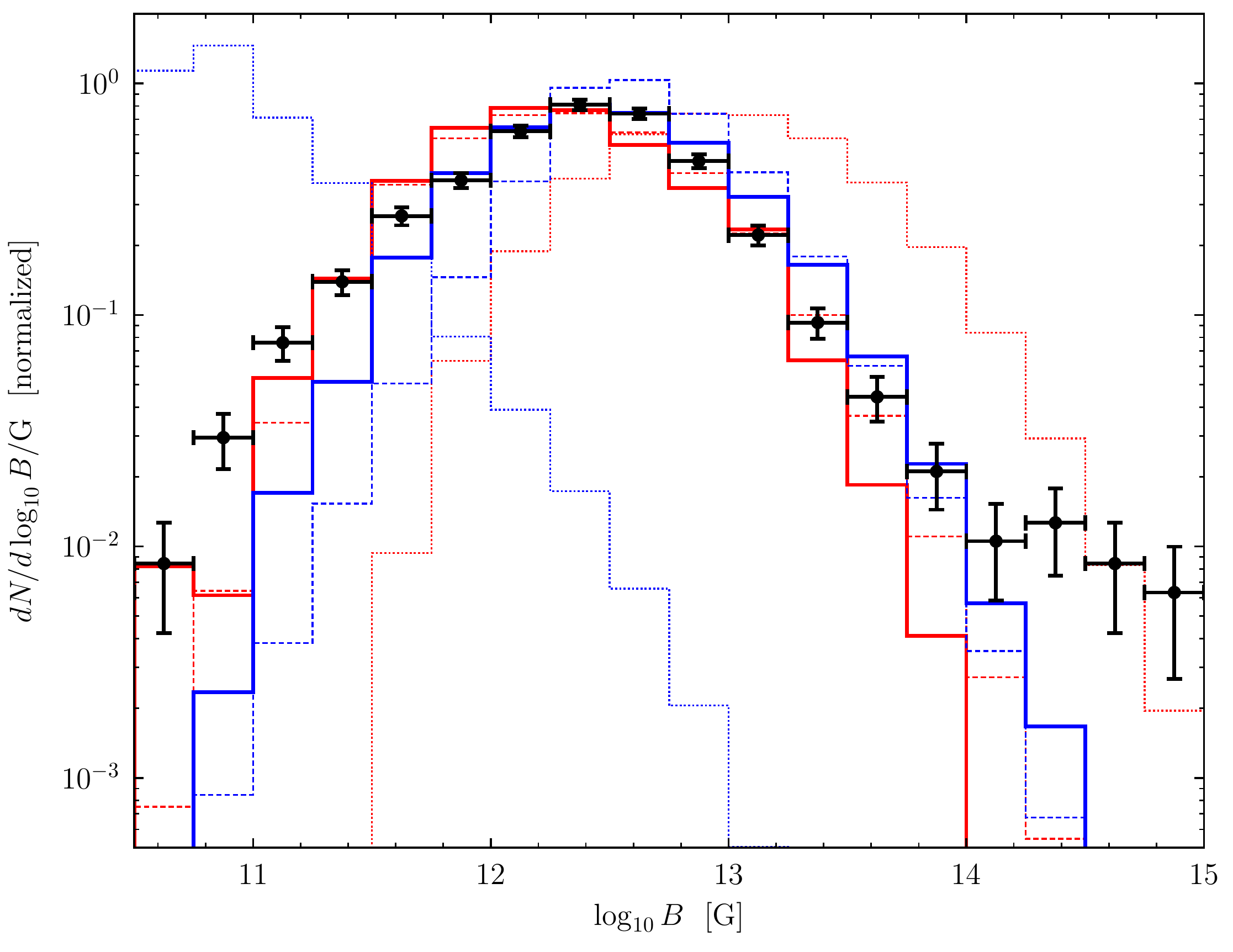}
\caption{(Left) Histogram of the NS periods for the ATNF catalog and example simulations.  The data values include Poisson 1$\sigma$ error bars from finite statistics.  We compare the data to simulated ATNF catalogs, for both NS models.  The simulated catalogs are a good match to the observed pulsar periods.  On the other hand, we also show the period distributions for the whole populations for active pulsars and dead NSs.  These distributions are centered at much higher periods than those for the observed pulsars.  (Right) As in the left panel, but for the magnetic fields at the poles.  The simulated catalogs are a reasonable match to the ATNF catalog, except that both models underpredict the number of high-field magnetars.  The biggest difference between NS models 1 and 2 is seen in the magnetic field distributions of the dead NSs; the distribution for model 1 extends to high field values, while that of model 2 is shifted to lower values as the result of field decay.
}
\label{fig: ATNF}
\end{figure*}
There are $\sim$1941 pulsars in the ATNF pulsar catalog that are not identified as belonging to binary systems and which are reported with positive periods.  In Fig.~\ref{fig: ATNF} we histogram their observed periods and magnetic field values, where we note that the magnetic field values at the poles are inferred using~\eqref{B_to_PdotP}.  Note that we also display Poisson 1$\sigma$ error bars on the histogram values to account for the finite statistics in inferring the true underlying distribution of observed pulsars.  In those figures, we also show example simulated pulsar catalogs for models 1 and 2.  In performing those simulations, we did not attempt to reproduce the number of pulsars observed in the ATNF catalog.  Rather, we sequentially simulated NSs and their observations one at a time until the number of observed pulsars in our simulated catalogs matched the number in the ATNF catalog. 

In Fig.~\ref{fig: ATNF} we compare the real ATNF catalog to example simulated ATNF catalogs, the samples of all active pulsars, and the sample of all dead NSs, for both models 1 and 2.  The distributions of observed periods and magnetic fields are seen to be a reasonably good match to the simulated ATNF catalog for both models. 
In both models, the simulated observed pulsar periods are systematically lower than the periods for the simulated pulsar populations as a whole.  This is likely due to the fact that lower pulsar periods have a higher luminosity within these models and are thus more likely to be detected.  On the other hand, the observed pulsar magnetic fields appears to be a relatively unbiased tracer for the population of active pulsars as a whole, within these models.  With that said, both models underpredict the number of low and high-field pulsars.  The high-field tail in particular, which is observed in the data but not in our simulations, could be important, because these pulsars are the most efficient at axion-photon conversion.  It is unclear, however, whether the lack of observed magnetars predicted in our simulations arises from un-modeled detector bias, for example in the luminosity of such objects, or in the intrinsic distribution of magnetic fields.   

While the properties of the active pulsars are relatively similar between the two models, the properties of the dead NSs differ substantially between the two models and relative to the active pulsars. 
The dead NSs tend to have higher spin periods than the active pulsars in both models.  This is likely a results of the fact that the dead NSs have had more time to evolve to higher spin periods.  In model 1, the dead NSs typically have larger magnetic fields than the active pulsars.  This is because higher initial magnetic fields cause the NSs to die sooner.  On the other hand, in model 2 the dead NSs have low field values because after the NSs die, their fields continue to decay within this model.  Since we assume a constant birthrate within these models over $\sim$$13 \times 10^9$ years, almost all of the dead NSs have fields $\lesssim 10^{11}$ G, except for those which happen to be relatively young.

The differences between the dead NS populations for models 1 and 2 have signifiant implications for radio searches for axions.
We illustrate this point by calculating, for each NS, the maximum mass attainable through resonant conversion within the GJ model.  That is, taking $\theta = 0^\circ$ in~\eqref{rc}, we solve for the $m_a$ such that $r_c = r_0$, with $B_0$ and $P$ fixed to the parameters for the NS under consideration.  Explicitly, the maximum mass detectable through resonant conversion within the GJ model is 
\es{GJ-maximum-mass}{
m_a^\text{max.} \approx 9.9 \times 10^{-5} \sqrt{ {B_0 \over 10^{14} \, \, \text{G}} {1 \, \, \text{s} \over P} } \, \, \text{eV}\,.
}     

In Fig.~\ref{fig: ma_max}, we illustrate example distributions of $m_a^\text{max.}$ over the dead NS and active pulsar populations found in realizations of models 1 and 2.
\begin{figure}[htb]
\includegraphics[width = 0.5\textwidth]{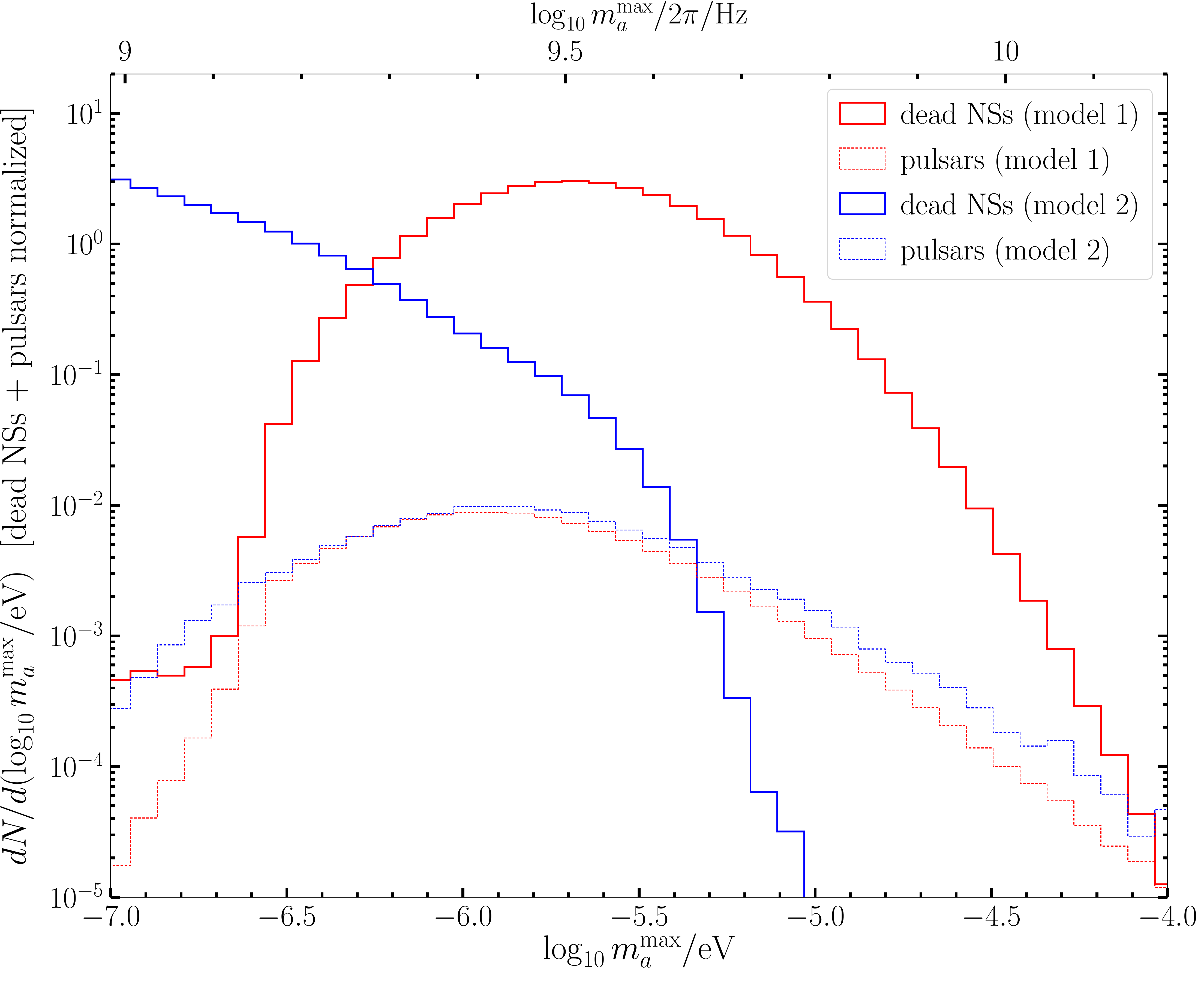}
\caption{Distributions of maximum attainable axion masses $m_a^\text{max.}$ through resonant conversion within the GJ model for NS models 1 and 2.  We show both the active and dead NS populations, and unlike in Fig.~\ref{fig: ATNF}, here we normalize the combination of both populations, with the relative normalization taken to be $\sim$$0.4$\%, which is the fraction of NSs expected to be active pulsars in these models with a constant birthrate over the age of the galaxy.  While the active pulsar properties are similar between the models, the dead NS population in model 2 only probes low-mass axions.
}
\label{fig: ma_max}
\end{figure}
The normalization of the distributions is such that the integrated distribution of active pulsars and dead NSs adds to unity, but the relative difference in normalizations between the active and dead NS populations reflects the fraction of NSs that are active versus dead.  In particular, for both models we find that $\sim$$0.4$\% of the NSs are active pulsars, though we emphasize that this fraction depends sensitively on our assumption that the NS-formation rate has been constant over the past $\sim$$13\times10^9$ years.

The distributions of $m_a^\text{max.}$ agree well across the entire mass range for the active pulsars between models 1 and 2.  This is perhaps not surprising, since the NS models were tuned to match the properties of the active pulsars.  However, the distributions of $m_a^\text{max.}$ for the dead NS populations are seen to behave significantly different between the two models.  In model 1, the distribution of $m_a^\text{max.}$ is maximum around $\log_{10} m_a^\text{max} / \text{eV} \approx -5.5$, and the distribution extends all the way to masses $m_a^\text{max} \approx 10^{-4}$ eV.  On the other hand, the dead NS $m_a^\text{max.}$ distribution for model 2 is skewed towards lower values.  This is because of the lower field values in NS model 2 for the dead NSs.  The NSs with higher values of $m_a^\text{max.}$ not only allow for higher-mass axion models to be tested, but also these NSs tend to dominate the sensitivity at lower masses, due to the increased magnetic fields.

\subsection{Impact of NS models on axion dark matter sensitivity}

In this subsection, we quantify how the different NS models affect the sensitivity to axion DM.  
To estimate the sensitivity of a radio telescope to a potential axion signal, it is not sufficient to only know the flux but also we need to know the width of the signal in frequency space.  This question will be addressed in more detail later in this work, but for now we follow~\cite{Hook:2018iia} and assume that the width of the signal from an individual NS is $\delta f / f \sim (v_0 / c)^2$, where $v_0$ is of order the local virial velocity.  Given the bandwidth $\delta f$, we may calculate the flux density $S \equiv F/\delta f$, where $F = dP/d\Omega / d^2$ is the flux and $d$ is the distance between the NS and Earth.  
  
  To illustrate the difference between the NS models, we take a reference NS located at $d = 250$ pc from Earth and assume $g_{a \gamma \gamma} = 10^{-12}$ GeV$^{-1}$, along with $\delta f / f \approx (200 \, \, \text{km}/\text{s} / c)^2 \approx 5 \times 10^{-7}$.  Note that the central frequency of the signal is given by $f = m_a / (2 \pi)$. 
  We take our reference NS to have the average properties of the NS population.  That is, we generate a large number of NSs and calculate the average flux density over the population.  Of course this is not physical for a single NS located 250 pc away from Earth, but having the flux density for such a reference NS allows us to easily rescale the flux density to NS populations located, for example, at the Galactic Center or in other galaxies.  
  
  The results for the flux density as computed in NS models 1 and 2 are shown in Fig.~\ref{fig: model-comp}. 
\begin{figure*}[htb]
\includegraphics[width = 0.49\textwidth]{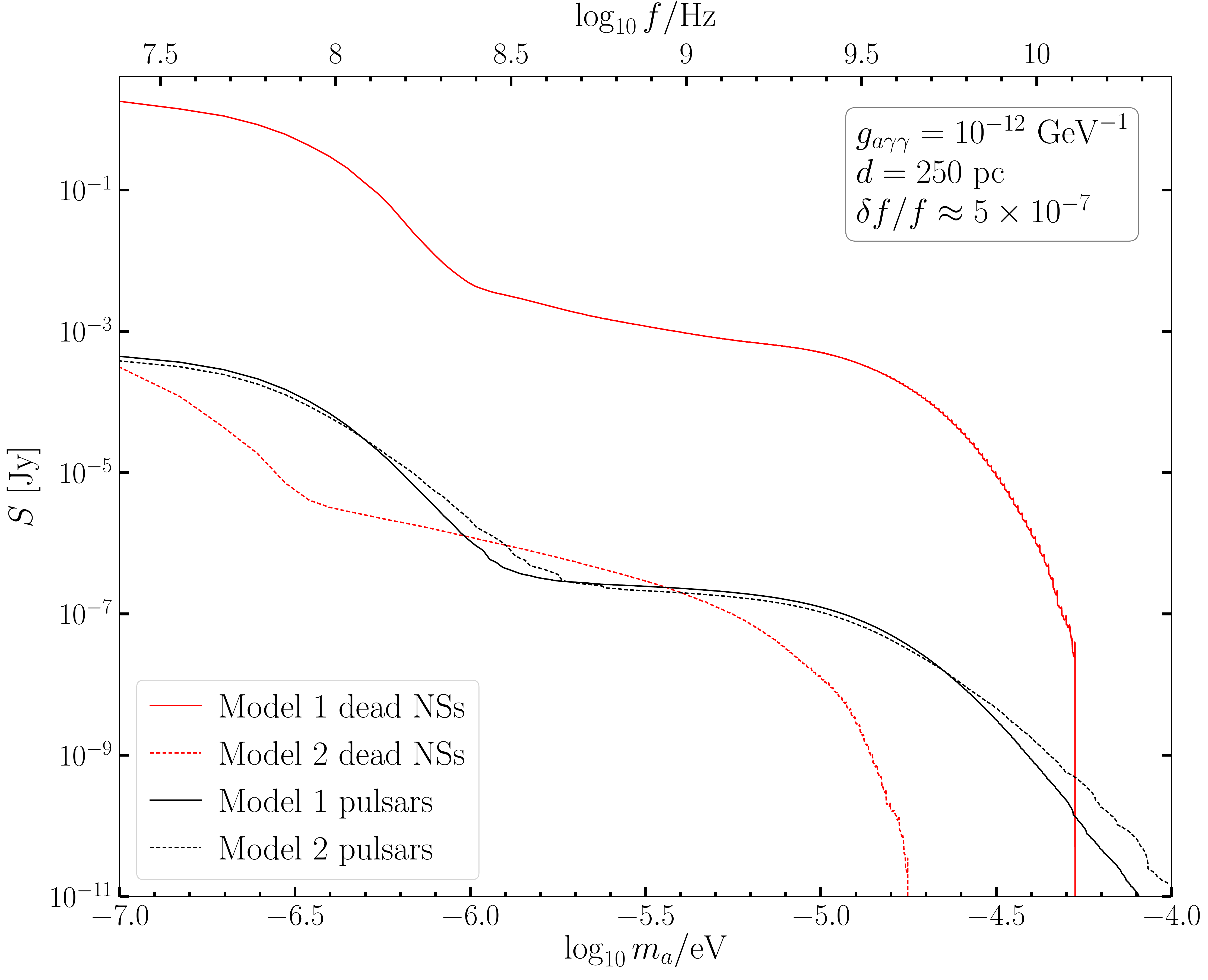}
\includegraphics[width = 0.49\textwidth]{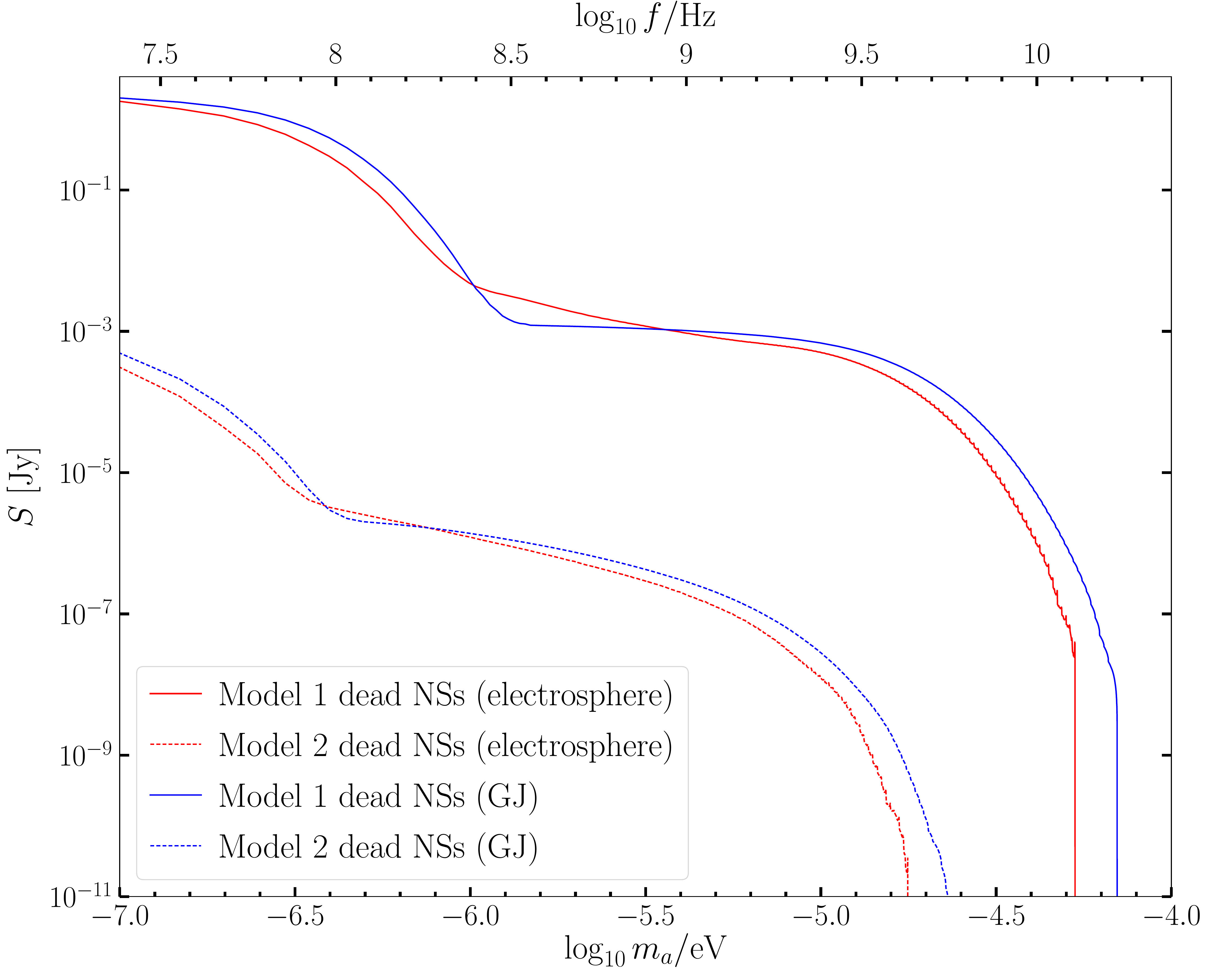}
\caption{ (Left) The flux density from a reference NS, with properties indicated, as a function of the putative axion mass $m_a$ in both NS models 1 and 2.  Note that this flux density is for a reference NS with properties given by averaging over the entire NS population; we show the contributions from the active pulsars and dead NSs separately.  In NS model 1, the dead NSs dominate the flux, while in model 2 the active and dead NSs produce similar flux levels at lower masses, with the active NSs dominating at higher masses.  (Right) As in the left panel, but here we only show the dead NS populations, and we compare the fluxes predicted within the GJ and electrosphere models for the magnetospheres.  The differences between the flux densities is minimal compared to the uncertainty between {\it e.g.} NS models 1 and 2 themselves. 
}
\label{fig: model-comp}
\end{figure*}
Note that we show the flux density for both the dead NS population and the pulsar population, with the relative normalization between the two fixed at the predicted $\sim$$0.4$\%.  That is, in averaging over the NS properties, we have assumed that the dead NSs and active pulsars are created in the predicted relative ratios assuming a constant birthrate over $\sim$$13 \times 10^9$ years.  
The two different NS models yield similar results for the active NS population.
On the other hand, the two different NS models yield qualitatively different results for the dead NS population.  In NS model 1, the dead NSs dominate the flux across almost all masses, while in NS model 2 it is instead the active pulsars that dominate the flux.  

In the left panel of Fig.~\ref{fig: model-comp} we have modeled the dead NSs by the numerical electrosphere solution and the active pulsars by the analytic GJ model.  In the right panel, we explore how sensitive the results are to this choice.  In particular, we compare how the dead NS flux-density curves change between the electrosphere and GJ models.  The two solutions are seen to give remarkably similar flux densities, when averaged over the NS populations, suggesting that the plasma-density profile is a sub-dominant source of uncertainty compared to, for example, the properties of the NS population model.  One noticeable difference between the two models is that the GJ model extends to higher frequencies than the electrosphere solution.  We caution that this may be a numerical artifact.  Due to numerical error, we are not able to extend the electrosphere all the way to the NS surface.  Thus, it is possible that the electrosphere plasma density continues to rise all the way to the NS surface, allowing for resonant conversion at the higher masses accessible by the GJ model.  Higher-resolution simulations would be needed to address this possibility.

\section{Neutron Stars and Dark Matter in the Milky Way}
\label{sec: NS-DM}
In this section we discuss the distribution of NSs and DM in the Milky Way for the purpose of computing the integrated signal from axion-photon conversion over the ensemble of NSs.  We then calculate the expected flux distribution within the inner Galaxy and the frequency spectrum of the putative signal.

\subsection{The Neutron Star Distribution}

We model three components of the NS distribution within the Milky Way: (i) the disk component, (ii) the bulge component, and (iii) the nuclear star cluster (NSC) component.  Of these contributions, we will see that the bulge component is the most important, unless there is a central DM spike, in which case the NSC can also play an important role.

Most of the NSs in the Galaxy are in the disk and bulge.  We assume that $\sim$$10^9$ NSs have been created in the Milky Way over its lifetime, with $\sim$60\% created in the bulge and $\sim$40\% in the disk~\cite{2009PASP..121..814O,2010A&A...510A..23S}.  The fact that the bulge contains less stellar mass than the disk but produced more NSs can be understood from the observation that the bulge stellar mass function is skewed towards heavier stars than the disk's mass function, and heavier stars more readily form NSs~\cite{2009PASP..121..814O}.   

\subsubsection{The Galactic disk}

We model the number density of NSs in the disk $n_\text{disk}$ 
by a double exponential profile:
\es{disk}{
n_\text{disk}(r,z) = {N_{\text{disk}} \over 4 \pi \sigma_r^2 \langle |z| \rangle}  e^{-{ r^2 \over 2 \sigma_r^2}} e^{- { |z| \over \langle |z| \rangle}} \,,
}
where $r$ is the cylindrical radius, $z$ is the height above the disk, and $N_\text{disk} \approx 4 \times 10^8$ is the total number of NSs in the disk.
Following~\cite{2010JCAP...01..005F}, which studied the distribution of millisecond pulsars in the disk, we take $\sigma_r = 5$ kpc and $\langle |z| \rangle = 1$ kpc.  Since millisecond pulsars are the oldest known pulsar population, their spatial distribution is likely a good proxy for the inactive and old NS populations.  It is possible that many of the isolated NSs born in the disk subsequently ended up in the stellar halo~\cite{2010A&A...510A..23S} due to natal velocity kicks during the supernova process, but we do not consider this possibility in detail because, as will be illustrated below, the disk contribution to the radio signal is subdominant compared to that from the bulge and the NSC.

\subsubsection{The Galactic bulge}

Given the lack of pulsars observed in the inner regions of the Milky Way, there is no direct measure of the distribution of isolated NSs in the bulge.  We thus simply assume that the isolated NS density distribution follows the distribution of matter in the bulge, though this need not be the case given the NS natal velocity kicks.  The bulge potential is assumed to take the simple, spherically-symmetric form~\cite{1990ApJ...356..359H} 
\es{bulge-pot}{
\Phi(R) = - {G M_\text{bulge} \over R + a} \,,
}
where $M_\text{bulge}$ is the bulge mass, $a \approx 0.6$ kpc is the bulge scale radius~\cite{1990ApJ...356..359H,2010A&A...510A..23S}, and $R$ is the distance from the Galactic Center.  The potential~\eqref{bulge-pot} is generated from a spherical density profile $\rho(R) \propto R^{-1} (R + a)^{-3}$, and so we assume that the NS density profile follows this same form:
\es{bulge}{
n_\text{bulge}(R) = {N_\text{bulge} \over 2 \pi } {a \over R} {1 \over (R + a)^3} \,,
}
with $N_\text{bulge} \approx 6 \times 10^8$ the total number of NSs in the bulge.

\subsubsection{The nuclear star cluster}

We model the Milky Way NSC using the results of the simulation in~\cite{Freitag:2006qf}.  The dynamics of stellar objects within the inner few pc of the Galactic Center are strongly influenced by the presence of the massive black hole Sgr A* at the center of the Galaxy, with a mass $M_\bullet \sim 3 - 4 \times 10^6$ $M_\odot$.  Bahcall and Wolf showed that within the sphere of influence of a central black hole, the stars in a NSC would approach the universal density profile $n(r) \sim r^{-7/4}$ if all of the stars have the same mass~\cite{1976ApJ...209..214B}.  Two-body dissipation between stars is the driving mechanism in this case that allows the stars to fall towards the center of the Galaxy.  The formation of the central Bahcall-Wolf cusp then takes place over the relaxation time, which is the time-scale for two-body interactions to take place and which can be significantly shorter than the age of the Galaxy.  However, real astrophysical systems are more complicated because the NSCs consist of multiple populations with different masses.  Ref.~\cite{1977ApJ...216..883B} solved the coupled Boltzmann system for stellar populations of different masses orbiting central massive black holes and showed that each population follows a cusped distribution $n(r) \sim r^{-\gamma}$, where the index $\gamma$ depends on the mass of the population.  The heavier populations lose energy to the lighter ones through two-body interactions and tend to form steeper profiles, with $\gamma \sim 1.85$.  The lighter populations, on the other hand, follow less-steep profiles with $\gamma \sim 1.5$.  

Ref.~\cite{Freitag:2006qf} performed numerical simulations of the evolution of the Milky Way NSC in the presence of the central black hole to quantify the properties of the inner stellar cusps for the different mass species.  They included main-sequence stars that were able to turn into compact remnants, including white dwarfs, NSs, and black holes, at the end of their lifetimes.  The white dwarfs were taken to have mass $0.6 M_\odot$, while the NSs and black holes had masses $1.4 M_\odot$ and $10 M_\odot$, respectively.  The NSs and black holes were given natal velocity kicks, as expected from the supernova process. In that work, it was shown that, for a variety of initial conditions, the black holes develop a steep cusp with $\gamma \sim 1.8$, while the other stellar remnants and main-sequence stars evolve to a flatter profile with $\gamma \sim 1.3$.  After $\sim$$13$ Gyr, there were approximately $\sim$$10^4$ NSs within 1 pc of the Galactic Center.  We expand the NS number density profile to arbitrary distances $R$ from the Galactic Center using the ``eta-models" for systems with a central object~\cite{1993MNRAS.265..250D,1994AJ....107..634T}:
\es{}{
n(R) = {\eta N_\text{tot} \over 4 \pi R_b^3} \left( {R \over R_b} \right)^{\eta - 3} \left( 1 + {R \over R_b} \right)^{- \eta - 1} \,,
}
where $\eta = 3 - \gamma$.  The radius of gravitational influence of the central black hole, defined as the radius where the velocity dispersion due to all enclosed matter is equal to twice the velocity dispersion due to the central black hole alone, is around 3 pc.  The break radius $R_b$ is simply related the radius of influence, and for the fiducial model in~\cite{Freitag:2006qf} it is given by $R_b \approx 28 \, \, \text{pc} \left(2 \eta - 1 \right)^{-1}$.  For our fiducial NS model, we take $\gamma = 1.3$, so that $\eta = 1.7$, $R_b \approx 12$ pc, and $N_\text{tot} = 7 \times 10^5$, implying that there are $\sim$$10^4$ NSs within a pc (and $\sim$$220$ NSs within 0.1 pc).        

It is also useful to model the velocity dispersion of the NS population as a function of radius from the Galactic Center.  We assume the velocity distribution is isotropic and takes the form 
\es{MB}{
f( {\bf v}) = {1 \over \pi^{3/2} v_{\text{NS},0}^3} {e^{- {\bf v}^2 / v_{\text{NS},0}^2}} \,,
}
where the 3-d velocity dispersion is given by $\sigma_v^2 = 3  v_{\text{NS},0}^2 / 2 $.  The subscript NS is used to differentiate the velocity dispersion of the NSs from that of the DM.  We define the circular velocity $V_c^2(R) = G M_\text{tot}(R) / R$, where $M_\text{tot}(R)$ is the mass enclosed within the radius $R$.  Moreover, we assume that both the density of the NSs and the circular velocity follow power-laws, with $n(R) \sim R^{-\gamma}$ and $V_c \sim R^\alpha$.  Then, we may solve the Jeans equation to find~\cite{Dehnen:2006cm}
\es{vel_dispersion}{
v_{\text{NS},0}^2 = {2 \over \gamma - 2 \alpha} V_c^2(R) \,.
}  
Within the sphere of influence of the central black hole $\alpha \approx -1/2$; taking $\gamma = -1.3$ and $M_\bullet = 3.5 \times 10^6 M_\odot$, this gives 
\es{}{
v_{\text{NS},0} \approx 97 \, \, {\text{km} \over \text{s} } \sqrt{ {1 \, \, \text{pc} \over R} }  \,, \qquad R \lesssim 3 \, \, \text{pc} \,.
}

\subsection{The Dark matter Distribution}

The DM density profile in the central regions of the Milky Way is highly uncertain.  We consider a variety of models for the DM profile to encompass the range of possibilities and understand how the DM profile uncertainty affects the radio signal.  Our fiducial model is the Navarro-Frenk-White (NFW)~\cite{Navarro:1995iw,Navarro:1996gj} profile:
\es{NFW}{
\rho_\text{NFW}(R) = {\rho_0 \over {R \over R_s} \left( 1 + {R \over R_s} \right)^2 } \,,
}
where $R_s$ is the scale radius, which we take to be 20 kpc.  Our secondary model is the cored Burkert profile~\cite{Burkert:1995yz}
\es{Burk}{
\rho_\text{Burk}(R) = {\rho_0 \over \left(1 +  {R \over R_c} \right) \left[ 1 + \left( {R \over R_c} \right)^2 \right] } \,,
}
where $R_c$ is the core radius.  The smaller the core radius, the larger the DM density near the Galactic Center.  To quantify the full uncertainty that could arise from a large DM core, we take a large core radius $R_c = 9$ kpc~\cite{Nesti:2013uwa}.  In both cases, we normalize the profile be requiring the local DM density to be $0.3$ GeV$/$cm$^3$.

Regardless of the DM distribution at large $R$, within the sphere of influence of Sgr A*, it is possible for the DM to develop a kinematic cusp with $\rho(R) \sim R^{- \gamma}$ and $\gamma \approx 1.5$ for the same reasons that the stellar populations may develop a density cusp.  Following~\cite{Merritt:2006mt}, we consider the possibility that within the radius of influence (3 pc), the DM density profile smoothly transitions to the power-law $\rho(R) \sim R^{-1.5}$.  Within 3 pc of the Galactic Center, the DM velocity dispersion may also be calculated from~\eqref{vel_dispersion}.  We refer to the possible DM kinematic cusp as the ``DM spike."

\subsection{Radio flux from axion-photon conversion}

In this subsection, we create a simulated population of NSs following the spatial distributions described above in order to describe the spatial morphology of the radio signal.  
It is useful to factorize the flux dependence from the NS's position from the flux dependence from the NS's internal properties.  Thus, in this subsection we only consider the dependence of the flux on factors arising from the spatial distributions of NSs and DM.

We normalize the flux distribution relative to the flux from a NS $d = 250$ pc away from Earth in a DM background with density $0.3$ GeV$/$cm$^{3}$ and velocity dispersion $v_0 = 200$ km/s.  We take the reference NS to be at rest within the frame in which the DM velocity dispersion is isotropic.  
 We will assume that all NSs are identical (or, more appropriately, that we average over a large population of NSs such that the variation of the ensemble-average properties does not vary significantly across the sky).  Then, the relative flux $F^\text{rel}$ from a NS at distance $d$ in DM density $\rho_\text{DM}$ relative to our reference NS is given by 
\es{rel_flux}{
F^\text{rel} &\equiv { F_{d, \rho_\text{DM}} \over F_{250 \, \, \text{pc }, \, 0.3 \, \, \text{GeV}/\text{cm}^3} } \\
&= \left( {250 \, \, \text{pc} \over d} \right)^2 { \rho_\text{DM} \over 0.3 \, \, {\text{GeV} \over \text{cm}^3} } \left( {200 \, \, \text{km}/\text{s} \over \sqrt{v_0^2 + v_{\text{NS},0}^2}} \right) \,. 
}
Above, it is important to differentiate the NS velocity dispersion parameter $v_{\text{NS},0}$ from the DM velocity dispersion parameter $v_{0}$.  The combination of the two appearing in~\eqref{rel_flux} is the relevant factor to account for the change in DM density at the conversion radius from gravitational focusing~\cite{Hook:2018iia}.  Note, however, that for the disk stars, we simply use the circular velocity instead of $v_{\text{NS},0}$, since the disk stars are to a good approximation following circular trajectories.  In addition to the relative fluxes across the sky, it is also important to understand how the frequency spectrum of the signal changes with sky location.  We will return to this question in the following subsection. 

\begin{figure}[htb]
\includegraphics[width =0.5\textwidth]{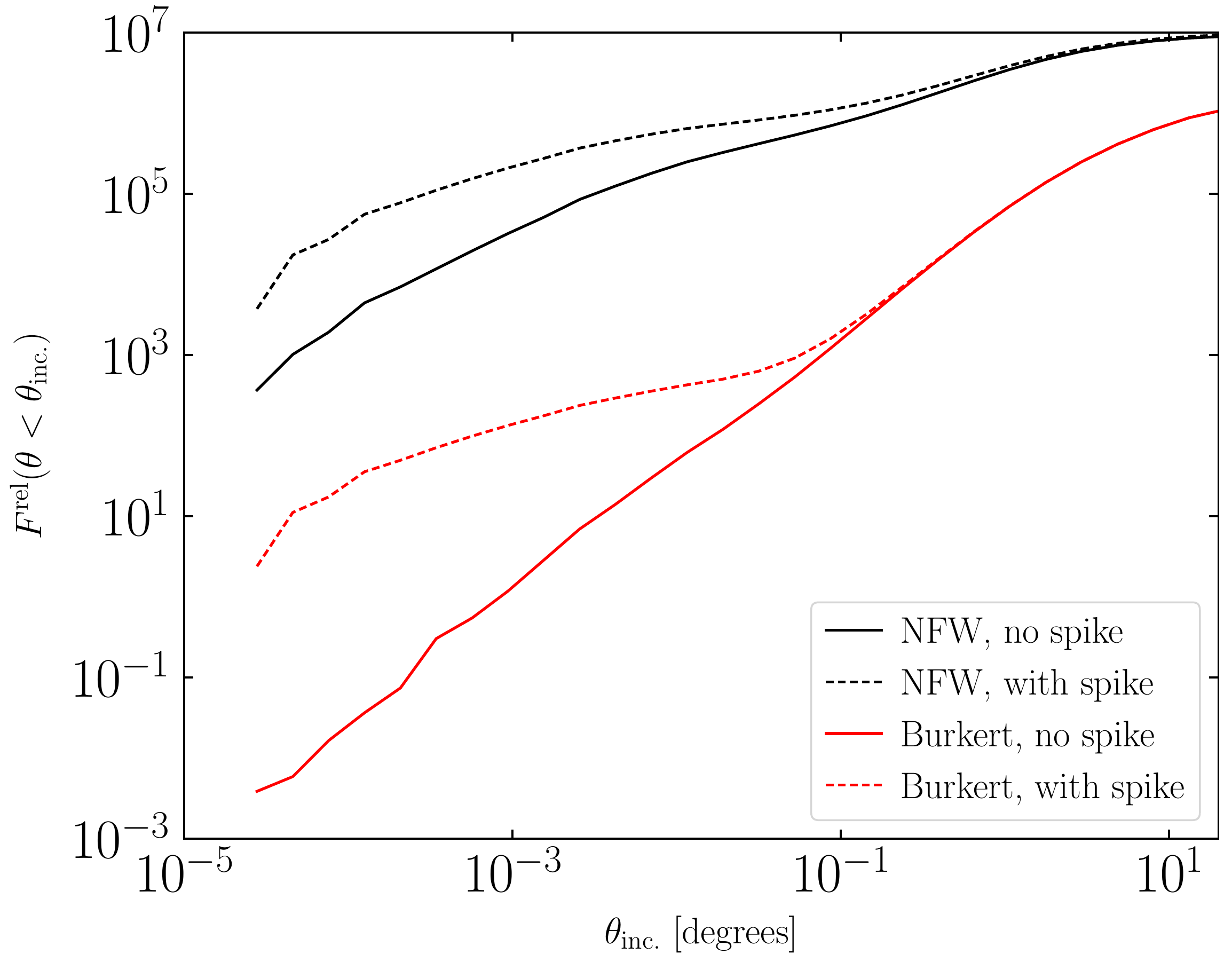}
 \vspace{-0.7cm}
\caption{The relative flux $F^\text{rel}$ contained within an angle $\theta_\text{inc.}$ of the Galactic Center for different DM models.  We consider NFW and Burkert DM profiles, and in each case we consider the possibility of a kinematic DM spike with $\gamma = 1.5$.
}
\label{fig: compare-GC}
\end{figure}

In Fig.~\ref{fig: compare-GC} we show the enclosed relative flux $F^\text{rel}(\theta < \theta_\text{inc.})$ when observing at the Galactic Center out to a radius $\theta_\text{inc.}$, as a function of $\theta_\text{inc.}$.  We illustrate the inclosed flux for both NFW and Burkert DM models, and we also show the variations to both scenarios in the case of a central DM spike.  Note that, roughly, we can consider the NSC to be the dominant source of NSs out to distances $\sim$1 pc away from the GC, corresponding to $\theta_\text{inc.} \approx 10^{-2}$ degrees. The bulge extends out to $\sim$1 kpc and thus dominates the flux out to $\theta_\text{inc.} \approx 5^\circ$.  At larger angles from the GC, the disk is most important. 

\subsection{Frequency Spectrum of Radio Emission}

In this subsection we describe the expected spectral morphology of the signal.
We begin by considering a single NS in its rest frame.  We assume that the DM velocity distribution is Maxwell-Boltzmann distributed with a velocity as in~\eqref{MB} with dispersion parameter $v_0$.  We take the NS to be boosted with respect to the frame in which the DM velocity distribution is isotropic by speed $v_B$.  Recall that in the NS frame, we may relate the frequency of outgoing radio waves, asymptotically far from the gravitational potential of the NS, to the energy of the in-falling axions by $\omega = m_a \sqrt{1 + v_\text{DM}^2}$, where $v_\text{DM}$ is the speed of the in-falling DM asymptotically far from the NS.  Thus, we may write down the following frequency distribution for the outgoing radiation in the NS frame, which we chose to do in terms of the quantity $\tilde \omega = \omega / m_a$~\cite{Hook:2018iia}:
\es{f_tilde_omega}{
f(\tilde \omega) = {2 \over \sqrt{\pi} v_B v_0} e^{ {2 - v_B^2 - 2 \tilde \omega \over v_0^2}} \sinh \left( {2 v_B \sqrt{2 \tilde \omega - 2} \over v_0^2} \right) \,.
} 
Importantly, in the NS frame $\tilde \omega > 0$.  However, to calculate the frequency spectrum in the observer's frame we need to apply the appropriate relativistic doppler shift.  For simplicity, let us assume that the observer's frame coincides with the frame in which the DM is isotropic, as if not this only induces a constant-frequency offset.  Then, the frequency $\tilde \omega_\text{obs}$ in the observer's frame is given by $\tilde \omega_\text{obs} = \tilde \omega \sqrt{ {1 - v_B^\text{L.O.S.} \over 1 + v_B^\text{L.O.S.}}}$, where $v_B^\text{L.O.S.}$ is the projection of the boost velocity along the line of sight, defined to be positive if the NS is moving away from the observer.  

The frequency dispersion in the radio signal from the individual NSs is of order $v_0^2$, as may be seen in~\eqref{f_tilde_omega}, while the frequency dispersion between different NSs is order the NS velocity dispersion, $v_{\text{NS},0}$, which determines the boosts $v_B$.  Thus, when averaging over an ensemble of NS targets, the dominant source of frequency dispersion comes from the velocity dispersion of the NSs and not the velocity dispersion of the DM.  Typically, in natural units within a system such as the Milky Way we expect $v_0 \sim v_{\text{NS},0} \sim 10^{-3}$.  This implies that the expected velocity dispersion over an ensemble of NSs is $\delta \omega / \omega \sim 10^{-3}$, while that from an individual NS is $\delta \omega / \omega \sim 10^{-6}$.  
Specifically, in the limit where the number of NSs is large so that the frequency distribution is continuous over scales $\delta \omega / \omega \sim v_{\text{NS,0}}$, we may describe the frequency distribution of the radio signal by 
 \es{}{
 f(\omega) = {1 \over \sqrt{2 \pi} \sigma_\omega} e^{-{ (\omega - m_a)^2  \over 2 \sigma_\omega^2}} \,,
 }
 where $\sigma_\omega^2 = m_a^2 v_{\text{NS},0}^2 / 2$.  
 
 \begin{figure}[htb]
\includegraphics[width = 0.49\textwidth]{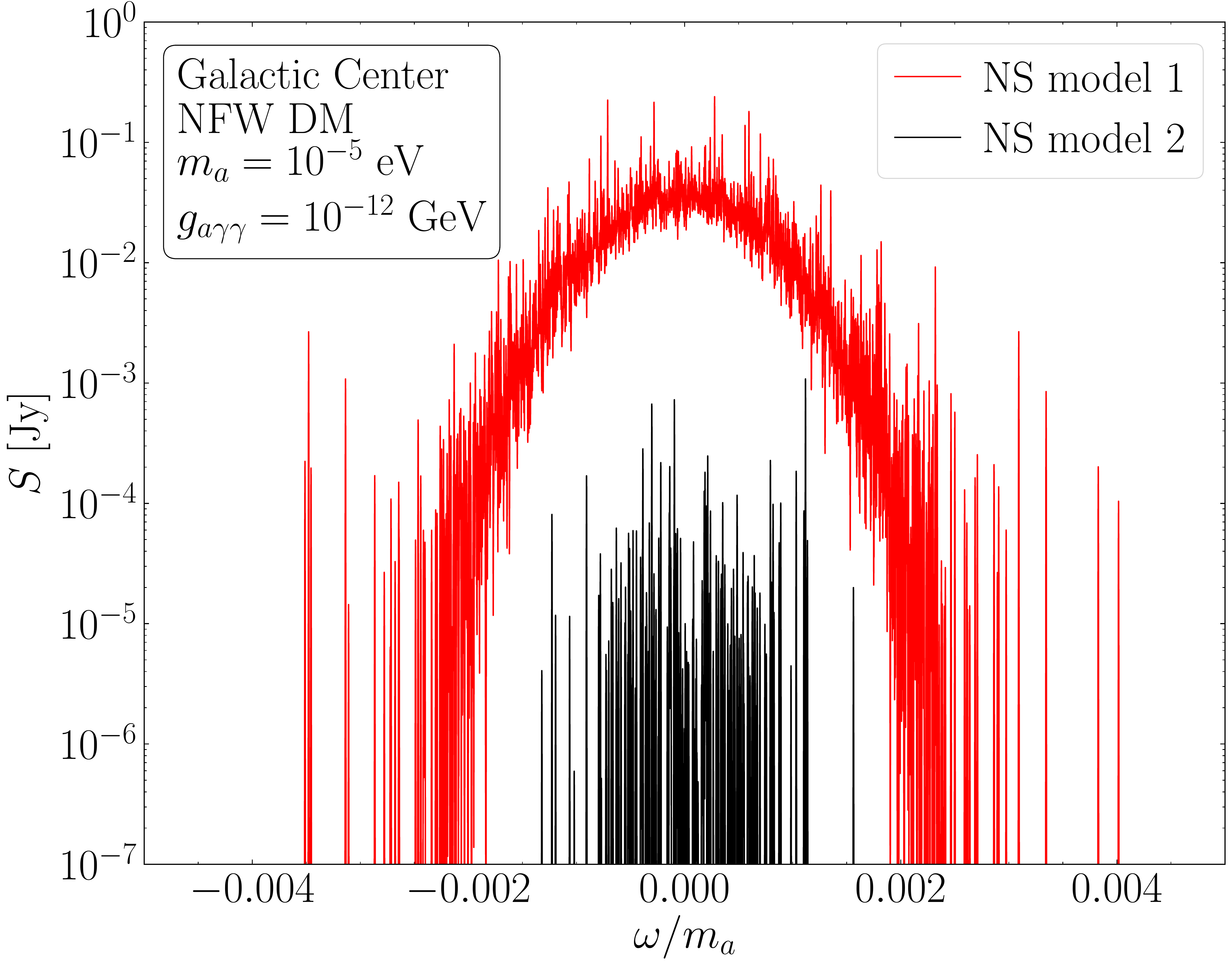}
\caption{
The simulated frequency spectra from the axion-induced signal contribution only (that is, not including other backgrounds) from an axion with parameters indicated and with a field of view of radius $2.5'$, which is appropriate for GBT at the relevant frequency $f \approx 2.4$ GHz corresponding to the example mass shown.  We display Monte Carlo realizations of the two NS models.   Bright, individual NSs arise from fortuitous spatial locations and from the NS's internal properties, which are drawn from the appropriate distributions.
}
\label{fig: freq_zoom}
\end{figure}

In Fig.~\ref{fig: freq_zoom} will illustrate simulated frequency spectra of an observation at the Galactic Center with an axion of mass $m_a = 10^{-5}$ eV, corresponding to $f \approx 2.4$ GHz, and $g_{a \gamma \gamma} = 10^{-12}$ GeV$^{-1}$.  Specifically, we show the flux density at Earth, $S$, which is defined as $dF/df$.  We illustrate both NS model 1 and model 2, with a field of view of radius $2.5'$, which, as will be discussed later, is consistent with that expected for GBT at this frequency.  To construct this simulation, we first distribute the NSs spatially and in velocity space.  Then, we use the formalism described in Sec.~\ref{sec: population-model} to attribute, randomly, intrinsic properties to each of the NSs.  To calculate the flux density, we also need to know the alignment angle between us and the NSs, which we assign randomly as well.  

We emphasize that the frequency spectra in Fig.~\ref{fig: freq_zoom} include the signal component only and not any backgrounds that may be present.  The inclusion of backgrounds and estimates for the sensitivity of radio telescopes to such a signal will be discussed later in this work.  Still, it is worth emphasizing that the unique frequency structure of the signal shown in Fig.~\ref{fig: freq_zoom} could be detected through different means.  For example, one search strategy might be to average $S$ over $\delta f / f \sim 10^{-3}$ and look for the broad emission from all of the NSs.  On the other hand, this strategy does not take advantage of the fact that the signal really consists of a forest of individual lines, many of which are much brighter than average due either to their fortuitous spatial location, intrinsic properties, or the alignment angle with respect to Earth.  Thus, another simplified strategy would instead be to search for the brightest individual line with $\delta f / f \sim 10^{-6}$.  Later, we will compute the sensitivity from each of these strategies.  Of course, the optimal strategy, which we do not pursue here, would likely construct a joint likelihood over the search for each individual line.  This is difficult, however, because we do not know a-priori where the lines are and which ones will be brightest.

\section{The axion-induced radio signal in M54}
\label{sec: M54}

The globular cluster M54 is unique in that it sits at the dynamical center of the nearby Sagittarius dwarf galaxy.  M54 should host many NSs, making the globular cluster an excellent target for axion-DM radio searches, since the Sagittarius dwarf should host a large and cold DM halo.  Below, we describe the models we use for the distribution of NSs and DM in Sagittarius, and then we calculate the projected radio flux from axion-DM conversion. 

\subsubsection{Neutron stars in M54} 

M54 has a total stellar mass of around  $\sim$$2 \times 10^6$ $M_\odot$ and is at a distance $d \approx 25$ kpc from Earth, at the heart of the Sagittarius dwarf galaxy~\cite{Monaco:2004ke,2009ApJ...699L.169I,Kunder:2009nb}.  
Ref.~\cite{Ivanova:2007bu} performed simulations of NS histories in globular clusters, for a variety of different metallicities, to address the question of how many NSs remain in the globular clusters, accounting for the natal velocity kicks that the NSs receive.  M54 is an old and metal-poor globular cluster, with $[\text{Fe}/\text{H}] \approx -1.55$~\cite{1999AJ....118.1245B}.  This corresponds most closely to the $Z = 0.0005$ metal-poor simulation in~\cite{Ivanova:2007bu}, though we note that the differences between the $Z = 0.0005$ through $z = 0.02$ simulations are small for our purposes ($\sim$20\%)~\cite{Ivanova:2007bu}.  Core collapse supernovae only account for a small fractions of the retained NSs, due to the high velocity kicks that the resulting NSs receive;~\cite{Ivanova:2007bu} (see also~\cite{Kuranov:2006kw}) found that most retained NSs arise from electron-capture supernovae processes, since these have smaller velocity kicks.  The metal-poor simulations in~\cite{Ivanova:2007bu} find that $\sim$$114$ NSs per $2 \times 10^5$ $M_\odot$ are retained within the core of the globular cluster, while $\sim$$126$ NSs per $2 \times 10^5$ $M_\odot$ are retained within the outer halo of the cluster.  This implies that for M54, we expect $\sim$$2400$ NSs retained within the cluster, with $\sim$$48$\% within the core.  

M54 appears to have a relatively flat core with a core radius $\sim$$6''$, although there are indications that the density could rise towards the inner $1''$ due to the possible presence of a central black hole~\cite{2009ApJ...699L.169I,Kunder:2009nb}.  We do not account for the possibility of a central density cusp and instead assume a constant-density NS core with core radius $a \sim 0.7$ pc.  We populate the inner core with 1152 NS targets.  Outside of the inner core, for radii $r>a$, we model the halo NSs by a Plummer sphere model; the halo is populated with 1248 NS targets.  In practice, the NSs within the core are more important since they reside in a region of higher DM density.

\subsubsection{Dark matter in Sagittarius}

The DM density profile in Sagittarius is highly uncertain, in part because the density profiles in dwarf galaxies are uncertain in general but also because Sagittarius in particular is disrupted from its recent merger with the Milky Way.
We consider both cored and cusped DM profiles within the Sagittarius dwarf, whose center is assumed to coincide with the center of M54.  For the cusped DM profile, we take the NFW profile in~\eqref{NFW} with scale radius $R_s \approx 0.2$ kpc and density parameter $R_s^3 \rho_0 \approx 3.3 \times 10^7$ $M_\odot$~\cite{Aharonian:2007km}.  
For the cored DM profile, we take the isothermal sphere model
\es{}{
\rho_\text{core}(R) = {v_a^2 \over 4 \pi G} {3 R_c^2 + R^2 \over (R_c^2 + R^2)^2} \,,
}
where $G$ is Newton's constant and $v_a \approx 13.4$ km$/$s is related to the central velocity dispersion~\cite{Aharonian:2007km}.  Here, $R_c$ is the core radius.  Ref.~\cite{Aharonian:2007km} assumed a small core radius $R_c \approx 1.5$ pc on the order of the radius of M54 itself.  However, some simulations suggest that DM cores form in dwarf galaxies at the scale of the half-light radius of the stellar component~\cite{Read:2015sta} because of feedback from, {\it e.g.}, supernovae.  Assuming Sagittarius has similar properties to the Draco dwarf galaxy, we take $R_c \approx 230$ pc~\cite{PhysRevD.69.123501}.  It is important to emphasize that the core DM density varies by over 4 orders of magnitude when switching between the two core radii discussed above; we assume the more conservative choice to better understand the possible systematics from the DM profile.  The DM density at the location of the M54 is uncertain, and this, in turn, is a significant source of uncertainty in the calculation of the resulting radio flux from axion-photon conversion.  To bracket the range of possibilities, we will consider both the cusped NFW DM profile and the cored profile with $R_c \approx 230$ pc.  However, it is important to point out that because Sagittarius is disrupted, it is possible that the DM density at the location of M54 is even lower than the cored value we assume.    

\subsubsection{Radio signal from axion-photon conversion in M54}

M54 is centered at the sky location $\ell_0 \approx 5.6072^\circ$ and $b_0 \approx -14.0872^\circ$~\cite{2009ApJ...699L.169I}.  We use the definition of the relative flux density, defined in~\eqref{rel_flux}, to construct the relative flux $F^\text{rel}$ contained within an angle $\theta_\text{inc}$ of the Globular Cluster center, shown in Fig.~\ref{fig: compare-M54}.  We illustrate both the NFW and cored DM profiles.
\begin{figure}[htb]
\includegraphics[width =0.5\textwidth]{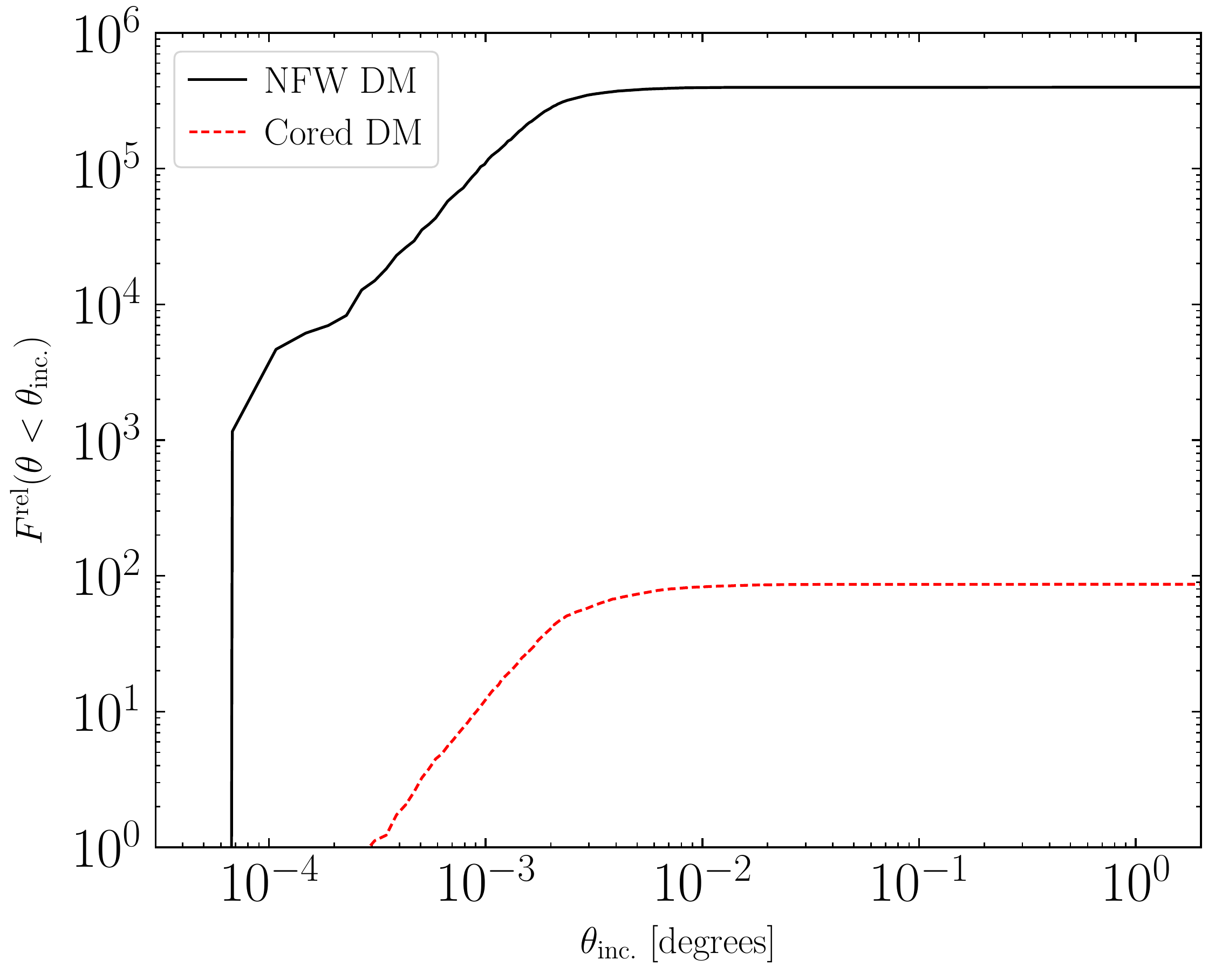}
 \vspace{-0.7cm}
\caption{As in Fig.~\ref{fig: compare-GC} except for an observation centered on the Globular Cluster M54 in the Sagittarius dwarf galaxy.  The cored DM profile has a large core radius $R_c \approx 230$ pc, leading to a significantly smaller DM density at the location of the Globular Cluster and thus a much smaller radio signal.
}
\label{fig: compare-M54}
\end{figure}

 \begin{figure}[htb]
\includegraphics[width = 0.49\textwidth]{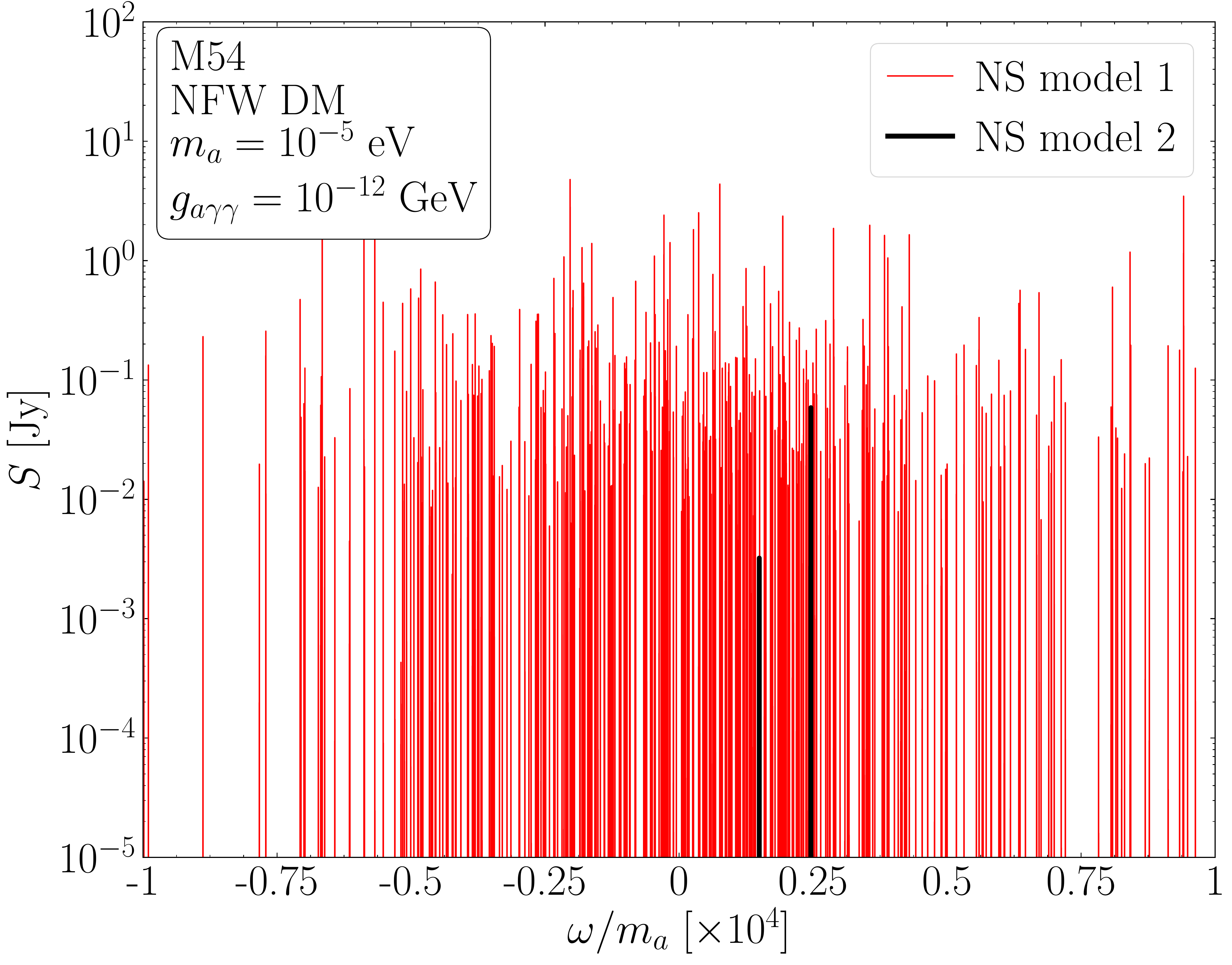}
\caption{
As in Fig.~\ref{fig: freq_zoom} except for the NFW DM profile in the M54 Globular Cluster, within the Sagittarius dwarf galaxy. The width of each NS signal is $\delta f/f \sim 10^{-8}$ and the width of the ensemble of NSs is $\delta f / f \sim 10^{-4}$.  In the case of NS model 2, there are only two participating NSs within the field view in this particular Monte Carlo realization.
}
\label{fig: freq_zoom_M54}
\end{figure}

The frequency spectra of the radio signals from M54 are qualitatively different from those expected in observations of the Galactic Center because of the fact that there are fewer NSs in M54 and they inhabit a region with a smaller DM and NS velocity dispersion. 
The projected velocity dispersion profile of M54 rises towards the center of the galaxy from values $\sim$5 km$/$s at distances of order $\sim$$35''$ to values $\sim$$10$ km$/$s at distances $\sim$$3.5'$ from the center~\cite{2009ApJ...699L.169I}.  
The velocity dispersion profile appears to continue to rise towards the center of the cluster, potentially indicating a density cusp from a central intermediate-mass black hole~\cite{2009ApJ...699L.169I}, though we do not consider this possibility here.  We will take the simplifying assumption that both the DM and NS populations have homogeneous and isotropic Maxwell-Boltzmann velocity distributions with $v_0 = 11$ km$/$s (corresponding to a projected velocity dispersion $\sim$$8$ km$/$s).  While it is possible that the NS population has a slightly higher velocity dispersion due to the natal velocity kicks, given that the NSs we consider are still bound to the globular cluster, the velocities cannot be that much greater.   

In Fig.~\ref{fig: freq_zoom_M54} we show an example of a simulation of the radio signal from M54 for an NFW DM profile, for NS models 1 and 2 and assuming the same axion properties and observational setup as in Fig.~\ref{fig: freq_zoom}.  In particular, notice that at this frequency ($m_a = 10^{-5}$ eV, $f \approx 2.4$ GHz), there are only two NSs, for this particular realization, within the field of view for NS model 2 that are participating in axion-photon conversion.  While the flux from these two NSs is comparable to the flux from the individual NSs in NS model 1, there are significantly more participating NSs in model 1 than in model 2.  In both models, the spread of the signal in frequency space is significantly narrower than in the Galactic Center observations.

\section{Andromeda galaxy}
\label{sec: M31}

The Galactic Center of the Milky Way has the advantage, from the point of view of the putative radio signal from axion-photon conversion, of being the closest galactic center.  However, as we will see in the subsequent section, while this means that the DM-induced flux from the Milky Way center is high, it does not necessarily ensure that the Galactic Center has the highest signal-to-noise radio at fixed $g_{a \gamma \gamma}$.  This is because the sky temperature at the center of the Milky Way is significantly higher than the average sky position due to normal radio emission from the central regions of the Galaxy. 
 This leads us to consider the possibility that other galaxies, beyond the Milky Way and its dwarfs, could be promising targets, especially at low frequencies where the Galactic Center signal is limited by astrophysical emission arising from the same location.

\subsection{Neutron stars and dark matter in M31}

A natural starting point is consider the Andromeda Galaxy (M31), which is the largest galaxy in the Local Group and relatively nearby at $d \approx 780$ kpc from Earth.  For the purpose of this explorative analysis, we take a simplistic model for M31 that mostly involves rescaling the Milky Way parameters to match those expected in M31.  The stellar masses of the stellar disk and stellar bulge in M31 are expected to be $\sim$$5.6 \times 10^{10}$ $M_\odot$ and $\sim$$3.1 \times 10^{10}$ $M_\odot$, respectively~\cite{2012A&A...546A...4T}.
If we assume that the stellar mass to NS ratio is the same when comparing M31 to the Milky Way, this would lead us to expect $N_\text{disk} \approx 4 \times 10^8$ ($N_\text{bulge} \approx 2 \times 10^9$)
 in the disk (bulge).  Note that here we have used the disk and bulge mass estimates for the Milky Way from~\cite{2015ApJ...806...96L}.
 
The M31 disk and bulge are modeled using the same functional forms as used for the Milky Way.  The bulge core radius is taken to be $a \approx 0.6$ kpc, following~\cite{Banerjee:2008kt}.  Note that this is the same core radius that we assumed for the Milky Way bulge in~\eqref{bulge}.  We assume that the disk NSs follow the thick disk of M31, which has been measured to have a vertical scale height $\sim$$2.8$ kpc and a radial scale length $\sim$$8.0$ kpc~\cite{2011MNRAS.413.1548C}.  We thus model the disk NSs using~\eqref{disk} with $\langle |z| \rangle = 2.8$ kpc and $\sigma_r = 5.7$ kpc.  

The nucleus of M31 is both more complicated than that of the Milky Way and also less well studied, given that it is further away.  Unlike the stellar nucleus of the Milky Way, which appears to be mostly spherical and dynamically centered on the central black hole, the nucleus of M31 appears to consist of multiple primary components.  The two main clusters consist of P1, which is offset from the nucleus by $\sim$$1.8$ pc, and P2, which is at the dynamical center of the Galaxy; additionally, there appears to be a disk of young stars embedded within P2 and centered around the central black hole called P3~\cite{Bender:2005rq}.  From observations of P3 with the {\it Hubble Space Telescope}, the mass of the central black hole is estimated at $M_\bullet \approx 1.4 \times 10^8$ $M_\odot$~\cite{Bender:2005rq}.  

The currently-accepted model for the P1-P2 system is that both nuclei are part of the same eccentric disk, with P2 centered around the black hole, with a rising stellar density towards the black hole, and P1 centered around the apocenter.  The increased brightness at P1 results from an accumulation of stars at the apocenter~\cite{1995AJ....110..628T}.  The P1-P2 system may be modeled as an eccentric disk with a radius $\sim$8 pc and a mass $M_\text{NSC} \approx 2 \times 10^7$ $M_\odot$~\cite{1995AJ....110..628T,Bender:2005rq}.  Estimating the mass of the Milky Way NSC to be $\sim$$7 \times 10^7$ $M_\odot$ from~\cite{Freitag:2006qf} and rescaling the number of NSs within the MW NSC we estimate $N_\text{NSC} \approx 2.3 \times 10^5$ NSs within the M31 nuclear disk.  For the purpose of our simplified analysis, we model the spatial distribution of NSs in the P1-P2 system by a thin exponential disk with a scale radius of 8 pc.  The line-of-sight velocity dispersion around the central black hole, measured in {\it e.g.}~\cite{1995AJ....110..628T,Bender:2005rq}, is roughly consistent with rescaling the circular velocity by a constant, radius-independent factor $\sim$$1/2$.  That is, the line-of-sight velocity dispersion $\sigma_\omega$, which determines the width of the signal in frequency space, is taken to be
\es{}{
\sigma_\omega \approx \kappa \sqrt{ {G M_\bullet \over R} } \,, \qquad \kappa \approx 0.5 \,.
}

We consider two models for the DM distribution within M31, the NFW DM model~\eqref{NFW} and the Burkert DM model~\eqref{Burk}, and for each model we also consider the possibility of a kinematic DM spike near the central black hole.  For the NFW model, we take the scale radius to be $R_s \approx 16.5$ kpc, while for the Burkert model the core radius is assumed to be similar to that of the MW: $R_c \approx 9.1$ kpc~\cite{2012A&A...546A...4T}.  To calculate the effect of a possible kinematic DM spike, which we define as the possibility that the DM density profile rises as $\rho_\text{DM} \sim r^{-1.5}$ for $r < r_\text{infl.}$, we need to know the radius of influence $r_\text{infl.}$ of the M31 black hole.  Using the estimate $M_\bullet \approx 1.4 \times 10^8$ $M_\odot$, we obtain $r_\text{infl.} \approx 25$ pc~\cite{Bender:2005rq}, which is significantly larger than the radius of influence of the Milky Way's black hole.  However, we note that it is non-trivial to determine whether the possible DM spike would survive the presence of the asymmetric eccentric nuclear disk; such studies are beyond the scope of the current work.

\subsection{Axion-induced radio flux from M31}
\begin{figure}[htb]
\includegraphics[width =0.5\textwidth]{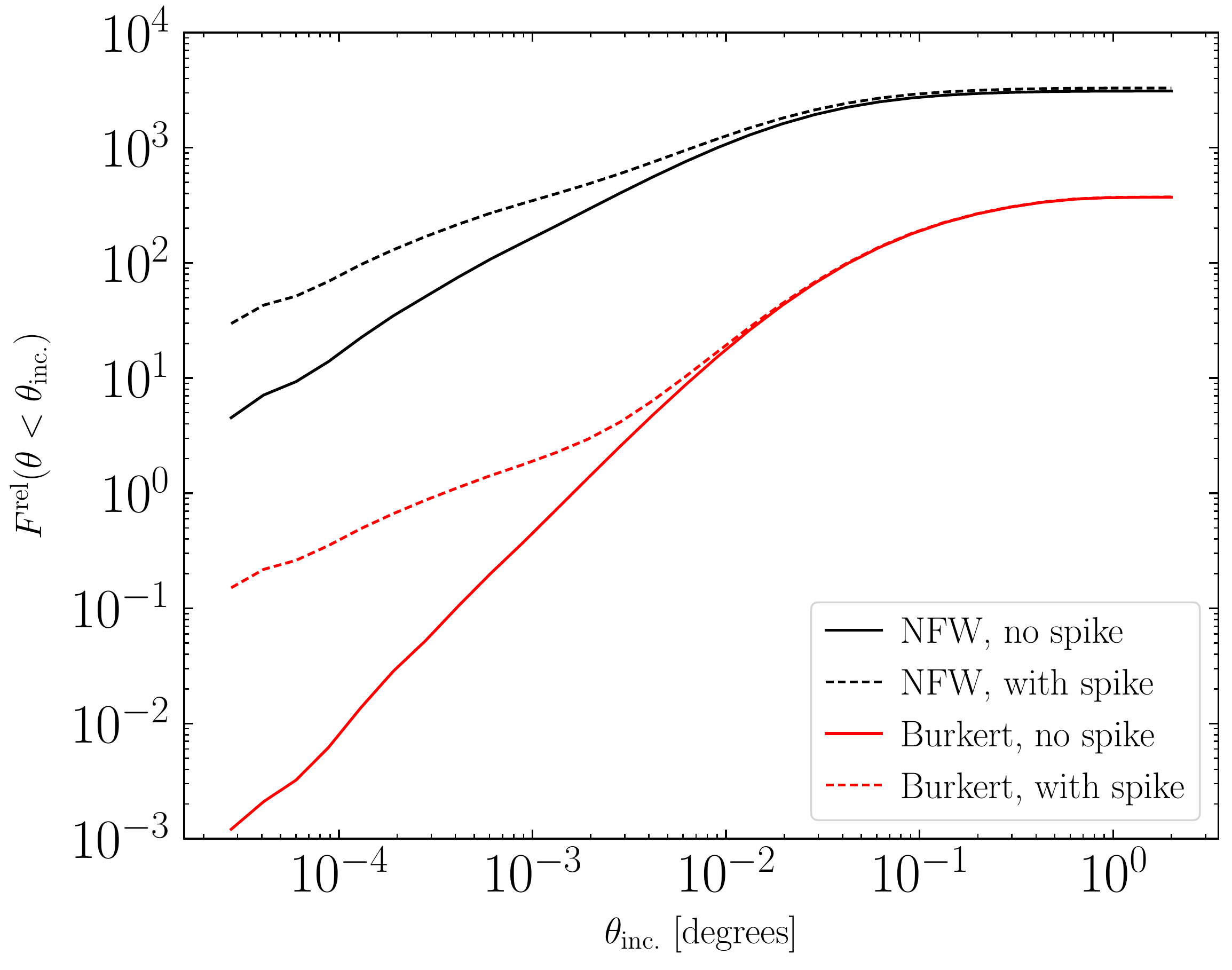}
 \vspace{-0.7cm}
\caption{As in  Fig.~\ref{fig: compare-GC}, except for an observation centered on M31.
}
\label{flux: M31}
\end{figure}
In Fig.~\ref{flux: M31} we show the simulated radio flux distribution for M31.  In particular, we illustrate the relative flux $F^\text{rel}$ included within an angle $\theta_\text{inc.}$ from the center of M31, as in Fig.~\ref{fig: compare-GC}, for various assumptions for the DM density.  Compared with the Milky Way observations, we see that the included flux approaches a constant at smaller $\theta_\text{inc.}$ since M31 is further away.  This also means that there are more NSs per beam relative to the Milky Way observations.  For example, for an observation with $r_\text{beam} = 10'$ centered on the center of M31, there are $\sim$$2 \times 10^9$ NSs within the field of view.  The frequency distribution of the signal is also similar to that expected for the Milky Way, with $\delta f /f \sim 10^{-3}$ for the ensemble of NSs and $\delta f / f \sim 10^{-6}$ for the individual NSs.

\section{Radio Observations}
\label{sec: radio}

In this section, we calculate the projected sensitivity from radio observations of astrophysical populations of NSs.  
 First, we discuss how to combine the formalisms developed in the previous sections for predicting the axion-induced signal with the properties of radio telescopes to project sensitivities to axion DM.  
 Then, we use this formalism to project sensitivities to axion DM from observations of the Galactic Center, M54, and M31 with GBT, VLA, and SKA.

\subsection{Observational details}

In this subsection, we outline how we estimate the sensitivity of radio telescopes to axion-induced lines using single-dishes and interferometer arrays.

\subsubsection{Single-dish telescopes}
Before calculating sensitivities for specific observations, we discuss some relevant properties for single-dish radio telescopes such as the GBT.  
  It is common when describing radio signals to work with the flux density $S \equiv F / B$, where $B$ is the bandwidth used in the analysis and $F$ is the flux contained within that bandwidth.  Note that we previously defined $S = dF /df$, where $f$ is frequency, but for observational purposes it is more appropriate to work with the finite version of the derivative, where the flux is averaged over the bandwidth $B$.
    The minimal detectable flux density, at a given signal-to-noise ratio SNR, is then  
\es{Smin}{
S_\text{min} = \text{SNR} {\text{SEFD} \over \sqrt{n_\text{pol} B   \Delta t_\text{obs}} } \,,
} 
where $n_\text{pol}$ is the number of polarization channels, $\Delta t_\text{obs}$ is the observation time, and $\text{SEFD}$ is the system-equivalent flux density, which is a property of the radio telescope in the given band and which will be described in more detail below.

  The SEFD may be written as 
\es{SEFD}{
\text{SEFD} ={ T_\text{sys}(f) \over G} \,,
}
where $T_\text{sys}(f)$ is the frequency-dependent system temperature and $G$ is the telescope gain.  The system temperature receives contributions from astrophysical backgrounds, the atmosphere, and the telescope electronics.  It is convenient to combine the latter two contributions into one telescope-specific temperature $T_R(f)$, such that we may write
\es{}{
T_\text{sys}(f) = T_\text{astro}(f) + T_R(f) \,,
}
where $T_\text{astro}(f)$ is the astrophysical temperature, which will be discussed more later.

The gain is defined by $G \equiv A_\text{eff} / (2 k)$, where $k$ is Boltzmann's constant, and $A_\text{eff}$ is the effective telescope area.  It is defined by $A_\text{eff} = \eta_a A_p$, where $A_p$ is the real projected area and $\eta_a$ is the aperture efficiency.  Note that for GBT, the gain is typically $\sim$$2$ K$/$Jy.  At GBT, the temperature $T_R(f)$ is typically between $\sim$$10$ -- $30$ K, depending on weather conditions and frequency band, for frequencies less than $\sim$$2$ GHz; $T_R(f)$ rises to $\sim$$100$ K at $f \sim 40$ GHz.\footnote{\url{https://science.nrao.edu/facilities/gbt/proposing/GBTpg.pdf}.}  Since the focus of this work is primarily at frequencies $f \lesssim 2$ GHz, we will assume a constant $T_R(f) = 25$ K. 

The beam width of a radio telescope is proportional to the wavelength of radiation under consideration, with the FWHM $\theta_b$ obeying the ratio $\theta_b \approx 1.25 \lambda / D$, where $D$ is the telescope diameter.  In general, we may approximate
\es{beam_width}{
\theta_b \approx 12.5' \left( {1 \, \, \text{GHz} \over f} \right) \left({100 \, \, \text{m} \over D} \right) \,.
}

\subsubsection{Simplified likelihood for radio arrays}
\label{sec: array}

The sensitivity calculation is more complicated when surveying an extended region, such as the Galactic Center region, with a radio array.  This is because there are multiple synthesized beams within the primary beam area, and one must construct a joint likelihood over these different regions.  In this subsection we outline a simplified approach to constructing such a joint likelihood.  
We will focus in this section on the combined signal from all the NSs within the field of view instead of analyses that try to resolve the individual NSs in frequency space.  That is, within each synthesized beam we will use a broad bandwidth determined by the velocity dispersion of the NSs within that beam.
With that said, accounting for the narrower bandwidth of the individual, resolved NSs within the synthesized beams could increase the sensitivity to the axion DM signal dramatically, and constructing such a likelihood and search strategy would be an interesting direction for future work.

First we note that the individual beam elements, consisting of dishes of diameter $D$, have approximate gains
\es{}{
G(D) \approx 2 \, \,{\text{K} \over \text{Jy}} \, \, \left( { D \over 100 \, \, \text{m}} \right)^2 \,,
}
and approximate beam diameters as given in~\eqref{beam_width}.
 In our simplified treatment, the primary beam radius is simply $r_\text{prim} \approx \theta_b/2$.  On the other hand, the synthesized beam diameter $\theta_\text{synth} = 2 r_\text{synth}$ is determined by the maximum baseline $B_\text{max}$ of the array:
\es{}{
\theta_\text{synth} \approx 50'' \left( {1 \, \, \text{GHz} \over f} \right)  \left( { 1 \, \, \text{km} \over B_\text{max}} \right)  \,.
}

We treat the signal as a collection of synthesized beams with beam radius $r_\text{synth}$ within the primary beam radius $r_\text{prim}$.  The number of synthesized beams is $N_\text{synth} \approx r_\text{prim}^2 / r_\text{synth}^2$.  The combined chi-square distribution is then simply 
\es{chi-square-combined}{
\chi^2(g_{a\gamma \gamma})  = \sum_{i=1}^{N_\text{synth}} { \left( S_i - S_{b,i} - S_{\text{sig,i}} \right)^2 \over \sigma_{S_{b,i}}^2 } \,,
}
where $S_i$ ($\sigma_{S_{b,i}}^2$) is the observed flux density (flux density variance) in each of the synthesized beams, labeled by $i$, $S_{b,i}$ is the predicted background flux density within the synthesized beam area, and $S_{\text{sig,i}}$ is the predicted signal flux density, which depends on the coupling $g_{a\gamma \gamma}$.  Note that the chi-square distribution is a function of the signal coupling $g_{a\gamma\gamma}$; the best-fit $g_{a\gamma\gamma}$ is that which minimizes the chi-square distribution, and the significance of the detection of axion DM may be computed from the difference of the chi-square distribution between the best-fit signal coupling and the null value $g_{a\gamma\gamma}= 0$.  This quantity is defined as $\delta \chi^2$. 

We may estimate $\sigma_{S_{b,i}}$ by 
\es{}{
\sigma_{S_{b,i}} = {\text{SEFD}_i \over \sqrt{n_\text{pol} B_i \Delta t_\text{obs}}} \,,
} 
where $\text{SEFD}_i$ is the SEFD for that particular synthesized beam, depending on the sky location and the frequency, and $B_i$ is the bandwidth of the signal.  Note that we are using the convention where $\text{SEFD}_i = T_i / G_\text{array}$, where $T_i$ is the temperature and $G_\text{array} = G \times \sqrt{N(N-1)}$ is the array gain, which is related to the gain $G$ from the individual antennas and the number $N$ of array elements.

We may estimate the sensitivity to a putative axion signal by assuming that the data is given by the background plus the signal: $S_i = S_{b,i} + S_{\text{sig},i}$~\cite{Cowan:2010js}. 
Thus,
\es{chi2}{
\delta \chi^2 = {n_\text{pol} \Delta t_\text{obs}} G_\text{array}^2 \sum_{i=1}^{N_\text{synth}} {F_i^2 \over B_i T_i^2} \,,
}
where $F_i$ is the predicted signal flux within each synthesized beam.  The significance of the detection is given by $\sqrt{\delta \chi^2}$; note that in the single-beam case, the significance is simply the SNR as given in~\eqref{Smin}.

Typically single dishes are better suited for searching for the axion-induced signal from the ensemble of NSs than interferometer arrays.  This is because if the signal is extended beyond the synthesized beam area, then one loses is sensitivity by having to construct the joint likelihood in~\eqref{chi2} over the different regions.  To illustrate this point, suppose that the fluxes $F_i$, bandwidths $B_i$, and the temperatures $T_i$ entering into~\eqref{chi2} are constant over the primary beam area.  Then we can write $F_i = F_\text{prim} / N_\text{synth}$, where $F_\text{prim}$ is the total flux over the primary beam area.  This implies that $\delta \chi^2 \propto G^2 {N^2 \over N_\text{synth}}  F_\text{prim}^2$, where $G$ is again the gain for the individual array elements and where we have assumed $N \gg 1$.  Thus, the larger  $ N_\text{synth}$, or equivalently the larger the baseline, the smaller $\delta \chi^2$.

\subsubsection{Searches for brightest individual neutron star}

The discussion in the previous subsection relates to searches for the combined emission from all NSs.  However, it is also possible to search for the individual NSs themselves by focusing on narrower frequency bandwidths.  Consider Fig.~\ref{fig: freq_zoom}, which shows an example frequency spectrum from a signal at the Galactic Center.  While most of the power in that case is distributed across a bandwidth of order $\delta f / f \sim 10^{-3}$, the brightest individual NSs are clearly visible, with relative bandwidths $\delta f / f \sim 10^{-6}$.  The signal-to-noise ratio SNR for the flux density scales like $F/ \sqrt{B}$, where $F$ is the signal flux and $B$ is the bandwidth.  That is, even if the flux is one and a half orders of magnitude smaller for an individual NS compared to the sum over all NSs, the individual NS can have the same SNR as the broader signal, for the case of the Galactic Center.  

One additional advantage of searches for the brightest, individual NSs compared to searches for the broader signal from the combination of NSs is that the former is better suited for searches with radio interferometer arrays, since the target of interest is a point source and one does not lose in sensitivity for increasing baseline.  

\subsubsection{Benchmark telescope configurations}

To estimate the sensitivity of radio observations to axion DM, we will assume a few simplified telescope configurations based on the GBT, a single 25-m dish, VLA (in configuration D), and the future SKA2.  The assumed properties of these configurations are summarized in Tab.~\ref{tab: telescopes}.  Note that for SKA2, we have calculated the number of array elements by requiring the total telescope area to be equal to a square kilometer, since precise designs do not yet exist for SKA2.  Similarly, the synthesized beam area is not known.  However, we will only project sensitivities with SKA2 for searches for the brightest individual NSs, for which the synthesized beam area is not important.  This is because the brightest NSs are point sources.  The 25-m dish is also a hypothetical telescope, though it could be one of the single VLA dishes.  The point of including this smaller telescope in the list is to illustrate how the sensitivity changes with dish size for a single dish.

\begin{table*}[htb]
  \centering
  \caption{List of benchmark telescope configurations used in this work.  We consider two single dishes (GBT and a 25-m dish) and two arrays (VLA and SKA2).  Note that for SKA2 we do not specify the baseline because precise plans do not yet exist and because it is not important, at leading order, for the point source analysis we consider with SKA2.  Note that $G$ refers to the gain from the individual array elements, $N$ is the number of elements in the arrays, and the primary and synthesized beam radii are normalized at $f = 1$ GHz.}
  \label{tab: telescopes}
  \begin{tabular}{l|c|c|c|c|c|c}
    Name & $d_\text{prim}$ [m] & G [K$/$Jy] & N & $r_\text{prim}^{1 \, \, \text{GHz}}$ & $r_\text{synth}^{1 \, \, \text{GHz}}$  & $T_R$ [K] \\
    \hline
    \hline
    GBT & 100 & 2.0 & 1 & 6.3' & - & 25 \\
    \hline
    25m-dish & 25 & 0.13 & 1 & 25' & - & 25 \\
    \hline
    VLA Config. D & 25& 0.13 & 27 & 25' & 0.58' & 25 \\
    \hline
    SKA2 & 15 & 0.045 & 5659 & 42' & \text{TBD} & 25 \\
    \hline
  \end{tabular}
\end{table*}

\subsubsection{Background astrophysical temperature}

We estimate the background astrophysical sky temperature from the Haslam map~\cite{1981A&A...100..209H,1982A&AS...47....1H,2015MNRAS.451.4311R}, which is illustrated in Fig.~\ref{haslam-label}.  Note that in that map we also show the sky locations of the three targets of interest discussed in this work.  The Haslam map is a 408 MHz all sky radio map that combines data taken by multiple radio surveys.  We use the reprocessed version~\cite{2015MNRAS.451.4311R} that is source-subtracted and destriped.  The map is smoothed to an angular resolution $\sim$$51'$.
\begin{figure}[htb]
\includegraphics[width =0.5\textwidth]{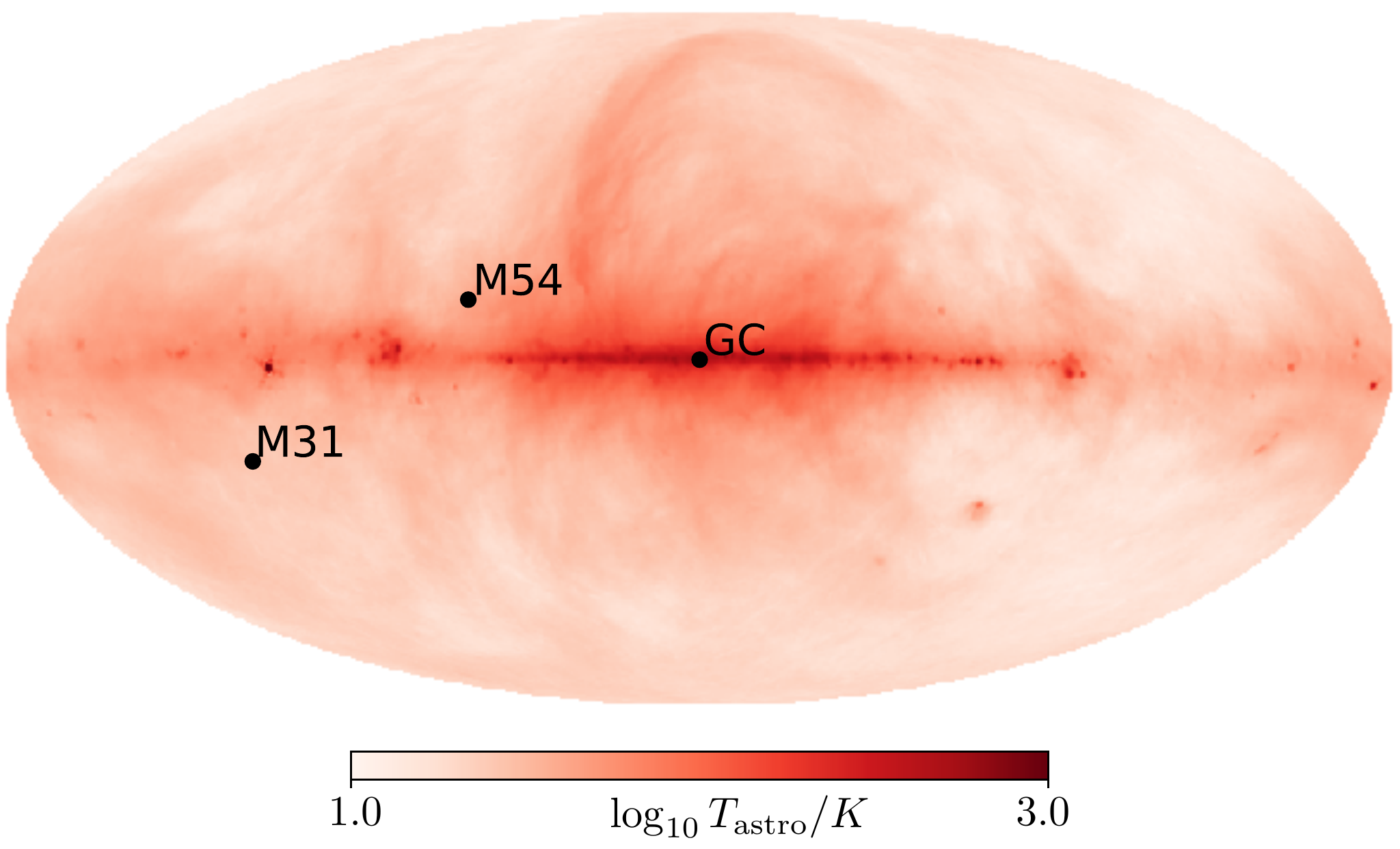}
 \vspace{-0.7cm}
\caption{The Haslam map in Galactic coordinates, with the Galactic Center indicated at the center, showing the sky temperature at 408 MHz.  In addition to the Galactic Center, we also show the locations of M54 and M31. 
}
\label{haslam-label}
\end{figure}

The radio emission is dominated by Galactic synchrotron radiation.  To infer the sky temperature beyond 408 MHz, we assume that the spectrum is a power-law $T_\text{astro}(f) = T_{408} (408 \, \, \text{MHz} / f)^\beta$.  The index $\beta$ has been measured to be $\beta \approx 2.76$ in the frequency range 0.4 - 7.5 GHz, averaged across the full sky~\cite{Platania:1997zn}.  We note that this index is similar to that found by the ARCADE collaboration~\cite{2011ApJ...734....5F}, which analyzed the extragalactic radio background between 22 MHz and 10 GHz and found an index $\beta \approx 2.62$.  We will assume $\beta = 2.76$ when extrapolating beyond 408 MHz, though we note that small variations to this index have little effect on our sensitivity projections.  

We also note that while the angular resolution of the Haslam map is sufficient for sensitivity projections in most regions of the sky, which lack bright point sources, it is likely insufficient to describe the region around the Galactic Center.  For example,~\cite{Macquart:2015jfa} found that the radio flux from the Galactic Center falls off at angular scales $\sim$20' from the center, which is a significantly smaller scale than the $\sim$51' resolution of the Haslam map.  Thus we caution that while we use the Haslam map for background temperature estimates in the Galactic Center region, these should be revisited with higher-resolution studies before interpreting the results of radio searches in this region.  

On a related note, we also do not consider possible sources of attenuation of the radio signals in this work.  Possible sources of attenuation could come from free-free absorption and scattering in the dense interstellar medium of the Galactic Center, though using results from previous studies ({\it e.g.},~\cite{Rajwade:2016cto}) that looked into these effects for pulsar surveys in the Inner Galaxy, we estimate that attenuation should not be important over the frequency ranges of interest and the targets of interest.

\subsection{Projected sensitivities}

In this subsection, we use the formalism described in the previous subsection to project the sensitivity to axion DM from radio observations with GBT, VLA, and the future SKA.
\subsubsection{The Galactic Center}

\begin{figure*}[htb]
\includegraphics[width = 0.49\textwidth]{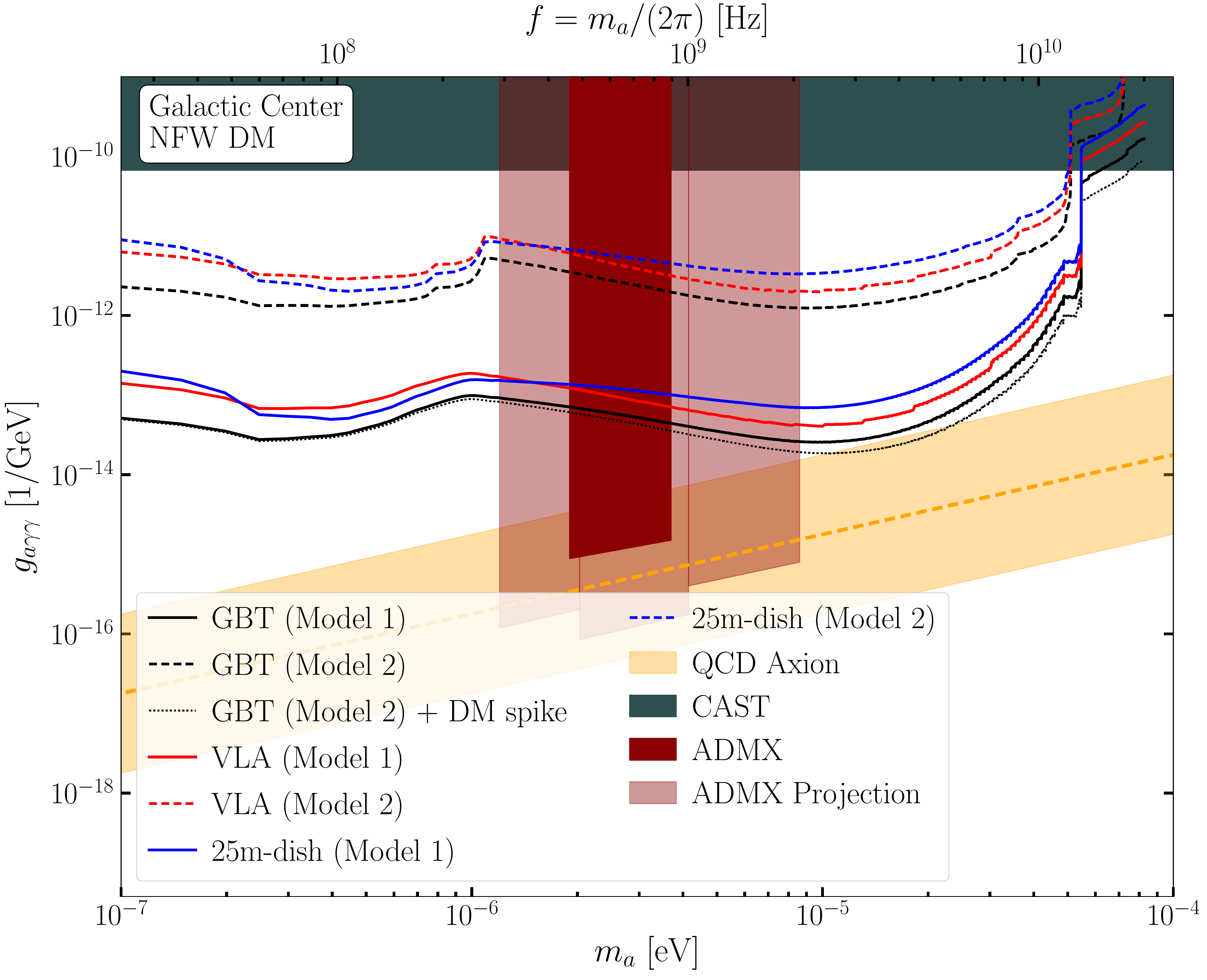} 
\includegraphics[width = 0.49\textwidth]{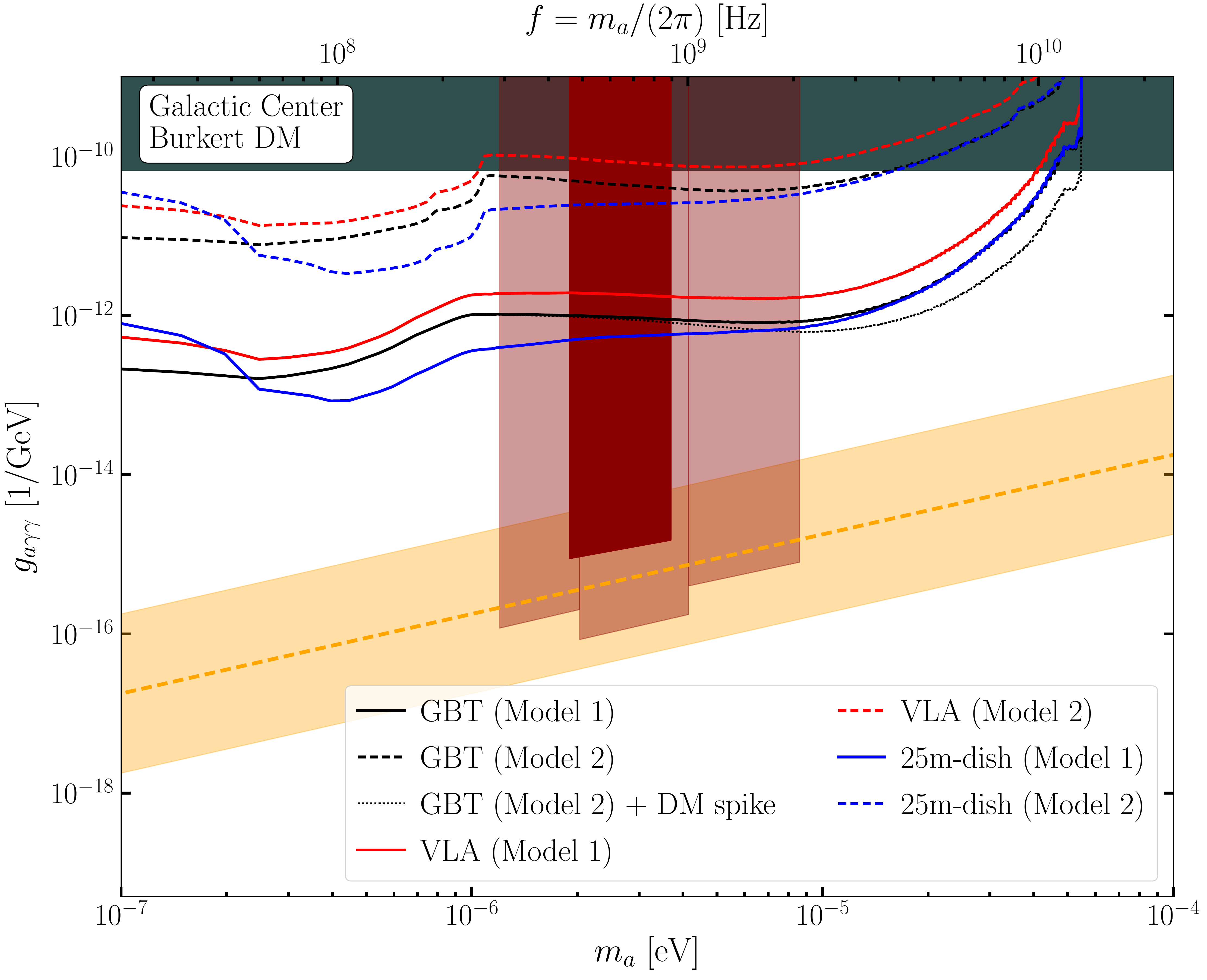} \\
\includegraphics[width = 0.49\textwidth]{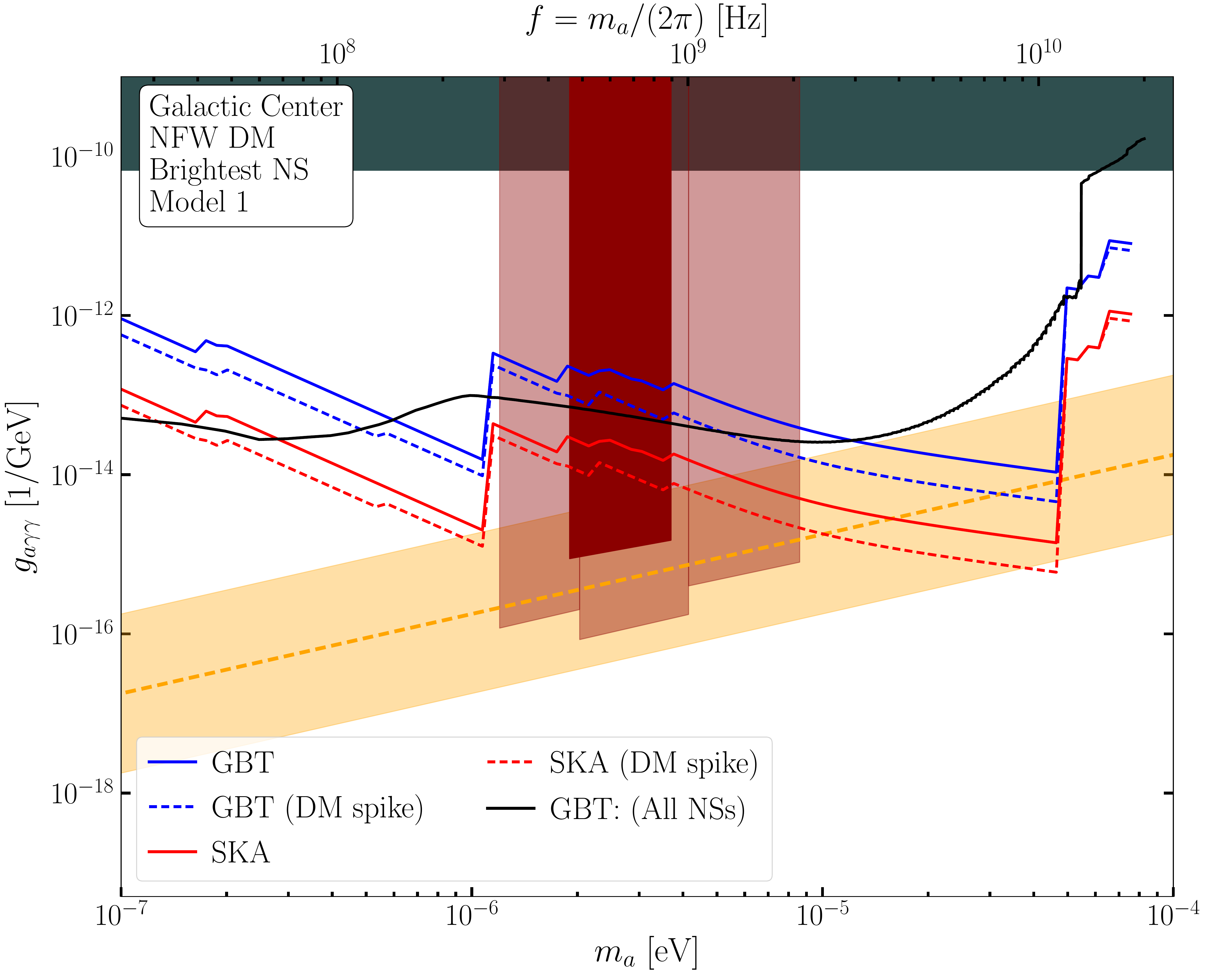} 
\includegraphics[width = 0.49\textwidth]{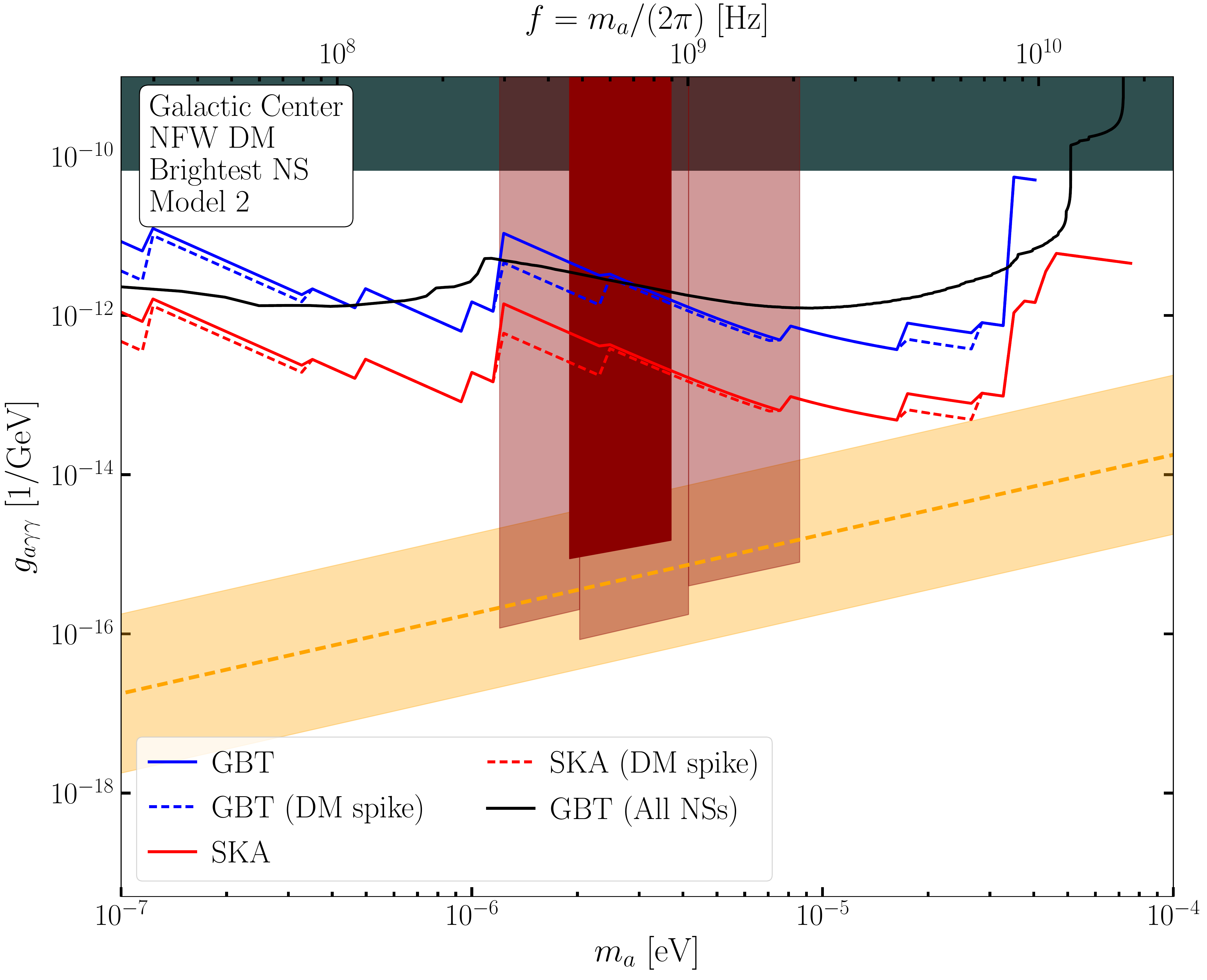}
\caption{
Projected sensitivities to axion DM from 24 hrs of observation time towards the Galactic Center with the telescopes indicated and requiring $5\sigma$ detection.  We consider variations to the DM profile, with NFW DM (top left) and Burkert DM (top right), as well as a possible DM spike at the Galactic Center.  The top two panels are for the analysis that looks for the broad ($\delta f / f \sim 10^{-3}$) emission from the ensemble of NSs, while the bottom two panels compare the sensitivity of this technique to that which looks only for the brightest individual NS, with $\delta f / f \sim 10^{-6}$.  In addition to variations in the DM density profile, we also show the uncertainty arising in these projections from the NS population models by switching between models 1 and 2.  All sensitivity curves exhibit two noticeable features: a slight increase in sensitivity below $\sim$$10^{-6}$ eV and a sharp decrease in sensitivity at high $m_a$.  The former feature is due, in part, to the transition to masses small enough that axions may resonantly convert in the ion-dominated part of the plasma.  The latter feature is due to the fact that at high masses the conditions for resonant conversion are not obtained, and the axion-photon transition occurs non-resonantly.  Note that in addition to the projected sensitivities, we also display the motivated region for the QCD axion, current constraints from the CAST experiment, and current and projected constraints from the ADMX experiment.
}
\label{fig: GC}
\end{figure*}

We begin by considering the Galactic Center region.  We assume 24 hr of observation time and require a SNR of $5.0$, corresponding to $\delta \chi^2 = 25$ or 5$\sigma$ detection.  Using the sky-temperature for the Galactic Center and the parameters described above, we find the projected sensitivities to $g_{a \gamma \gamma}$ shown in Fig.~\ref{fig: GC}, as functions of the axion mass $m_a$.  The top left and right panels, which we discuss first, are for searches for the broad-spectrum ($\delta f / f \sim 10^{-3}$) emission from the collection of NSs within the field of view.  The top left panel assumes an NFW DM profile while the top right takes the Burkert DM profile.  In both panels, we show the sensitivities calculated for NS models 1 and 2 and for three different telescope configurations, as described in Tab.~\ref{tab: telescopes}: GBT, VLA, and a 25-m dish.  Additionally, for GBT we show how the sensitivity changes, for NS model 1, if there is a DM spike at the center of the Galaxy.  The ranges between the NS models 1 and 2 curves can be thought of as an estimate for the current systematic uncertainty on the sensitivities arising from the NS model assumptions.  The differences between the two panels, on the other hand, is an estimate for the systematic coming from the DM density distribution model assumptions.  Except at the highest frequencies, the dominant uncertainty is the NS population model in this case.

In Fig.~\ref{fig: GC} we compare the projected sensitivities to the limit set by CAST~\cite{Arik:2013nya,Arik:2015cjv,Anastassopoulos:2017ftl}, the current (dark red) and possible future (light red) limits by ADMX~\cite{Shokair:2014rna,Rosenberg:2015kxa}, as well as the parameter space (roughly indicated in orange) where the QCD axion is expected to live.  In the case of NFW DM and NS model 1, searches around a few GHz with GBT could approach sensitivity to QCD axion DM with this search strategy.  However, we stress that any sensitivity curve passing below the CAST limit and outside of the ADMX limit would probe uncharted parameter space for ALP DM. 

One takeaway point from Fig.~\ref{fig: GC} is that a telescope array such as VLA is not the ideal instrument for searching for the extended signal.  This should not be too surprising, given the discussion in Sec.~\ref{sec: array} for why telescope arrays lose in sensitivity compared to single dishes of the same total area when the source size is larger than the synthesized beam.  This is especially clear when comparing the VLA array sensitivity to the hypothetical 25-m dish sensitivity, since VLA itself is an array of 25-m dishes.  In the case of a Burkert DM profile, where the inner regions of the galaxy are not necessarily brighter than the outer regions of the stellar bulge in terms of the DM-induced signal, a single 25-m dish can actually outperform an array of such dishes.  This is because, recalling the discussion in Sec.~\ref{sec: array}, the significance in favor of detection increases with the number of array elements but decreases with the number of synthesized beams, which is a function of the array baseline.  In fact, in this case the 25-m single dish can even outperform a 100-m single dish, when pointed at the Galactic Center.  This is due to the fact that the regions further away from the Galactic Center have a higher signal-to-noise ratio, due to the decreased background temperature, compared to the Galactic Center itself.  In this case, it would be better to point the 100-m telescope slightly off the Galactic Center, though we do not consider this possibility in detail since it is a small perturbation to the sensitivities shown in Fig.~\ref{fig: GC}. 

Lastly, we note that the data from the VLA array need not be combined in the interferometer mode, but rather could be combined in the naive fashion ({\it i.e.}, in series) to effectively increase the integration time by the number of array elements, with the gain at the single-element level.  This would be the best use of VLA in this case; the sensitivity to $g_{a\gamma\gamma}$ may simply be estimated by dividing the 25-m dish sensitivity by $27^{1/4}$.

In the bottom two panels in Fig.~\ref{fig: GC} we compare the sensitivity estimated using GBT and searching for the ensemble signal over all NSs (solid black) to the estimated sensitivities for the searches for the brightest individual NSs.  As a reminder of this strategy, we refer back to Fig.~\ref{fig: freq_zoom}.  In that figure, the emission over all NSs spans a relative frequency bandwidth $\delta f / f \sim 10^{-3}$, while the individual NSs have $\delta f / f \sim 10^{-6}$.  The search for the single brightest NS can have increased sensitivity compared to the search for the combined emission if a significant portion of the combined emission actually comes from a single NS, since one gains in sensitivity by having a narrower bandwidth $\delta f$.  Generally, as we go to higher frequencies the number of NSs contributing to the axion-induced signal shrinks, since the beam area shrinks and the fraction of NSs with appropriate $m_a^\text{max}$ decreases as well.  This behavior is seen in the bottom right panel of Fig.~\ref{fig: GC}, for example, which takes the NFW DM profile and the NS model 2.  The solid blue curve is the estimated sensitivity with GBT, which can be directly compared to the solid black curve.  Note that the dotted curves estimate the sensitivity including the possible DM spike, which can play a more important role here since it may enhance the flux of the brightest NS.  The red curves show the estimated sensitivity using the future SKA2, as described in Tab.~\ref{tab: telescopes}.

For NS model 1, as illustrated in the bottom left panel of Fig.~\ref{fig: GC}, the brightest NS search is seen to be highly variable over the entire frequency range, and even at low frequencies there are regions where the single-NS GBT search is more sensitive than the GBT search for the ensemble signal.  This variability is due to the fact that for NS model 1, even at low frequencies the signal is dominated by a few NSs with high $m_a^\text{max}$; these are NSs with very high magnetic fields.  In the case of SKA2 and NS model 1, this means that sensitivity to the QCD axion could be reached from frequencies as low as $m_a \sim 10^{-6}$ eV all the way to $m_a \sim 5 \times 10^{-5}$ eV.  However, it is important to note that the sensitivities in Fig.~\ref{fig: GC} for the single NS searches appear jagged because these are from individual simulations of the NSs at the Galactic Center.  That is, performing a different simulation, with the same initial conditions, would give rise to a slightly different looking sensitivity curve.  The ensemble NS curves are smoother, in part, because these have been averaged over many different simulations.

\subsubsection{M54}

\begin{figure*}[htb]
\includegraphics[width = 0.49\textwidth]{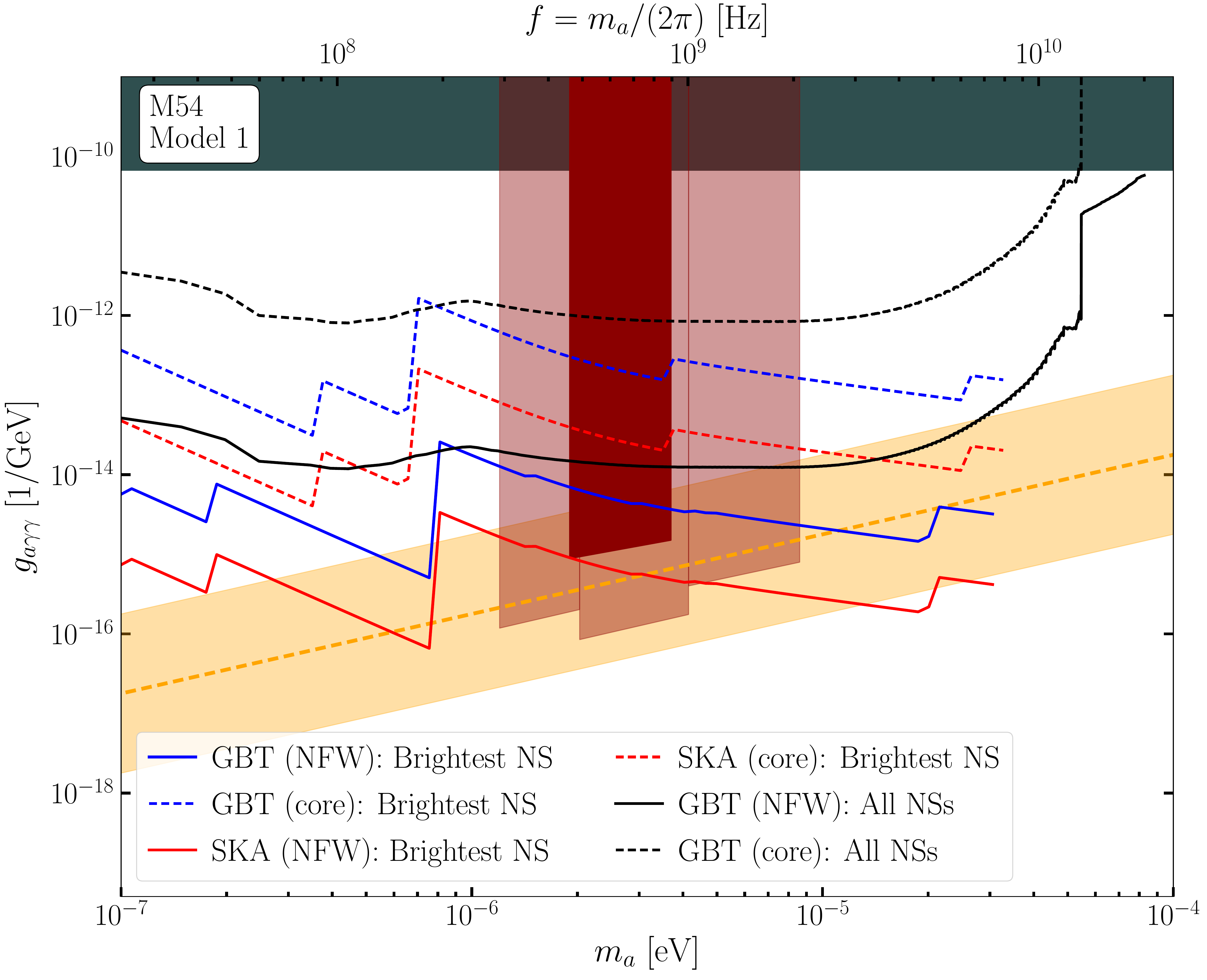} 
\includegraphics[width = 0.49\textwidth]{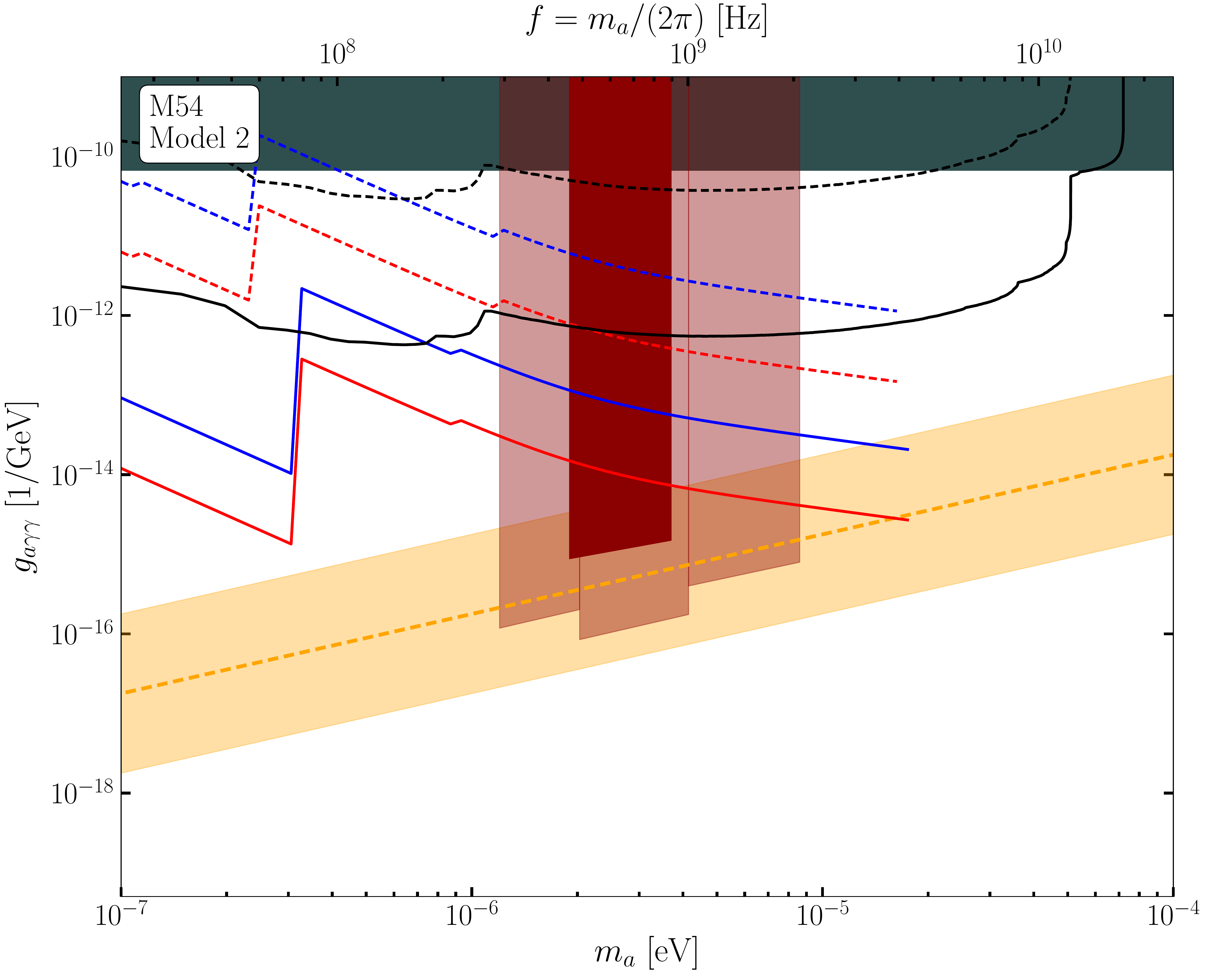} 
\caption{As in Fig.~\ref{fig: GC}, except for observations of the globular cluster M54.  The left panel assumes NS model 1, while the right takes NS model 2.  The solid black curves illustrate the sensitivity for the broad search for the ($\delta f / f \sim 10^{-4}$) signal from the ensemble of NSs, while the colored curves project the sensitivity to the single brightest NS (with $\delta f /f \sim 10^{-8}$) for one illustrative Monte Carlo realization.
}
\label{fig: M54}
\end{figure*}
In this subsection we consider the projected sensitivity from observations of M54.  
 The projected sensitivities for 24 hr of observation with GBT-like and SKA2-like telescopes, with analogous requirements to those taken for the Galactic Center examples, are shown in Fig.~\ref{fig: M54}.   In the left panel we use NS model 1, while in the right panel we instead take NS model 2, which has less high $m_a^\text{max}$ and high-$B$ NSs.  In both panels, we consider the NFW and cored DM profiles in solid and dashed, respectively.  Also, we estimate the sensitivity for the search for all NSs, with $\delta f / f \sim 10^{-4}$, and the search for the brightest individual NS, with $\delta f / f \sim 10^{-8}$.  In this case, unlike for the Galactic Center observation, the brightest NS search outperforms the search for the ensemble signal over almost the entire frequency range.  This is because the relative width of the individual NSs is much narrower, in the case of M54, and because there are simply less NSs to begin with in the beam area.  The M54 observations appear very promising; in the case of an NFW DM profile, GBT and SKA observations could have sensitivity to significant regions of the QCD axion parameter space.  Even in the case of a cored DM profile, if NS model 1 is correct then the SKA observations would still reach the QCD axion region at high frequencies.  M54 appears to be one of the most promising targets for future radio-line observations of axion DM.

\subsubsection{M31}

Radio observations of the nearby galaxy M31, which is similar in size to the Milky Way, will contain more NSs within the field of view relative to Milky Way Galactic Center observations because of the increased distance to the source.  
\begin{figure}[htb]
\includegraphics[width = 0.49\textwidth]{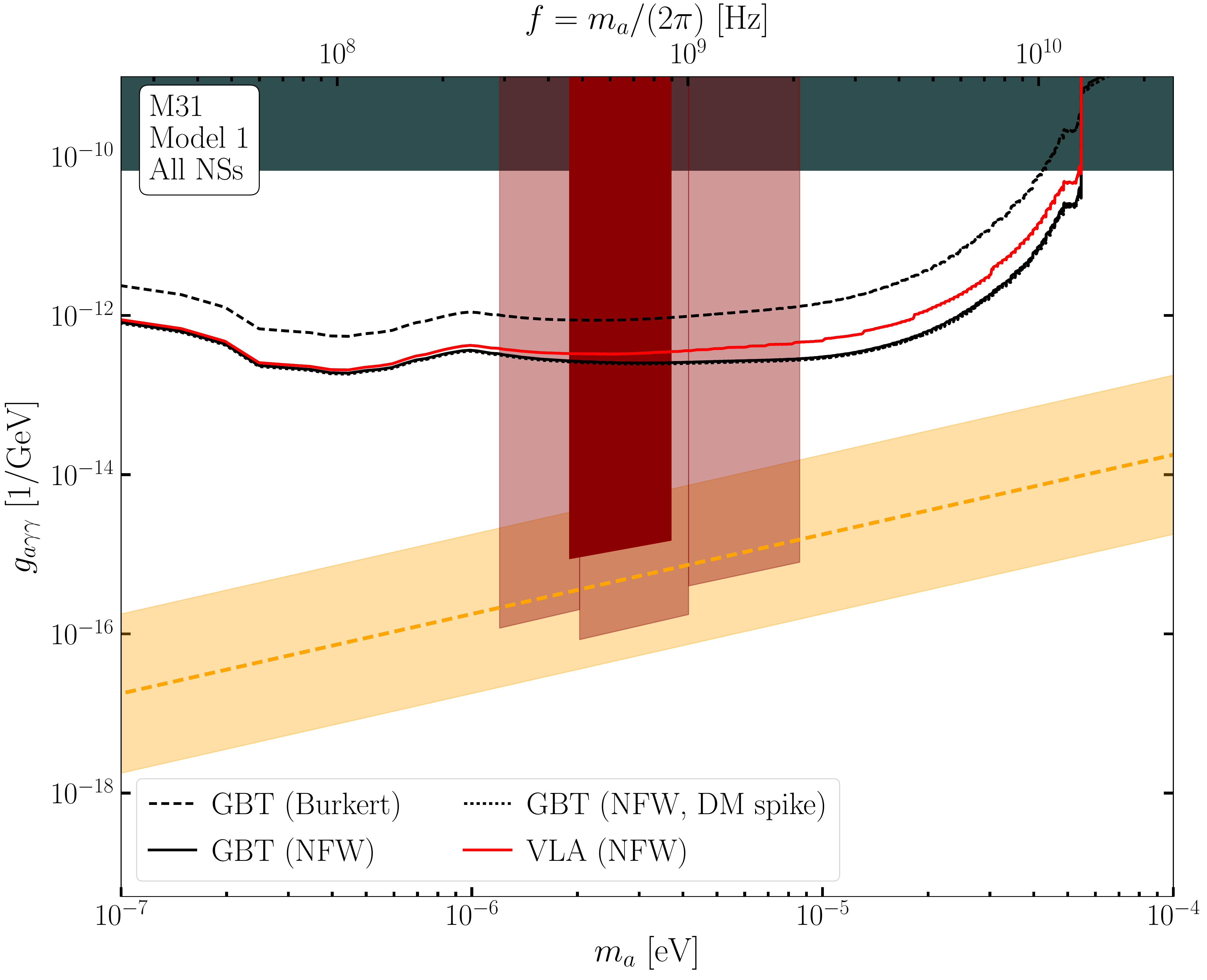}  
\caption{As in Fig.~\ref{fig: GC}, except for observations of the nearby galaxy M31.  Here, we only consider the search for the ensemble of NSs, since there are many more NSs within the field of view for the case of M31 relative to the Galactic Center.  The frequency spread of this signal is expected to be $\delta f / f \sim 10^{-3}$.  We illustrate the sensitivity for NS model 1 and for the Burkert and NFW DM profiles.
}
\label{fig: M31}
\end{figure}
Fig.~\ref{fig: M31} shows the projected sensitivities of observations of M31, assuming the same observational setups as used in the previous subsections.  For illustration, we use NS model 1 for all the projected sensitivities and we also only consider the search for the ensemble of NSs, with $\delta f / f \sim 10^{-3}$.  The search for the brightest individual NS is not competitive in this case, since each individual NS is quite dim but there are many more NSs within the field of view compared to even {\it e.g.} the Milky Way Galactic Center case.  For the NFW DM profile, we show both the GBT and VLA projected sensitivities.  While again GBT outperforms VLA, the difference is smaller compared to the Galactic Center observations since M31 is a smaller target on the sky.  Changing between DM profiles, we see that the difference between the Burkert and NFW DM profiles is also less pronounced than in the Galactic Center case.  This is again related to the fact that there are more NSs within the field of view, so the signal is dominated by NSs further away from the Galactic Center of M31, where the differences in DM density between the two models is less severe.  Similarly, including the central DM spike of M31 makes virtually no difference to the sensitivity, despite the large central black hole of M31, because the contributions of the inner NSs are subdominant compared to the contributions of the ensemble of NSs further away from the center of the galaxy.

Observations of M31 appear less promising than Galactic Center and M54 observations.  On the other hand, the M31 observations appear to be remarkably robust, owing to the fact that there are more NSs within the field of view.

\section{Discussion}
\label{sec: discussion}

In this work, we have outlined a framework for estimating the sensitivity of radio observations of astrophysical systems containing NSs to axion DM.  This work expands upon the recent work in~\cite{Hook:2018iia}, which outlined how to calculate the radio flux from axion-photon conversion in an individual NS, in four main directions: (i) we consider the expected spatial distributions of NSs in galactic systems, in conjunction with the spatial distribution of DM, (ii) we model the distribution of NS properties using NS models that have been tuned to observable pulsar data, (iii) we model the inactive NS magnetosphere using state-of-the-art numerical simulations of the electrosphere solution, and (iv) we precisely model the sensitivity, including backgrounds, for specific telescopes such as GBT and for specific astrophysical targets, such as the Galactic Center, M54, and M31.

We find that observations with modest observing times $\sim$24 hrs of the targets listed above would probe uncharted regions of ALP DM parameter space over a large range of possible ALP masses.  Observations of the Galactic Center and M54 have the potential to reach sensitivity to the QCD axion scenario, even with GBT.  However, this conclusion is subject to systematic uncertainties from the DM profiles within these systems and the distributions of NS properties.  The former systematic, related to the distribution of DM within galaxies and dwarf galaxies, is one that affects all indirect DM searches.  Improving upon this systematic will likely require better hydrodynamic simulations of galaxies and their DM halos, in addition to better measurements of kinematic tracers of the gravitational potentials within these systems.  The latter systematic, concerning the distribution of NS properties, is unique to this search strategy.  
Depending on the evolution of magnetic fields of NSs on timescales of order the age of the Galaxy, the large population of dead NSs could dominate the axion-induced flux.  On the other hand, if those fields decay on these long timescales, then the younger active pulsars still provide a significant source of axion-induced radio flux that can be used to constrain axion DM. 

Understanding the evolution of magnetic fields of dead NSs at late times is crucial for making more accurate predictions for the sensitivity of radio searches to axion DM.  Additionally, more work is needed to properly model the magnetar population and the magnetospheres of these systems, since such high-field NSs are an important contribution to the flux from the active population. 
We hope that this potential avenue towards axion DM detection can serve as motivation for dedicated efforts to further understand the properties of these systems. 

On the observational side, we outlined the advantages and disadvantages of using telescope arrays, such as VLA or the future SKA, versus large single dishes, like GBT, in searching for the axion-induced signal.  For the Galactic Center, for example, the emission from the ensemble of NSs is extended on angular scales typically much larger than the synthesized beam areas of telescope arrays, making single dishes more favorable instruments.  On the other hand, we showed that searches for the brightest individual NSs can give comparable or increased sensitivity, relative to searches for the broader-spectrum emission from the ensemble of NSs.  Searches for the narrow-band bright single NSs may be performed equally well with either telescope arrays or single dishes; in this case arrays having the advantage of larger primary beam areas over which to search.  However, there is likely an analysis framework between these two extremes that gives superior sensitivity.  For example, instead of looking for the brightest NS one could look for the brightest two NSs, or the brightest $N$ NSs.  However, accounting for the trials factor in such a search is non-trivial, given that the signals from the NSs are doppler shifted.  We leave the construction of such an analysis technique to future work.

Lastly, we emphasize the complementarity with direct detection efforts.  If a signal candidate is seen in an indirect search along the lines of that outlined here, it would likely be straightforward to verify that this candidate is in fact due to axion DM by performing a laboratory search.  This is because resonant cavity searches such as ADMX are difficult in part because they do not know the frequency of the particle they are looking for, and they must scan over millions of different possible frequencies, each time adjusting the cavity resonant frequency, to cover a single decade of mass.  If the mass of the axion is known ahead of time, for example by an indirect detection signal candidate, checking this frequency with direct detection would be straightforward because the cavity experiments would not have to scan over a large range of frequencies.  
\newline

\begin{acknowledgements}~We thank A. Hook, Y. Kahn, and A. Spitkovsky for collaboration on early stages of this project, and we thank R. Eatough, Y. Kahn, and N. Rodd for useful comments on the manuscript.  We also thank V. Kondratiev, W. Peters, C. Weniger, and K. Zurek for useful discussions. 
The work of B.R.S. and Z.S. was supported in part by the DOE Early Career Grant DE-SC0019225.  This research was supported in part through computational resources and
services provided by Advanced Research Computing at the University of
Michigan, Ann Arbor.
\end{acknowledgements}

\bibliography{Axion}

\begin{thebibliography}{104}%
\makeatletter
\providecommand \@ifxundefined [1]{%
 \@ifx{#1\undefined}
}%
\providecommand \@ifnum [1]{%
 \ifnum #1\expandafter \@firstoftwo
 \else \expandafter \@secondoftwo
 \fi
}%
\providecommand \@ifx [1]{%
 \ifx #1\expandafter \@firstoftwo
 \else \expandafter \@secondoftwo
 \fi
}%
\providecommand \natexlab [1]{#1}%
\providecommand \enquote  [1]{``#1''}%
\providecommand \bibnamefont  [1]{#1}%
\providecommand \bibfnamefont [1]{#1}%
\providecommand \citenamefont [1]{#1}%
\providecommand \href@noop [0]{\@secondoftwo}%
\providecommand \href [0]{\begingroup \@sanitize@url \@href}%
\providecommand \@href[1]{\@@startlink{#1}\@@href}%
\providecommand \@@href[1]{\endgroup#1\@@endlink}%
\providecommand \@sanitize@url [0]{\catcode `\\12\catcode `\$12\catcode
  `\&12\catcode `\#12\catcode `\^12\catcode `\_12\catcode `\%12\relax}%
\providecommand \@@startlink[1]{}%
\providecommand \@@endlink[0]{}%
\providecommand \url  [0]{\begingroup\@sanitize@url \@url }%
\providecommand \@url [1]{\endgroup\@href {#1}{\urlprefix }}%
\providecommand \urlprefix  [0]{URL }%
\providecommand \Eprint [0]{\href }%
\providecommand \doibase [0]{http://dx.doi.org/}%
\providecommand \selectlanguage [0]{\@gobble}%
\providecommand \bibinfo  [0]{\@secondoftwo}%
\providecommand \bibfield  [0]{\@secondoftwo}%
\providecommand \translation [1]{[#1]}%
\providecommand \BibitemOpen [0]{}%
\providecommand \bibitemStop [0]{}%
\providecommand \bibitemNoStop [0]{.\EOS\space}%
\providecommand \EOS [0]{\spacefactor3000\relax}%
\providecommand \BibitemShut  [1]{\csname bibitem#1\endcsname}%
\let\auto@bib@innerbib\@empty
%</preamble>
\bibitem [{\citenamefont {Hook}\ \emph {et~al.}(2018)\citenamefont {Hook},
  \citenamefont {Kahn}, \citenamefont {Safdi},\ and\ \citenamefont
  {Sun}}]{Hook:2018iia}%
  \BibitemOpen
  \bibfield  {author} {\bibinfo {author} {\bibfnamefont {Anson}\ \bibnamefont
  {Hook}}, \bibinfo {author} {\bibfnamefont {Yonatan}\ \bibnamefont {Kahn}},
  \bibinfo {author} {\bibfnamefont {Benjamin~R.}\ \bibnamefont {Safdi}}, \ and\
  \bibinfo {author} {\bibfnamefont {Zhiquan}\ \bibnamefont {Sun}},\ }\bibfield
  {title} {\enquote {\bibinfo {title} {{Radio Signals from Axion Dark Matter
  Conversion in Neutron Star Magnetospheres}},}\ }\href@noop {} {\  (\bibinfo
  {year} {2018})},\ \Eprint {http://arxiv.org/abs/1804.03145} {arXiv:1804.03145
  [hep-ph]} \BibitemShut {NoStop}%
%%CITATION = ARXIV:1804.03145;%%
\bibitem [{\citenamefont {Pshirkov}(2009)}]{Pshirkov:2007st}%
  \BibitemOpen
  \bibfield  {author} {\bibinfo {author} {\bibfnamefont {M.~S.}\ \bibnamefont
  {Pshirkov}},\ }\bibfield  {title} {\enquote {\bibinfo {title} {{Conversion of
  Dark matter axions to photons in magnetospheres of neutron stars}},}\ }\href
  {\doibase 10.1134/S1063776109030030} {\bibfield  {journal} {\bibinfo
  {journal} {J. Exp. Theor. Phys.}\ }\textbf {\bibinfo {volume} {108}},\
  \bibinfo {pages} {384--388} (\bibinfo {year} {2009})},\ \Eprint
  {http://arxiv.org/abs/0711.1264} {arXiv:0711.1264 [astro-ph]} \BibitemShut
  {NoStop}%
%%CITATION = ARXIV:0711.1264;%%
\bibitem [{\citenamefont {Huang}\ \emph {et~al.}(2018)\citenamefont {Huang},
  \citenamefont {Kadota}, \citenamefont {Sekiguchi},\ and\ \citenamefont
  {Tashiro}}]{Huang:2018lxq}%
  \BibitemOpen
  \bibfield  {author} {\bibinfo {author} {\bibfnamefont {Fa~Peng}\ \bibnamefont
  {Huang}}, \bibinfo {author} {\bibfnamefont {Kenji}\ \bibnamefont {Kadota}},
  \bibinfo {author} {\bibfnamefont {Toyokazu}\ \bibnamefont {Sekiguchi}}, \
  and\ \bibinfo {author} {\bibfnamefont {Hiroyuki}\ \bibnamefont {Tashiro}},\
  }\bibfield  {title} {\enquote {\bibinfo {title} {{The radio telescope search
  for the resonant conversion of cold dark matter axions from the magnetized
  astrophysical sources}},}\ }\href@noop {} {\  (\bibinfo {year} {2018})},\
  \Eprint {http://arxiv.org/abs/1803.08230} {arXiv:1803.08230 [hep-ph]}
  \BibitemShut {NoStop}%
%%CITATION = ARXIV:1803.08230;%%
\bibitem [{\citenamefont {Peccei}\ and\ \citenamefont
  {Quinn}(1977{\natexlab{a}})}]{Peccei:1977hh}%
  \BibitemOpen
  \bibfield  {author} {\bibinfo {author} {\bibfnamefont {R.~D.}\ \bibnamefont
  {Peccei}}\ and\ \bibinfo {author} {\bibfnamefont {Helen~R.}\ \bibnamefont
  {Quinn}},\ }\bibfield  {title} {\enquote {\bibinfo {title} {{CP Conservation
  in the Presence of Instantons}},}\ }\href {\doibase
  10.1103/PhysRevLett.38.1440} {\bibfield  {journal} {\bibinfo  {journal}
  {Phys. Rev. Lett.}\ }\textbf {\bibinfo {volume} {38}},\ \bibinfo {pages}
  {1440--1443} (\bibinfo {year} {1977}{\natexlab{a}})}\BibitemShut {NoStop}%
%%CITATION = PRLTA,38,1440;%%
\bibitem [{\citenamefont {Peccei}\ and\ \citenamefont
  {Quinn}(1977{\natexlab{b}})}]{Peccei:1977ur}%
  \BibitemOpen
  \bibfield  {author} {\bibinfo {author} {\bibfnamefont {R.~D.}\ \bibnamefont
  {Peccei}}\ and\ \bibinfo {author} {\bibfnamefont {Helen~R.}\ \bibnamefont
  {Quinn}},\ }\bibfield  {title} {\enquote {\bibinfo {title} {{Constraints
  Imposed by CP Conservation in the Presence of Instantons}},}\ }\href
  {\doibase 10.1103/PhysRevD.16.1791} {\bibfield  {journal} {\bibinfo
  {journal} {Phys. Rev.}\ }\textbf {\bibinfo {volume} {D16}},\ \bibinfo {pages}
  {1791--1797} (\bibinfo {year} {1977}{\natexlab{b}})}\BibitemShut {NoStop}%
%%CITATION = PHRVA,D16,1791;%%
\bibitem [{\citenamefont {Weinberg}(1978)}]{Weinberg:1977ma}%
  \BibitemOpen
  \bibfield  {author} {\bibinfo {author} {\bibfnamefont {Steven}\ \bibnamefont
  {Weinberg}},\ }\bibfield  {title} {\enquote {\bibinfo {title} {{A New Light
  Boson?}}}\ }\href {\doibase 10.1103/PhysRevLett.40.223} {\bibfield  {journal}
  {\bibinfo  {journal} {Phys.Rev.Lett.}\ }\textbf {\bibinfo {volume} {40}},\
  \bibinfo {pages} {223--226} (\bibinfo {year} {1978})}\BibitemShut {NoStop}%
%%CITATION = PRLTA,40,223;%%
\bibitem [{\citenamefont {Wilczek}(1978)}]{Wilczek:1977pj}%
  \BibitemOpen
  \bibfield  {author} {\bibinfo {author} {\bibfnamefont {Frank}\ \bibnamefont
  {Wilczek}},\ }\bibfield  {title} {\enquote {\bibinfo {title} {{Problem of
  Strong P and T Invariance in the Presence of Instantons}},}\ }\href {\doibase
  10.1103/PhysRevLett.40.279} {\bibfield  {journal} {\bibinfo  {journal}
  {Phys.Rev.Lett.}\ }\textbf {\bibinfo {volume} {40}},\ \bibinfo {pages}
  {279--282} (\bibinfo {year} {1978})}\BibitemShut {NoStop}%
%%CITATION = PRLTA,40,279;%%
\bibitem [{\citenamefont {Preskill}\ \emph {et~al.}(1983)\citenamefont
  {Preskill}, \citenamefont {Wise},\ and\ \citenamefont
  {Wilczek}}]{Preskill:1982cy}%
  \BibitemOpen
  \bibfield  {author} {\bibinfo {author} {\bibfnamefont {John}\ \bibnamefont
  {Preskill}}, \bibinfo {author} {\bibfnamefont {Mark~B.}\ \bibnamefont
  {Wise}}, \ and\ \bibinfo {author} {\bibfnamefont {Frank}\ \bibnamefont
  {Wilczek}},\ }\bibfield  {title} {\enquote {\bibinfo {title} {{Cosmology of
  the Invisible Axion}},}\ }\href {\doibase 10.1016/0370-2693(83)90637-8}
  {\bibfield  {journal} {\bibinfo  {journal} {Phys. Lett.}\ }\textbf {\bibinfo
  {volume} {B120}},\ \bibinfo {pages} {127--132} (\bibinfo {year}
  {1983})}\BibitemShut {NoStop}%
%%CITATION = PHLTA,B120,127;%%
\bibitem [{\citenamefont {Abbott}\ and\ \citenamefont
  {Sikivie}(1983)}]{Abbott:1982af}%
  \BibitemOpen
  \bibfield  {author} {\bibinfo {author} {\bibfnamefont {L.~F.}\ \bibnamefont
  {Abbott}}\ and\ \bibinfo {author} {\bibfnamefont {P.}~\bibnamefont
  {Sikivie}},\ }\bibfield  {title} {\enquote {\bibinfo {title} {{A Cosmological
  Bound on the Invisible Axion}},}\ }\href {\doibase
  10.1016/0370-2693(83)90638-X} {\bibfield  {journal} {\bibinfo  {journal}
  {Phys. Lett.}\ }\textbf {\bibinfo {volume} {B120}},\ \bibinfo {pages}
  {133--136} (\bibinfo {year} {1983})}\BibitemShut {NoStop}%
%%CITATION = PHLTA,B120,133;%%
\bibitem [{\citenamefont {Dine}\ and\ \citenamefont
  {Fischler}(1983)}]{Dine:1982ah}%
  \BibitemOpen
  \bibfield  {author} {\bibinfo {author} {\bibfnamefont {Michael}\ \bibnamefont
  {Dine}}\ and\ \bibinfo {author} {\bibfnamefont {Willy}\ \bibnamefont
  {Fischler}},\ }\bibfield  {title} {\enquote {\bibinfo {title} {{The Not So
  Harmless Axion}},}\ }\href {\doibase 10.1016/0370-2693(83)90639-1} {\bibfield
   {journal} {\bibinfo  {journal} {Phys. Lett.}\ }\textbf {\bibinfo {volume}
  {B120}},\ \bibinfo {pages} {137--141} (\bibinfo {year} {1983})}\BibitemShut
  {NoStop}%
%%CITATION = PHLTA,B120,137;%%
\bibitem [{\citenamefont {Dine}\ \emph {et~al.}(1981)\citenamefont {Dine},
  \citenamefont {Fischler},\ and\ \citenamefont {Srednicki}}]{Dine:1981rt}%
  \BibitemOpen
  \bibfield  {author} {\bibinfo {author} {\bibfnamefont {Michael}\ \bibnamefont
  {Dine}}, \bibinfo {author} {\bibfnamefont {Willy}\ \bibnamefont {Fischler}},
  \ and\ \bibinfo {author} {\bibfnamefont {Mark}\ \bibnamefont {Srednicki}},\
  }\bibfield  {title} {\enquote {\bibinfo {title} {{A Simple Solution to the
  Strong CP Problem with a Harmless Axion}},}\ }\href {\doibase
  10.1016/0370-2693(81)90590-6} {\bibfield  {journal} {\bibinfo  {journal}
  {Phys. Lett.}\ }\textbf {\bibinfo {volume} {B104}},\ \bibinfo {pages} {199}
  (\bibinfo {year} {1981})}\BibitemShut {NoStop}%
%%CITATION = PHLTA,B104,199;%%
\bibitem [{\citenamefont {Zhitnitsky}(1980)}]{Zhitnitsky:1980tq}%
  \BibitemOpen
  \bibfield  {author} {\bibinfo {author} {\bibfnamefont {A.~R.}\ \bibnamefont
  {Zhitnitsky}},\ }\bibfield  {title} {\enquote {\bibinfo {title} {{On Possible
  Suppression of the Axion Hadron Interactions. (In Russian)}},}\ }\href@noop
  {} {\bibfield  {journal} {\bibinfo  {journal} {Sov. J. Nucl. Phys.}\ }\textbf
  {\bibinfo {volume} {31}},\ \bibinfo {pages} {260} (\bibinfo {year} {1980})},\
  \bibinfo {note} {[Yad. Fiz.31,497(1980)]}\BibitemShut {NoStop}%
%%CITATION = SJNCA,31,260;%%
\bibitem [{\citenamefont {Kim}(1979)}]{Kim:1979if}%
  \BibitemOpen
  \bibfield  {author} {\bibinfo {author} {\bibfnamefont {Jihn~E.}\ \bibnamefont
  {Kim}},\ }\bibfield  {title} {\enquote {\bibinfo {title} {{Weak Interaction
  Singlet and Strong CP Invariance}},}\ }\href {\doibase
  10.1103/PhysRevLett.43.103} {\bibfield  {journal} {\bibinfo  {journal} {Phys.
  Rev. Lett.}\ }\textbf {\bibinfo {volume} {43}},\ \bibinfo {pages} {103}
  (\bibinfo {year} {1979})}\BibitemShut {NoStop}%
%%CITATION = PRLTA,43,103;%%
\bibitem [{\citenamefont {Shifman}\ \emph {et~al.}(1980)\citenamefont
  {Shifman}, \citenamefont {Vainshtein},\ and\ \citenamefont
  {Zakharov}}]{Shifman:1979if}%
  \BibitemOpen
  \bibfield  {author} {\bibinfo {author} {\bibfnamefont {Mikhail~A.}\
  \bibnamefont {Shifman}}, \bibinfo {author} {\bibfnamefont {A.~I.}\
  \bibnamefont {Vainshtein}}, \ and\ \bibinfo {author} {\bibfnamefont
  {Valentin~I.}\ \bibnamefont {Zakharov}},\ }\bibfield  {title} {\enquote
  {\bibinfo {title} {{Can Confinement Ensure Natural CP Invariance of Strong
  Interactions?}}}\ }\href {\doibase 10.1016/0550-3213(80)90209-6} {\bibfield
  {journal} {\bibinfo  {journal} {Nucl. Phys.}\ }\textbf {\bibinfo {volume}
  {B166}},\ \bibinfo {pages} {493} (\bibinfo {year} {1980})}\BibitemShut
  {NoStop}%
%%CITATION = NUPHA,B166,493;%%
\bibitem [{\citenamefont {Sikivie}(1983)}]{Sikivie:1983ip}%
  \BibitemOpen
  \bibfield  {author} {\bibinfo {author} {\bibfnamefont {P.}~\bibnamefont
  {Sikivie}},\ }\bibfield  {title} {\enquote {\bibinfo {title} {{Experimental
  Tests of the Invisible Axion}},}\ }\bibfield  {booktitle} {\emph {\bibinfo
  {booktitle} {{11th International Symposium on Lepton and Photon Interactions
  at High Energies Ithaca, New York, August 4-9, 1983}}},\ }\href {\doibase
  10.1103/PhysRevLett.51.1415} {\bibfield  {journal} {\bibinfo  {journal}
  {Phys. Rev. Lett.}\ }\textbf {\bibinfo {volume} {51}},\ \bibinfo {pages}
  {1415--1417} (\bibinfo {year} {1983})},\ \bibinfo {note} {[Erratum: Phys.
  Rev. Lett.52,695(1984)]}\BibitemShut {NoStop}%
%%CITATION = PRLTA,51,1415;%%
\bibitem [{\citenamefont {Raffelt}\ and\ \citenamefont
  {Stodolsky}(1988)}]{Raffelt:1987im}%
  \BibitemOpen
  \bibfield  {author} {\bibinfo {author} {\bibfnamefont {Georg}\ \bibnamefont
  {Raffelt}}\ and\ \bibinfo {author} {\bibfnamefont {Leo}\ \bibnamefont
  {Stodolsky}},\ }\bibfield  {title} {\enquote {\bibinfo {title} {{Mixing of
  the Photon with Low Mass Particles}},}\ }\href {\doibase
  10.1103/PhysRevD.37.1237} {\bibfield  {journal} {\bibinfo  {journal} {Phys.
  Rev.}\ }\textbf {\bibinfo {volume} {D37}},\ \bibinfo {pages} {1237} (\bibinfo
  {year} {1988})}\BibitemShut {NoStop}%
%%CITATION = PHRVA,D37,1237;%%
\bibitem [{\citenamefont {Svrcek}\ and\ \citenamefont
  {Witten}(2006)}]{Svrcek:2006yi}%
  \BibitemOpen
  \bibfield  {author} {\bibinfo {author} {\bibfnamefont {Peter}\ \bibnamefont
  {Svrcek}}\ and\ \bibinfo {author} {\bibfnamefont {Edward}\ \bibnamefont
  {Witten}},\ }\bibfield  {title} {\enquote {\bibinfo {title} {{Axions In
  String Theory}},}\ }\href {\doibase 10.1088/1126-6708/2006/06/051} {\bibfield
   {journal} {\bibinfo  {journal} {JHEP}\ }\textbf {\bibinfo {volume} {06}},\
  \bibinfo {pages} {051} (\bibinfo {year} {2006})},\ \Eprint
  {http://arxiv.org/abs/hep-th/0605206} {arXiv:hep-th/0605206 [hep-th]}
  \BibitemShut {NoStop}%
%%CITATION = HEP-TH/0605206;%%
\bibitem [{\citenamefont {Faucher-Giguere}\ and\ \citenamefont
  {Kaspi}(2006)}]{FaucherGiguere:2005ny}%
  \BibitemOpen
  \bibfield  {author} {\bibinfo {author} {\bibfnamefont {Claude-Andre}\
  \bibnamefont {Faucher-Giguere}}\ and\ \bibinfo {author} {\bibfnamefont
  {Victoria~M.}\ \bibnamefont {Kaspi}},\ }\bibfield  {title} {\enquote
  {\bibinfo {title} {{Birth and evolution of isolated radio pulsars}},}\ }\href
  {\doibase 10.1086/501516} {\bibfield  {journal} {\bibinfo  {journal}
  {Astrophys. J.}\ }\textbf {\bibinfo {volume} {643}},\ \bibinfo {pages}
  {332--355} (\bibinfo {year} {2006})},\ \Eprint
  {http://arxiv.org/abs/astro-ph/0512585} {arXiv:astro-ph/0512585 [astro-ph]}
  \BibitemShut {NoStop}%
%%CITATION = ASTRO-PH/0512585;%%
\bibitem [{\citenamefont {{Popov}}\ \emph {et~al.}(2010)\citenamefont
  {{Popov}}, \citenamefont {{Pons}}, \citenamefont {{Miralles}}, \citenamefont
  {{Boldin}},\ and\ \citenamefont {{Posselt}}}]{2010MNRAS.401.2675P}%
  \BibitemOpen
  \bibfield  {author} {\bibinfo {author} {\bibfnamefont {S.~B.}\ \bibnamefont
  {{Popov}}}, \bibinfo {author} {\bibfnamefont {J.~A.}\ \bibnamefont {{Pons}}},
  \bibinfo {author} {\bibfnamefont {J.~A.}\ \bibnamefont {{Miralles}}},
  \bibinfo {author} {\bibfnamefont {P.~A.}\ \bibnamefont {{Boldin}}}, \ and\
  \bibinfo {author} {\bibfnamefont {B.}~\bibnamefont {{Posselt}}},\ }\bibfield
  {title} {\enquote {\bibinfo {title} {{Population synthesis studies of
  isolated neutron stars with magnetic field decay}},}\ }\href {\doibase
  10.1111/j.1365-2966.2009.15850.x} {\bibfield  {journal} {\bibinfo  {journal}
  {MNRAS}\ }\textbf {\bibinfo {volume} {401}},\ \bibinfo {pages} {2675--2686}
  (\bibinfo {year} {2010})},\ \Eprint {http://arxiv.org/abs/0910.2190}
  {arXiv:0910.2190 [astro-ph.HE]} \BibitemShut {NoStop}%
\bibitem [{\citenamefont {Spitkovsky}(2006)}]{Spitkovsky:2006np}%
  \BibitemOpen
  \bibfield  {author} {\bibinfo {author} {\bibfnamefont {Anatoly}\ \bibnamefont
  {Spitkovsky}},\ }\bibfield  {title} {\enquote {\bibinfo {title}
  {{Time-dependent force-free pulsar magnetospheres: axisymmetric and oblique
  rotators}},}\ }\href {\doibase 10.1086/507518} {\bibfield  {journal}
  {\bibinfo  {journal} {Astrophys. J.}\ }\textbf {\bibinfo {volume} {648}},\
  \bibinfo {pages} {L51--L54} (\bibinfo {year} {2006})},\ \Eprint
  {http://arxiv.org/abs/astro-ph/0603147} {arXiv:astro-ph/0603147 [astro-ph]}
  \BibitemShut {NoStop}%
%%CITATION = ASTRO-PH/0603147;%%
\bibitem [{\citenamefont {{Pons}}\ \emph {et~al.}(2009)\citenamefont {{Pons}},
  \citenamefont {{Miralles}},\ and\ \citenamefont
  {{Geppert}}}]{2009A&A...496..207P}%
  \BibitemOpen
  \bibfield  {author} {\bibinfo {author} {\bibfnamefont {J.~A.}\ \bibnamefont
  {{Pons}}}, \bibinfo {author} {\bibfnamefont {J.~A.}\ \bibnamefont
  {{Miralles}}}, \ and\ \bibinfo {author} {\bibfnamefont {U.}~\bibnamefont
  {{Geppert}}},\ }\bibfield  {title} {\enquote {\bibinfo {title}
  {{Magneto-thermal evolution of neutron stars}},}\ }\href {\doibase
  10.1051/0004-6361:200811229} {\bibfield  {journal} {\bibinfo  {journal}
  {A\&A}\ }\textbf {\bibinfo {volume} {496}},\ \bibinfo {pages} {207--216}
  (\bibinfo {year} {2009})},\ \Eprint {http://arxiv.org/abs/0812.3018}
  {arXiv:0812.3018 [Astrophysics]} \BibitemShut {NoStop}%
\bibitem [{\citenamefont {{Boldin}}\ and\ \citenamefont
  {{Popov}}(2010)}]{2010MNRAS.407.1090B}%
  \BibitemOpen
  \bibfield  {author} {\bibinfo {author} {\bibfnamefont {P.~A.}\ \bibnamefont
  {{Boldin}}}\ and\ \bibinfo {author} {\bibfnamefont {S.~B.}\ \bibnamefont
  {{Popov}}},\ }\bibfield  {title} {\enquote {\bibinfo {title} {{The evolution
  of isolated neutron stars until accretion: the role of the initial magnetic
  field}},}\ }\href {\doibase 10.1111/j.1365-2966.2010.16910.x} {\bibfield
  {journal} {\bibinfo  {journal} {MNRAS}\ }\textbf {\bibinfo {volume} {407}},\
  \bibinfo {pages} {1090--1097} (\bibinfo {year} {2010})},\ \Eprint
  {http://arxiv.org/abs/1004.4805} {arXiv:1004.4805 [astro-ph.HE]} \BibitemShut
  {NoStop}%
\bibitem [{\citenamefont {Philippov}\ \emph {et~al.}(2014)\citenamefont
  {Philippov}, \citenamefont {Tchekhovskoy},\ and\ \citenamefont
  {Li}}]{Philippov:2013aha}%
  \BibitemOpen
  \bibfield  {author} {\bibinfo {author} {\bibfnamefont {Alexander}\
  \bibnamefont {Philippov}}, \bibinfo {author} {\bibfnamefont {Alexander}\
  \bibnamefont {Tchekhovskoy}}, \ and\ \bibinfo {author} {\bibfnamefont
  {Jason~G.}\ \bibnamefont {Li}},\ }\bibfield  {title} {\enquote {\bibinfo
  {title} {{Time evolution of pulsar obliquity angle from 3D simulations of
  magnetospheres}},}\ }\href {\doibase 10.1093/mnras/stu591} {\bibfield
  {journal} {\bibinfo  {journal} {Mon. Not. Roy. Astron. Soc.}\ }\textbf
  {\bibinfo {volume} {441}},\ \bibinfo {pages} {1879--1887} (\bibinfo {year}
  {2014})},\ \Eprint {http://arxiv.org/abs/1311.1513} {arXiv:1311.1513
  [astro-ph.HE]} \BibitemShut {NoStop}%
%%CITATION = ARXIV:1311.1513;%%
\bibitem [{\citenamefont {Johnston}\ and\ \citenamefont
  {Karastergiou}(2017)}]{Johnston:2017wgm}%
  \BibitemOpen
  \bibfield  {author} {\bibinfo {author} {\bibfnamefont {Simon}\ \bibnamefont
  {Johnston}}\ and\ \bibinfo {author} {\bibfnamefont {Aris}\ \bibnamefont
  {Karastergiou}},\ }\bibfield  {title} {\enquote {\bibinfo {title} {{Pulsar
  braking and the P--$\dot{P}$ diagram}},}\ }\href {\doibase
  10.1093/mnras/stx377} {\bibfield  {journal} {\bibinfo  {journal} {Mon. Not.
  Roy. Astron. Soc.}\ }\textbf {\bibinfo {volume} {467}},\ \bibinfo {pages}
  {3493--3499} (\bibinfo {year} {2017})},\ \Eprint
  {http://arxiv.org/abs/1702.03616} {arXiv:1702.03616 [astro-ph.HE]}
  \BibitemShut {NoStop}%
%%CITATION = ARXIV:1702.03616;%%
\bibitem [{\citenamefont {Zhang}\ \emph {et~al.}(2000)\citenamefont {Zhang},
  \citenamefont {Harding},\ and\ \citenamefont {Muslimov}}]{Zhang:2000rd}%
  \BibitemOpen
  \bibfield  {author} {\bibinfo {author} {\bibfnamefont {Bing}\ \bibnamefont
  {Zhang}}, \bibinfo {author} {\bibfnamefont {Alice~K.}\ \bibnamefont
  {Harding}}, \ and\ \bibinfo {author} {\bibfnamefont {Alexander~G.}\
  \bibnamefont {Muslimov}},\ }\bibfield  {title} {\enquote {\bibinfo {title}
  {{Radio pulsar death line revisited: Is PSR J2144-3933 anomalous?}}}\ }\href
  {\doibase 10.1086/312542} {\bibfield  {journal} {\bibinfo  {journal}
  {Astrophys. J.}\ }\textbf {\bibinfo {volume} {531}},\ \bibinfo {pages}
  {L135--L138} (\bibinfo {year} {2000})},\ \Eprint
  {http://arxiv.org/abs/astro-ph/0001341} {arXiv:astro-ph/0001341 [astro-ph]}
  \BibitemShut {NoStop}%
%%CITATION = ASTRO-PH/0001341;%%
\bibitem [{\citenamefont {Tan}\ \emph {et~al.}(2018)\citenamefont {Tan} \emph
  {et~al.}}]{Tan:2018rhg}%
  \BibitemOpen
  \bibfield  {author} {\bibinfo {author} {\bibfnamefont {C.~M.}\ \bibnamefont
  {Tan}} \emph {et~al.},\ }\bibfield  {title} {\enquote {\bibinfo {title}
  {{LOFAR Discovery of a 23.5 s Radio Pulsar}},}\ }\href {\doibase
  10.3847/1538-4357/aade88} {\bibfield  {journal} {\bibinfo  {journal}
  {Astrophys. J.}\ }\textbf {\bibinfo {volume} {866}},\ \bibinfo {pages} {54}
  (\bibinfo {year} {2018})},\ \Eprint {http://arxiv.org/abs/1809.00965}
  {arXiv:1809.00965 [astro-ph.HE]} \BibitemShut {NoStop}%
%%CITATION = ARXIV:1809.00965;%%
\bibitem [{\citenamefont {{Goldreich}}\ and\ \citenamefont
  {{Julian}}(1969)}]{1969ApJ...157..869G}%
  \BibitemOpen
  \bibfield  {author} {\bibinfo {author} {\bibfnamefont {P.}~\bibnamefont
  {{Goldreich}}}\ and\ \bibinfo {author} {\bibfnamefont {W.~H.}\ \bibnamefont
  {{Julian}}},\ }\bibfield  {title} {\enquote {\bibinfo {title} {{Pulsar
  Electrodynamics}},}\ }\href {\doibase 10.1086/150119} {\bibfield  {journal}
  {\bibinfo  {journal} {\apj}\ }\textbf {\bibinfo {volume} {157}},\ \bibinfo
  {pages} {869} (\bibinfo {year} {1969})}\BibitemShut {NoStop}%
\bibitem [{\citenamefont {{Krause-Polstorff}}\ and\ \citenamefont
  {{Michel}}(1985)}]{1985MNRAS.213P..43K}%
  \BibitemOpen
  \bibfield  {author} {\bibinfo {author} {\bibfnamefont {J.}~\bibnamefont
  {{Krause-Polstorff}}}\ and\ \bibinfo {author} {\bibfnamefont {F.~C.}\
  \bibnamefont {{Michel}}},\ }\bibfield  {title} {\enquote {\bibinfo {title}
  {{Electrosphere of an aligned magnetized neutron star}},}\ }\href {\doibase
  10.1093/mnras/213.1.43P} {\bibfield  {journal} {\bibinfo  {journal} {MNRAS}\
  }\textbf {\bibinfo {volume} {213}},\ \bibinfo {pages} {43P--49P} (\bibinfo
  {year} {1985})}\BibitemShut {NoStop}%
\bibitem [{\citenamefont {Spitkovsky}\ and\ \citenamefont
  {Arons}(2002)}]{Spitkovsky:2002wg}%
  \BibitemOpen
  \bibfield  {author} {\bibinfo {author} {\bibfnamefont {Anatoly}\ \bibnamefont
  {Spitkovsky}}\ and\ \bibinfo {author} {\bibfnamefont {Jonathan}\ \bibnamefont
  {Arons}},\ }\bibfield  {title} {\enquote {\bibinfo {title} {{Simulations of
  pulsar wind formation}},}\ }\bibfield  {booktitle} {\emph {\bibinfo
  {booktitle} {{Proceedings, Conference on Neutron Stars in Supernova Remnants:
  Boston, Massachusetts, August 14-17, 2001}}},\ }\href@noop {} {\bibfield
  {journal} {\bibinfo  {journal} {ASP Conf. Ser.}\ }\textbf {\bibinfo {volume}
  {271}},\ \bibinfo {pages} {81} (\bibinfo {year} {2002})},\ \Eprint
  {http://arxiv.org/abs/astro-ph/0201360} {arXiv:astro-ph/0201360 [astro-ph]}
  \BibitemShut {NoStop}%
%%CITATION = ASTRO-PH/0201360;%%
\bibitem [{\citenamefont {{P{\'e}tri}}\ \emph {et~al.}(2002)\citenamefont
  {{P{\'e}tri}}, \citenamefont {{Heyvaerts}},\ and\ \citenamefont
  {{Bonazzola}}}]{2002A&A...387..520P}%
  \BibitemOpen
  \bibfield  {author} {\bibinfo {author} {\bibfnamefont {J.}~\bibnamefont
  {{P{\'e}tri}}}, \bibinfo {author} {\bibfnamefont {J.}~\bibnamefont
  {{Heyvaerts}}}, \ and\ \bibinfo {author} {\bibfnamefont {S.}~\bibnamefont
  {{Bonazzola}}},\ }\bibfield  {title} {\enquote {\bibinfo {title} {{Diocotron
  instability in pulsar electrospheres. I. Linear analysis}},}\ }\href
  {\doibase 10.1051/0004-6361:20020442} {\bibfield  {journal} {\bibinfo
  {journal} {A\&A}\ }\textbf {\bibinfo {volume} {387}},\ \bibinfo {pages}
  {520--530} (\bibinfo {year} {2002})}\BibitemShut {NoStop}%
\bibitem [{\citenamefont {Cerutti}\ and\ \citenamefont
  {Beloborodov}(2017)}]{Cerutti:2016ttn}%
  \BibitemOpen
  \bibfield  {author} {\bibinfo {author} {\bibfnamefont {Beno{\^\i}t}\
  \bibnamefont {Cerutti}}\ and\ \bibinfo {author} {\bibfnamefont {Andrei}\
  \bibnamefont {Beloborodov}},\ }\bibfield  {title} {\enquote {\bibinfo {title}
  {{Electrodynamics of pulsar magnetospheres}},}\ }\href {\doibase
  10.1007/s11214-016-0315-7} {\bibfield  {journal} {\bibinfo  {journal} {Space
  Sci. Rev.}\ }\textbf {\bibinfo {volume} {207}},\ \bibinfo {pages} {111--136}
  (\bibinfo {year} {2017})},\ \Eprint {http://arxiv.org/abs/1611.04331}
  {arXiv:1611.04331 [astro-ph.HE]} \BibitemShut {NoStop}%
%%CITATION = ARXIV:1611.04331;%%
\bibitem [{\citenamefont {Shokair}\ \emph {et~al.}(2014)\citenamefont {Shokair}
  \emph {et~al.}}]{Shokair:2014rna}%
  \BibitemOpen
  \bibfield  {author} {\bibinfo {author} {\bibfnamefont {T.~M.}\ \bibnamefont
  {Shokair}} \emph {et~al.},\ }\bibfield  {title} {\enquote {\bibinfo {title}
  {{Future Directions in the Microwave Cavity Search for Dark Matter
  Axions}},}\ }\href {\doibase 10.1142/S0217751X14430040} {\bibfield  {journal}
  {\bibinfo  {journal} {Int. J. Mod. Phys.}\ }\textbf {\bibinfo {volume}
  {A29}},\ \bibinfo {pages} {1443004} (\bibinfo {year} {2014})},\ \Eprint
  {http://arxiv.org/abs/1405.3685} {arXiv:1405.3685 [physics.ins-det]}
  \BibitemShut {NoStop}%
%%CITATION = ARXIV:1405.3685;%%
\bibitem [{\citenamefont {Rosenberg}(2015)}]{Rosenberg:2015kxa}%
  \BibitemOpen
  \bibfield  {author} {\bibinfo {author} {\bibfnamefont {Leslie~J.}\
  \bibnamefont {Rosenberg}},\ }\bibfield  {title} {\enquote {\bibinfo {title}
  {{Dark-matter QCD-axion searches}},}\ }in\ \href {\doibase
  10.1073/pnas.1308788112} {\emph {\bibinfo {booktitle} {{Sackler Colloquium:
  Dark Matter Universe: On the Threshhold of Discovery Irvine, USA, October
  18-20, 2012}}}}\ (\bibinfo {year} {2015})\BibitemShut {NoStop}%
%%CITATION = INSPIRE-1355312;%%
\bibitem [{\citenamefont {Du}\ \emph {et~al.}(2018)\citenamefont {Du} \emph
  {et~al.}}]{Du:2018uak}%
  \BibitemOpen
  \bibfield  {author} {\bibinfo {author} {\bibfnamefont {N.}~\bibnamefont {Du}}
  \emph {et~al.} (\bibinfo {collaboration} {ADMX}),\ }\bibfield  {title}
  {\enquote {\bibinfo {title} {{A Search for Invisible Axion Dark Matter with
  the Axion Dark Matter Experiment}},}\ }\href {\doibase
  10.1103/PhysRevLett.120.151301} {\bibfield  {journal} {\bibinfo  {journal}
  {Phys. Rev. Lett.}\ }\textbf {\bibinfo {volume} {120}},\ \bibinfo {pages}
  {151301} (\bibinfo {year} {2018})},\ \Eprint
  {http://arxiv.org/abs/1804.05750} {arXiv:1804.05750 [hep-ex]} \BibitemShut
  {NoStop}%
%%CITATION = ARXIV:1804.05750;%%
\bibitem [{\citenamefont {Brubaker}\ \emph
  {et~al.}(2017{\natexlab{a}})\citenamefont {Brubaker} \emph
  {et~al.}}]{Brubaker:2016ktl}%
  \BibitemOpen
  \bibfield  {author} {\bibinfo {author} {\bibfnamefont {B.~M.}\ \bibnamefont
  {Brubaker}} \emph {et~al.},\ }\bibfield  {title} {\enquote {\bibinfo {title}
  {{First results from a microwave cavity axion search at 24 $\mu$eV}},}\
  }\href {\doibase 10.1103/PhysRevLett.118.061302} {\bibfield  {journal}
  {\bibinfo  {journal} {Phys. Rev. Lett.}\ }\textbf {\bibinfo {volume} {118}},\
  \bibinfo {pages} {061302} (\bibinfo {year} {2017}{\natexlab{a}})},\ \Eprint
  {http://arxiv.org/abs/1610.02580} {arXiv:1610.02580 [astro-ph.CO]}
  \BibitemShut {NoStop}%
%%CITATION = ARXIV:1610.02580;%%
\bibitem [{\citenamefont {Al~Kenany}\ \emph {et~al.}(2017)\citenamefont
  {Al~Kenany} \emph {et~al.}}]{Kenany:2016tta}%
  \BibitemOpen
  \bibfield  {author} {\bibinfo {author} {\bibfnamefont {S.}~\bibnamefont
  {Al~Kenany}} \emph {et~al.},\ }\bibfield  {title} {\enquote {\bibinfo {title}
  {{Design and operational experience of a microwave cavity axion detector for
  the 20--100 $\mu$eV range}},}\ }\href {\doibase 10.1016/j.nima.2017.02.012}
  {\bibfield  {journal} {\bibinfo  {journal} {Nucl. Instrum. Meth.}\ }\textbf
  {\bibinfo {volume} {A854}},\ \bibinfo {pages} {11--24} (\bibinfo {year}
  {2017})},\ \Eprint {http://arxiv.org/abs/1611.07123} {arXiv:1611.07123
  [physics.ins-det]} \BibitemShut {NoStop}%
%%CITATION = ARXIV:1611.07123;%%
\bibitem [{\citenamefont {Brubaker}\ \emph
  {et~al.}(2017{\natexlab{b}})\citenamefont {Brubaker}, \citenamefont {Zhong},
  \citenamefont {Lamoreaux}, \citenamefont {Lehnert},\ and\ \citenamefont {van
  Bibber}}]{Brubaker:2017rna}%
  \BibitemOpen
  \bibfield  {author} {\bibinfo {author} {\bibfnamefont {B.~M.}\ \bibnamefont
  {Brubaker}}, \bibinfo {author} {\bibfnamefont {L.}~\bibnamefont {Zhong}},
  \bibinfo {author} {\bibfnamefont {S.~K.}\ \bibnamefont {Lamoreaux}}, \bibinfo
  {author} {\bibfnamefont {K.~W.}\ \bibnamefont {Lehnert}}, \ and\ \bibinfo
  {author} {\bibfnamefont {K.~A.}\ \bibnamefont {van Bibber}},\ }\bibfield
  {title} {\enquote {\bibinfo {title} {{The HAYSTAC Axion Search Analysis
  Procedure}},}\ }\href@noop {} {\  (\bibinfo {year} {2017}{\natexlab{b}})},\
  \Eprint {http://arxiv.org/abs/1706.08388} {arXiv:1706.08388 [astro-ph.IM]}
  \BibitemShut {NoStop}%
%%CITATION = ARXIV:1706.08388;%%
\bibitem [{\citenamefont {Majorovits}\ and\ \citenamefont
  {Redondo}(2017)}]{Majorovits:2016yvk}%
  \BibitemOpen
  \bibfield  {author} {\bibinfo {author} {\bibfnamefont {B{\'e}la}\
  \bibnamefont {Majorovits}}\ and\ \bibinfo {author} {\bibfnamefont {Javier}\
  \bibnamefont {Redondo}} (\bibinfo {collaboration} {MADMAX Working Group}),\
  }\bibfield  {title} {\enquote {\bibinfo {title} {{MADMAX: A new Dark Matter
  Axion Search using a Dielectric Haloscope}},}\ }in\ \href {\doibase
  10.3204/DESY-PROC-2009-03/Majorovits_Bela} {\emph {\bibinfo {booktitle}
  {{Proceedings, 12th Patras Workshop on Axions, WIMPs and WISPs (PATRAS 2016):
  Jeju Island, South Korea, June 20-24, 2016}}}}\ (\bibinfo {year} {2017})\
  pp.\ \bibinfo {pages} {94--97},\ \Eprint {http://arxiv.org/abs/1611.04549}
  {arXiv:1611.04549 [astro-ph.IM]} \BibitemShut {NoStop}%
%%CITATION = ARXIV:1611.04549;%%
\bibitem [{\citenamefont {Kahn}\ \emph {et~al.}(2016)\citenamefont {Kahn},
  \citenamefont {Safdi},\ and\ \citenamefont {Thaler}}]{Kahn:2016aff}%
  \BibitemOpen
  \bibfield  {author} {\bibinfo {author} {\bibfnamefont {Yonatan}\ \bibnamefont
  {Kahn}}, \bibinfo {author} {\bibfnamefont {Benjamin~R.}\ \bibnamefont
  {Safdi}}, \ and\ \bibinfo {author} {\bibfnamefont {Jesse}\ \bibnamefont
  {Thaler}},\ }\bibfield  {title} {\enquote {\bibinfo {title} {{Broadband and
  Resonant Approaches to Axion Dark Matter Detection}},}\ }\href {\doibase
  10.1103/PhysRevLett.117.141801} {\bibfield  {journal} {\bibinfo  {journal}
  {Phys. Rev. Lett.}\ }\textbf {\bibinfo {volume} {117}},\ \bibinfo {pages}
  {141801} (\bibinfo {year} {2016})},\ \Eprint
  {http://arxiv.org/abs/1602.01086} {arXiv:1602.01086 [hep-ph]} \BibitemShut
  {NoStop}%
%%CITATION = ARXIV:1602.01086;%%
\bibitem [{\citenamefont {Foster}\ \emph {et~al.}(2017)\citenamefont {Foster},
  \citenamefont {Rodd},\ and\ \citenamefont {Safdi}}]{Foster:2017hbq}%
  \BibitemOpen
  \bibfield  {author} {\bibinfo {author} {\bibfnamefont {Joshua~W.}\
  \bibnamefont {Foster}}, \bibinfo {author} {\bibfnamefont {Nicholas~L.}\
  \bibnamefont {Rodd}}, \ and\ \bibinfo {author} {\bibfnamefont {Benjamin~R.}\
  \bibnamefont {Safdi}},\ }\bibfield  {title} {\enquote {\bibinfo {title}
  {{Revealing the Dark Matter Halo with Axion Direct Detection}},}\ }\href@noop
  {} {\  (\bibinfo {year} {2017})},\ \Eprint {http://arxiv.org/abs/1711.10489}
  {arXiv:1711.10489 [astro-ph.CO]} \BibitemShut {NoStop}%
%%CITATION = ARXIV:1711.10489;%%
\bibitem [{\citenamefont {Henning}\ \emph {et~al.}(2018)\citenamefont {Henning}
  \emph {et~al.}}]{Henning:2018ogd}%
  \BibitemOpen
  \bibfield  {author} {\bibinfo {author} {\bibfnamefont {Reyco}\ \bibnamefont
  {Henning}} \emph {et~al.} (\bibinfo {collaboration} {ABRACADABRA}),\
  }\bibfield  {title} {\enquote {\bibinfo {title} {{ABRACADABRA, A Search for
  Low-Mass Axion Dark Matter}},}\ }in\ \href {\doibase
  10.3204/DESY-PROC-2017-02/henning_reyco} {\emph {\bibinfo {booktitle}
  {{Proceedings, 13th Patras Workshop on Axions, WIMPs and WISPs, (PATRAS
  2017): Thessaloniki, Greece, 15 May 2017 - 19, 2017}}}}\ (\bibinfo {year}
  {2018})\ pp.\ \bibinfo {pages} {28--31}\BibitemShut {NoStop}%
%%CITATION = DESY-PROC-2017-02;%%
\bibitem [{\citenamefont {Ouellet}\ \emph {et~al.}(2018)\citenamefont {Ouellet}
  \emph {et~al.}}]{Ouellet:2018beu}%
  \BibitemOpen
  \bibfield  {author} {\bibinfo {author} {\bibfnamefont {Jonathan~L.}\
  \bibnamefont {Ouellet}} \emph {et~al.},\ }\bibfield  {title} {\enquote
  {\bibinfo {title} {{First Results from ABRACADABRA-10 cm: A Search for
  Sub-$\mu$eV Axion Dark Matter}},}\ }\href@noop {} {\  (\bibinfo {year}
  {2018})},\ \Eprint {http://arxiv.org/abs/1810.12257} {arXiv:1810.12257
  [hep-ex]} \BibitemShut {NoStop}%
%%CITATION = ARXIV:1810.12257;%%
\bibitem [{\citenamefont {Chaudhuri}\ \emph {et~al.}(2015)\citenamefont
  {Chaudhuri}, \citenamefont {Graham}, \citenamefont {Irwin}, \citenamefont
  {Mardon}, \citenamefont {Rajendran},\ and\ \citenamefont
  {Zhao}}]{Chaudhuri:2014dla}%
  \BibitemOpen
  \bibfield  {author} {\bibinfo {author} {\bibfnamefont {Saptarshi}\
  \bibnamefont {Chaudhuri}}, \bibinfo {author} {\bibfnamefont {Peter~W.}\
  \bibnamefont {Graham}}, \bibinfo {author} {\bibfnamefont {Kent}\ \bibnamefont
  {Irwin}}, \bibinfo {author} {\bibfnamefont {Jeremy}\ \bibnamefont {Mardon}},
  \bibinfo {author} {\bibfnamefont {Surjeet}\ \bibnamefont {Rajendran}}, \ and\
  \bibinfo {author} {\bibfnamefont {Yue}\ \bibnamefont {Zhao}},\ }\bibfield
  {title} {\enquote {\bibinfo {title} {{Radio for hidden-photon dark matter
  detection}},}\ }\href {\doibase 10.1103/PhysRevD.92.075012} {\bibfield
  {journal} {\bibinfo  {journal} {Phys. Rev.}\ }\textbf {\bibinfo {volume}
  {D92}},\ \bibinfo {pages} {075012} (\bibinfo {year} {2015})},\ \Eprint
  {http://arxiv.org/abs/1411.7382} {arXiv:1411.7382 [hep-ph]} \BibitemShut
  {NoStop}%
%%CITATION = ARXIV:1411.7382;%%
\bibitem [{\citenamefont {Silva-Feaver}\ \emph {et~al.}(2016)\citenamefont
  {Silva-Feaver} \emph {et~al.}}]{Silva-Feaver:2016qhh}%
  \BibitemOpen
  \bibfield  {author} {\bibinfo {author} {\bibfnamefont {Maximiliano}\
  \bibnamefont {Silva-Feaver}} \emph {et~al.},\ }\bibfield  {title} {\enquote
  {\bibinfo {title} {{Design Overview of the DM Radio Pathfinder
  Experiment}},}\ }\bibfield  {booktitle} {\emph {\bibinfo {booktitle}
  {{Proceedings, Applied Superconductivity Conference (ASC 2016): Denver,
  Colorado, September 4-9, 2016}}},\ }\href {\doibase
  10.1109/TASC.2016.2631425} {\bibfield  {journal} {\bibinfo  {journal} {IEEE
  Trans. Appl. Supercond.}\ }\textbf {\bibinfo {volume} {27}},\ \bibinfo
  {pages} {1400204} (\bibinfo {year} {2016})},\ \Eprint
  {http://arxiv.org/abs/1610.09344} {arXiv:1610.09344 [astro-ph.IM]}
  \BibitemShut {NoStop}%
%%CITATION = ARXIV:1610.09344;%%
\bibitem [{\citenamefont {Budker}\ \emph {et~al.}(2014)\citenamefont {Budker},
  \citenamefont {Graham}, \citenamefont {Ledbetter}, \citenamefont
  {Rajendran},\ and\ \citenamefont {Sushkov}}]{Budker:2013hfa}%
  \BibitemOpen
  \bibfield  {author} {\bibinfo {author} {\bibfnamefont {Dmitry}\ \bibnamefont
  {Budker}}, \bibinfo {author} {\bibfnamefont {Peter~W.}\ \bibnamefont
  {Graham}}, \bibinfo {author} {\bibfnamefont {Micah}\ \bibnamefont
  {Ledbetter}}, \bibinfo {author} {\bibfnamefont {Surjeet}\ \bibnamefont
  {Rajendran}}, \ and\ \bibinfo {author} {\bibfnamefont {Alex}\ \bibnamefont
  {Sushkov}},\ }\bibfield  {title} {\enquote {\bibinfo {title} {{Proposal for a
  Cosmic Axion Spin Precession Experiment (CASPEr)}},}\ }\href {\doibase
  10.1103/PhysRevX.4.021030} {\bibfield  {journal} {\bibinfo  {journal} {Phys.
  Rev.}\ }\textbf {\bibinfo {volume} {X4}},\ \bibinfo {pages} {021030}
  (\bibinfo {year} {2014})},\ \Eprint {http://arxiv.org/abs/1306.6089}
  {arXiv:1306.6089 [hep-ph]} \BibitemShut {NoStop}%
%%CITATION = ARXIV:1306.6089;%%
\bibitem [{\citenamefont {{Chen}}\ and\ \citenamefont
  {{Beloborodov}}(2014)}]{2014ApJ...795L..22C}%
  \BibitemOpen
  \bibfield  {author} {\bibinfo {author} {\bibfnamefont {A.~Y.}\ \bibnamefont
  {{Chen}}}\ and\ \bibinfo {author} {\bibfnamefont {A.~M.}\ \bibnamefont
  {{Beloborodov}}},\ }\bibfield  {title} {\enquote {\bibinfo {title}
  {{Electrodynamics of Axisymmetric Pulsar Magnetosphere with Electron-Positron
  Discharge: A Numerical Experiment}},}\ }\href {\doibase
  10.1088/2041-8205/795/1/L22} {\bibfield  {journal} {\bibinfo  {journal}
  {ApJL}\ }\textbf {\bibinfo {volume} {795}},\ \bibinfo {eid} {L22} (\bibinfo
  {year} {2014})},\ \Eprint {http://arxiv.org/abs/1406.7834} {arXiv:1406.7834
  [astro-ph.HE]} \BibitemShut {NoStop}%
\bibitem [{\citenamefont {Kaplan}\ and\ \citenamefont {van
  Kerkwijk}(2009)}]{Kaplan:2009ce}%
  \BibitemOpen
  \bibfield  {author} {\bibinfo {author} {\bibfnamefont {D.~L.}\ \bibnamefont
  {Kaplan}}\ and\ \bibinfo {author} {\bibfnamefont {M.~H.}\ \bibnamefont {van
  Kerkwijk}},\ }\bibfield  {title} {\enquote {\bibinfo {title} {{Constraining
  the Spin-down of the Nearby Isolated Neutron Star RX J0806.4-4123, and
  Implications for the Population of Nearby Neutron Stars}},}\ }\href {\doibase
  10.1088/0004-637X/705/1/798} {\bibfield  {journal} {\bibinfo  {journal}
  {Astrophys. J.}\ }\textbf {\bibinfo {volume} {705}},\ \bibinfo {pages}
  {798--808} (\bibinfo {year} {2009})},\ \Eprint
  {http://arxiv.org/abs/0909.5218} {arXiv:0909.5218 [astro-ph.HE]} \BibitemShut
  {NoStop}%
%%CITATION = ARXIV:0909.5218;%%
\bibitem [{\citenamefont {{Manchester}}\ \emph {et~al.}(2005)\citenamefont
  {{Manchester}}, \citenamefont {{Hobbs}}, \citenamefont {{Teoh}},\ and\
  \citenamefont {{Hobbs}}}]{2005AJ....129.1993M}%
  \BibitemOpen
  \bibfield  {author} {\bibinfo {author} {\bibfnamefont {R.~N.}\ \bibnamefont
  {{Manchester}}}, \bibinfo {author} {\bibfnamefont {G.~B.}\ \bibnamefont
  {{Hobbs}}}, \bibinfo {author} {\bibfnamefont {A.}~\bibnamefont {{Teoh}}}, \
  and\ \bibinfo {author} {\bibfnamefont {M.}~\bibnamefont {{Hobbs}}},\
  }\bibfield  {title} {\enquote {\bibinfo {title} {{The Australia Telescope
  National Facility Pulsar Catalogue}},}\ }\href {\doibase 10.1086/428488}
  {\bibfield  {journal} {\bibinfo  {journal} {AJ}\ }\textbf {\bibinfo {volume}
  {129}},\ \bibinfo {pages} {1993--2006} (\bibinfo {year} {2005})},\ \Eprint
  {http://arxiv.org/abs/astro-ph/0412641} {astro-ph/0412641} \BibitemShut
  {NoStop}%
\bibitem [{\citenamefont {{Michel}}\ and\ \citenamefont
  {{Goldwire}}(1970)}]{1970ApL.....5...21M}%
  \BibitemOpen
  \bibfield  {author} {\bibinfo {author} {\bibfnamefont {F.~C.}\ \bibnamefont
  {{Michel}}}\ and\ \bibinfo {author} {\bibfnamefont {H.~C.}\ \bibnamefont
  {{Goldwire}}, \bibfnamefont {Jr.}},\ }\bibfield  {title} {\enquote {\bibinfo
  {title} {{Alignment of Oblique Rotators}},}\ }\href@noop {} {\bibfield
  {journal} {\bibinfo  {journal} {Astrophys.~Lett.}\ }\textbf {\bibinfo
  {volume} {5}},\ \bibinfo {pages} {21} (\bibinfo {year} {1970})}\BibitemShut
  {NoStop}%
\bibitem [{\citenamefont {{Bhattacharya}}\ \emph {et~al.}(1992)\citenamefont
  {{Bhattacharya}}, \citenamefont {{Wijers}}, \citenamefont {{Hartman}},\ and\
  \citenamefont {{Verbunt}}}]{1992A&A...254..198B}%
  \BibitemOpen
  \bibfield  {author} {\bibinfo {author} {\bibfnamefont {D.}~\bibnamefont
  {{Bhattacharya}}}, \bibinfo {author} {\bibfnamefont {R.~A.~M.~J.}\
  \bibnamefont {{Wijers}}}, \bibinfo {author} {\bibfnamefont {J.~W.}\
  \bibnamefont {{Hartman}}}, \ and\ \bibinfo {author} {\bibfnamefont
  {F.}~\bibnamefont {{Verbunt}}},\ }\bibfield  {title} {\enquote {\bibinfo
  {title} {{On the decay of the magnetic fields of single radio pulsars.}}}\
  }\href@noop {} {\bibfield  {journal} {\bibinfo  {journal} {A\&A}\ }\textbf
  {\bibinfo {volume} {254}},\ \bibinfo {pages} {198--212} (\bibinfo {year}
  {1992})}\BibitemShut {NoStop}%
\bibitem [{\citenamefont {{Goldreich}}\ and\ \citenamefont
  {{Reisenegger}}(1992)}]{1992ApJ...395..250G}%
  \BibitemOpen
  \bibfield  {author} {\bibinfo {author} {\bibfnamefont {P.}~\bibnamefont
  {{Goldreich}}}\ and\ \bibinfo {author} {\bibfnamefont {A.}~\bibnamefont
  {{Reisenegger}}},\ }\bibfield  {title} {\enquote {\bibinfo {title} {{Magnetic
  field decay in isolated neutron stars}},}\ }\href {\doibase 10.1086/171646}
  {\bibfield  {journal} {\bibinfo  {journal} {ApJ}\ }\textbf {\bibinfo {volume}
  {395}},\ \bibinfo {pages} {250--258} (\bibinfo {year} {1992})}\BibitemShut
  {NoStop}%
\bibitem [{\citenamefont {Cumming}\ \emph {et~al.}(2004)\citenamefont
  {Cumming}, \citenamefont {Arras},\ and\ \citenamefont
  {Zweibel}}]{Cumming:2004mf}%
  \BibitemOpen
  \bibfield  {author} {\bibinfo {author} {\bibfnamefont {Andrew}\ \bibnamefont
  {Cumming}}, \bibinfo {author} {\bibfnamefont {Phil}\ \bibnamefont {Arras}}, \
  and\ \bibinfo {author} {\bibfnamefont {Ellen~G.}\ \bibnamefont {Zweibel}},\
  }\bibfield  {title} {\enquote {\bibinfo {title} {{Magnetic field evolution in
  neutron star crusts due to the Hall effect and Ohmic decay}},}\ }\href
  {\doibase 10.1086/421324} {\bibfield  {journal} {\bibinfo  {journal}
  {Astrophys. J.}\ }\textbf {\bibinfo {volume} {609}},\ \bibinfo {pages}
  {999--1017} (\bibinfo {year} {2004})},\ \Eprint
  {http://arxiv.org/abs/astro-ph/0402392} {arXiv:astro-ph/0402392 [astro-ph]}
  \BibitemShut {NoStop}%
%%CITATION = ASTRO-PH/0402392;%%
\bibitem [{\citenamefont {Aguilera}\ \emph {et~al.}(2008)\citenamefont
  {Aguilera}, \citenamefont {Pons},\ and\ \citenamefont
  {Miralles}}]{Aguilera:2007xk}%
  \BibitemOpen
  \bibfield  {author} {\bibinfo {author} {\bibfnamefont {Deborah~N.}\
  \bibnamefont {Aguilera}}, \bibinfo {author} {\bibfnamefont {Jose~A.}\
  \bibnamefont {Pons}}, \ and\ \bibinfo {author} {\bibfnamefont {Juan~A.}\
  \bibnamefont {Miralles}},\ }\bibfield  {title} {\enquote {\bibinfo {title}
  {{2D Cooling of Magnetized Neutron Stars}},}\ }\href {\doibase
  10.1051/0004-6361:20078786} {\bibfield  {journal} {\bibinfo  {journal}
  {Astron. Astrophys.}\ }\textbf {\bibinfo {volume} {486}},\ \bibinfo {pages}
  {255--271} (\bibinfo {year} {2008})},\ \Eprint
  {http://arxiv.org/abs/0710.0854} {arXiv:0710.0854 [astro-ph]} \BibitemShut
  {NoStop}%
%%CITATION = ARXIV:0710.0854;%%
\bibitem [{\citenamefont {{Gourgouliatos}}\ and\ \citenamefont
  {{Cumming}}(2014)}]{2014MNRAS.438.1618G}%
  \BibitemOpen
  \bibfield  {author} {\bibinfo {author} {\bibfnamefont {K.~N.}\ \bibnamefont
  {{Gourgouliatos}}}\ and\ \bibinfo {author} {\bibfnamefont {A.}~\bibnamefont
  {{Cumming}}},\ }\bibfield  {title} {\enquote {\bibinfo {title} {{Hall effect
  in neutron star crusts: evolution, endpoint and dependence on initial
  conditions}},}\ }\href {\doibase 10.1093/mnras/stt2300} {\bibfield  {journal}
  {\bibinfo  {journal} {MNRAS}\ }\textbf {\bibinfo {volume} {438}},\ \bibinfo
  {pages} {1618--1629} (\bibinfo {year} {2014})},\ \Eprint
  {http://arxiv.org/abs/1311.7004} {arXiv:1311.7004 [astro-ph.SR]} \BibitemShut
  {NoStop}%
\bibitem [{\citenamefont {Beloborodov}\ and\ \citenamefont
  {Li}(2016)}]{Beloborodov:2016mmx}%
  \BibitemOpen
  \bibfield  {author} {\bibinfo {author} {\bibfnamefont {Andrei~M.}\
  \bibnamefont {Beloborodov}}\ and\ \bibinfo {author} {\bibfnamefont {Xinyu}\
  \bibnamefont {Li}},\ }\bibfield  {title} {\enquote {\bibinfo {title}
  {{Magnetar heating}},}\ }\href {\doibase 10.3847/1538-4357/833/2/261}
  {\bibfield  {journal} {\bibinfo  {journal} {Astrophys. J.}\ }\textbf
  {\bibinfo {volume} {833}},\ \bibinfo {pages} {261} (\bibinfo {year}
  {2016})},\ \Eprint {http://arxiv.org/abs/1605.09077} {arXiv:1605.09077
  [astro-ph.HE]} \BibitemShut {NoStop}%
%%CITATION = ARXIV:1605.09077;%%
\bibitem [{\citenamefont {Bransgrove}\ \emph {et~al.}(2018)\citenamefont
  {Bransgrove}, \citenamefont {Levin},\ and\ \citenamefont
  {Beloborodov}}]{Bransgrove:2017jzs}%
  \BibitemOpen
  \bibfield  {author} {\bibinfo {author} {\bibfnamefont {Ashley}\ \bibnamefont
  {Bransgrove}}, \bibinfo {author} {\bibfnamefont {Yuri}\ \bibnamefont
  {Levin}}, \ and\ \bibinfo {author} {\bibfnamefont {Andrei}\ \bibnamefont
  {Beloborodov}},\ }\bibfield  {title} {\enquote {\bibinfo {title} {{Magnetic
  field evolution of neutron stars -- I. Basic formalism, numerical techniques
  and first results}},}\ }\href {\doibase 10.1093/mnras/stx2508} {\bibfield
  {journal} {\bibinfo  {journal} {Mon. Not. Roy. Astron. Soc.}\ }\textbf
  {\bibinfo {volume} {473}},\ \bibinfo {pages} {2771--2790} (\bibinfo {year}
  {2018})},\ \Eprint {http://arxiv.org/abs/1709.09167} {arXiv:1709.09167
  [astro-ph.HE]} \BibitemShut {NoStop}%
%%CITATION = ARXIV:1709.09167;%%
\bibitem [{\citenamefont {{Shapiro}}\ and\ \citenamefont
  {{Teukolsky}}(1983)}]{1983bhwd.book.....S}%
  \BibitemOpen
  \bibfield  {author} {\bibinfo {author} {\bibfnamefont {S.~L.}\ \bibnamefont
  {{Shapiro}}}\ and\ \bibinfo {author} {\bibfnamefont {S.~A.}\ \bibnamefont
  {{Teukolsky}}},\ }\href@noop {} {\emph {\bibinfo {title} {Research supported
  by the National Science Foundation.~New York, Wiley-Interscience, 1983, 663
  p.}}}\ (\bibinfo {year} {1983})\BibitemShut {NoStop}%
\bibitem [{\citenamefont {{Jones}}(2006)}]{2006MNRAS.365..339J}%
  \BibitemOpen
  \bibfield  {author} {\bibinfo {author} {\bibfnamefont {P.~B.}\ \bibnamefont
  {{Jones}}},\ }\bibfield  {title} {\enquote {\bibinfo {title} {{Type II
  superconductivity and magnetic flux transport in neutron stars}},}\ }\href
  {\doibase 10.1111/j.1365-2966.2005.09724.x} {\bibfield  {journal} {\bibinfo
  {journal} {MNRAS}\ }\textbf {\bibinfo {volume} {365}},\ \bibinfo {pages}
  {339--344} (\bibinfo {year} {2006})},\ \Eprint
  {http://arxiv.org/abs/astro-ph/0510396} {astro-ph/0510396} \BibitemShut
  {NoStop}%
\bibitem [{\citenamefont {{Flowers}}\ and\ \citenamefont
  {{Ruderman}}(1977)}]{1977ApJ...215..302F}%
  \BibitemOpen
  \bibfield  {author} {\bibinfo {author} {\bibfnamefont {E.}~\bibnamefont
  {{Flowers}}}\ and\ \bibinfo {author} {\bibfnamefont {M.~A.}\ \bibnamefont
  {{Ruderman}}},\ }\bibfield  {title} {\enquote {\bibinfo {title} {{Evolution
  of pulsar magnetic fields}},}\ }\href {\doibase 10.1086/155359} {\bibfield
  {journal} {\bibinfo  {journal} {ApJ}\ }\textbf {\bibinfo {volume} {215}},\
  \bibinfo {pages} {302--310} (\bibinfo {year} {1977})}\BibitemShut {NoStop}%
\bibitem [{\citenamefont {Jones}(2001)}]{doi:10.1046/j.1365-8711.2001.03990.x}%
  \BibitemOpen
  \bibfield  {author} {\bibinfo {author} {\bibfnamefont {P.~B.}\ \bibnamefont
  {Jones}},\ }\bibfield  {title} {\enquote {\bibinfo {title} {First-principles
  point-defect calculations for solid neutron star matter},}\ }\href {\doibase
  10.1046/j.1365-8711.2001.03990.x} {\bibfield  {journal} {\bibinfo  {journal}
  {MNRAS}\ }\textbf {\bibinfo {volume} {321}},\ \bibinfo {pages} {167--175}
  (\bibinfo {year} {2001})}\BibitemShut {NoStop}%
\bibitem [{\citenamefont {Manchester}\ \emph {et~al.}(2001)\citenamefont
  {Manchester} \emph {et~al.}}]{Manchester:2001fp}%
  \BibitemOpen
  \bibfield  {author} {\bibinfo {author} {\bibfnamefont {R.~N.}\ \bibnamefont
  {Manchester}} \emph {et~al.},\ }\bibfield  {title} {\enquote {\bibinfo
  {title} {{The Parkes Multibeam Pulsar Survey. 1. Observing and data analysis
  systems, discovery and timing of 100 pulsars}},}\ }\href {\doibase
  10.1046/j.1365-8711.2001.04751.x} {\bibfield  {journal} {\bibinfo  {journal}
  {Mon. Not. Roy. Astron. Soc.}\ }\textbf {\bibinfo {volume} {328}},\ \bibinfo
  {pages} {17} (\bibinfo {year} {2001})},\ \Eprint
  {http://arxiv.org/abs/astro-ph/0106522} {arXiv:astro-ph/0106522 [astro-ph]}
  \BibitemShut {NoStop}%
%%CITATION = ASTRO-PH/0106522;%%
\bibitem [{\citenamefont {{Edwards}}\ \emph {et~al.}(2001)\citenamefont
  {{Edwards}}, \citenamefont {{Bailes}}, \citenamefont {{van Straten}},\ and\
  \citenamefont {{Britton}}}]{2001MNRAS.326..358E}%
  \BibitemOpen
  \bibfield  {author} {\bibinfo {author} {\bibfnamefont {R.~T.}\ \bibnamefont
  {{Edwards}}}, \bibinfo {author} {\bibfnamefont {M.}~\bibnamefont {{Bailes}}},
  \bibinfo {author} {\bibfnamefont {W.}~\bibnamefont {{van Straten}}}, \ and\
  \bibinfo {author} {\bibfnamefont {M.~C.}\ \bibnamefont {{Britton}}},\
  }\bibfield  {title} {\enquote {\bibinfo {title} {{The Swinburne
  intermediate-latitude pulsar survey}},}\ }\href {\doibase
  10.1046/j.1365-8711.2001.04637.x} {\bibfield  {journal} {\bibinfo  {journal}
  {MNRAS}\ }\textbf {\bibinfo {volume} {326}},\ \bibinfo {pages} {358--374}
  (\bibinfo {year} {2001})},\ \Eprint {http://arxiv.org/abs/astro-ph/0105126}
  {astro-ph/0105126} \BibitemShut {NoStop}%
\bibitem [{\citenamefont {Cordes}\ and\ \citenamefont
  {Lazio}(2002)}]{Cordes:2002wz}%
  \BibitemOpen
  \bibfield  {author} {\bibinfo {author} {\bibfnamefont {James~M.}\
  \bibnamefont {Cordes}}\ and\ \bibinfo {author} {\bibfnamefont {T.~J.~W.}\
  \bibnamefont {Lazio}},\ }\bibfield  {title} {\enquote {\bibinfo {title}
  {{NE2001. 1. A New model for the galactic distribution of free electrons and
  its fluctuations}},}\ }\href@noop {} {\  (\bibinfo {year} {2002})},\ \Eprint
  {http://arxiv.org/abs/astro-ph/0207156} {arXiv:astro-ph/0207156 [astro-ph]}
  \BibitemShut {NoStop}%
%%CITATION = ASTRO-PH/0207156;%%
\bibitem [{\citenamefont {{Ofek}}(2009)}]{2009PASP..121..814O}%
  \BibitemOpen
  \bibfield  {author} {\bibinfo {author} {\bibfnamefont {E.~O.}\ \bibnamefont
  {{Ofek}}},\ }\bibfield  {title} {\enquote {\bibinfo {title} {{Space and
  Velocity Distributions of Galactic Isolated Old Neutron Stars}},}\ }\href
  {\doibase 10.1086/605389} {\bibfield  {journal} {\bibinfo  {journal} {PASP}\
  }\textbf {\bibinfo {volume} {121}},\ \bibinfo {pages} {814} (\bibinfo {year}
  {2009})},\ \Eprint {http://arxiv.org/abs/0910.3684} {arXiv:0910.3684
  [astro-ph.GA]} \BibitemShut {NoStop}%
\bibitem [{\citenamefont {{Sartore}}\ \emph {et~al.}(2010)\citenamefont
  {{Sartore}}, \citenamefont {{Ripamonti}}, \citenamefont {{Treves}},\ and\
  \citenamefont {{Turolla}}}]{2010A&A...510A..23S}%
  \BibitemOpen
  \bibfield  {author} {\bibinfo {author} {\bibfnamefont {N.}~\bibnamefont
  {{Sartore}}}, \bibinfo {author} {\bibfnamefont {E.}~\bibnamefont
  {{Ripamonti}}}, \bibinfo {author} {\bibfnamefont {A.}~\bibnamefont
  {{Treves}}}, \ and\ \bibinfo {author} {\bibfnamefont {R.}~\bibnamefont
  {{Turolla}}},\ }\bibfield  {title} {\enquote {\bibinfo {title} {{Galactic
  neutron stars. I. Space and velocity distributions in the disk and in the
  halo}},}\ }\href {\doibase 10.1051/0004-6361/200912222} {\bibfield  {journal}
  {\bibinfo  {journal} {A\&A}\ }\textbf {\bibinfo {volume} {510}},\ \bibinfo
  {eid} {A23} (\bibinfo {year} {2010})},\ \Eprint
  {http://arxiv.org/abs/0908.3182} {arXiv:0908.3182} \BibitemShut {NoStop}%
\bibitem [{\citenamefont {{Faucher-Gigu{\`e}re}}\ and\ \citenamefont
  {{Loeb}}(2010)}]{2010JCAP...01..005F}%
  \BibitemOpen
  \bibfield  {author} {\bibinfo {author} {\bibfnamefont {C.-A.}\ \bibnamefont
  {{Faucher-Gigu{\`e}re}}}\ and\ \bibinfo {author} {\bibfnamefont
  {A.}~\bibnamefont {{Loeb}}},\ }\bibfield  {title} {\enquote {\bibinfo {title}
  {{The pulsar contribution to the gamma-ray background}},}\ }\href {\doibase
  10.1088/1475-7516/2010/01/005} {\bibfield  {journal} {\bibinfo  {journal}
  {JCAP}\ }\textbf {\bibinfo {volume} {1}},\ \bibinfo {eid} {005} (\bibinfo
  {year} {2010})},\ \Eprint {http://arxiv.org/abs/0904.3102} {arXiv:0904.3102
  [astro-ph.HE]} \BibitemShut {NoStop}%
\bibitem [{\citenamefont {{Hernquist}}(1990)}]{1990ApJ...356..359H}%
  \BibitemOpen
  \bibfield  {author} {\bibinfo {author} {\bibfnamefont {L.}~\bibnamefont
  {{Hernquist}}},\ }\bibfield  {title} {\enquote {\bibinfo {title} {{An
  analytical model for spherical galaxies and bulges}},}\ }\href {\doibase
  10.1086/168845} {\bibfield  {journal} {\bibinfo  {journal} {\apj}\ }\textbf
  {\bibinfo {volume} {356}},\ \bibinfo {pages} {359--364} (\bibinfo {year}
  {1990})}\BibitemShut {NoStop}%
\bibitem [{\citenamefont {Freitag}\ \emph {et~al.}(2006)\citenamefont
  {Freitag}, \citenamefont {Amaro-Seoane},\ and\ \citenamefont
  {Kalogera}}]{Freitag:2006qf}%
  \BibitemOpen
  \bibfield  {author} {\bibinfo {author} {\bibfnamefont {Marc}\ \bibnamefont
  {Freitag}}, \bibinfo {author} {\bibfnamefont {Pau}\ \bibnamefont
  {Amaro-Seoane}}, \ and\ \bibinfo {author} {\bibfnamefont {Vassiliki}\
  \bibnamefont {Kalogera}},\ }\bibfield  {title} {\enquote {\bibinfo {title}
  {{Stellar remnants in galactic nuclei: mass segregation}},}\ }\href {\doibase
  10.1086/506193} {\bibfield  {journal} {\bibinfo  {journal} {Astrophys. J.}\
  }\textbf {\bibinfo {volume} {649}},\ \bibinfo {pages} {91--117} (\bibinfo
  {year} {2006})},\ \Eprint {http://arxiv.org/abs/astro-ph/0603280}
  {arXiv:astro-ph/0603280 [astro-ph]} \BibitemShut {NoStop}%
%%CITATION = ASTRO-PH/0603280;%%
\bibitem [{\citenamefont {{Bahcall}}\ and\ \citenamefont
  {{Wolf}}(1976)}]{1976ApJ...209..214B}%
  \BibitemOpen
  \bibfield  {author} {\bibinfo {author} {\bibfnamefont {J.~N.}\ \bibnamefont
  {{Bahcall}}}\ and\ \bibinfo {author} {\bibfnamefont {R.~A.}\ \bibnamefont
  {{Wolf}}},\ }\bibfield  {title} {\enquote {\bibinfo {title} {{Star
  distribution around a massive black hole in a globular cluster}},}\ }\href
  {\doibase 10.1086/154711} {\bibfield  {journal} {\bibinfo  {journal}
  {Astrophys. J.}\ }\textbf {\bibinfo {volume} {209}},\ \bibinfo {pages}
  {214--232} (\bibinfo {year} {1976})}\BibitemShut {NoStop}%
\bibitem [{\citenamefont {{Bahcall}}\ and\ \citenamefont
  {{Wolf}}(1977)}]{1977ApJ...216..883B}%
  \BibitemOpen
  \bibfield  {author} {\bibinfo {author} {\bibfnamefont {J.~N.}\ \bibnamefont
  {{Bahcall}}}\ and\ \bibinfo {author} {\bibfnamefont {R.~A.}\ \bibnamefont
  {{Wolf}}},\ }\bibfield  {title} {\enquote {\bibinfo {title} {{The star
  distribution around a massive black hole in a globular cluster. II Unequal
  star masses}},}\ }\href {\doibase 10.1086/155534} {\bibfield  {journal}
  {\bibinfo  {journal} {Astrophys. J.}\ }\textbf {\bibinfo {volume} {216}},\
  \bibinfo {pages} {883--907} (\bibinfo {year} {1977})}\BibitemShut {NoStop}%
\bibitem [{\citenamefont {{Dehnen}}(1993)}]{1993MNRAS.265..250D}%
  \BibitemOpen
  \bibfield  {author} {\bibinfo {author} {\bibfnamefont {W.}~\bibnamefont
  {{Dehnen}}},\ }\bibfield  {title} {\enquote {\bibinfo {title} {{A Family of
  Potential-Density Pairs for Spherical Galaxies and Bulges}},}\ }\href
  {\doibase 10.1093/mnras/265.1.250} {\bibfield  {journal} {\bibinfo  {journal}
  {MNRAS}\ }\textbf {\bibinfo {volume} {265}},\ \bibinfo {pages} {250}
  (\bibinfo {year} {1993})}\BibitemShut {NoStop}%
\bibitem [{\citenamefont {{Tremaine}}\ \emph {et~al.}(1994)\citenamefont
  {{Tremaine}}, \citenamefont {{Richstone}}, \citenamefont {{Byun}},
  \citenamefont {{Dressler}}, \citenamefont {{Faber}}, \citenamefont
  {{Grillmair}}, \citenamefont {{Kormendy}},\ and\ \citenamefont
  {{Lauer}}}]{1994AJ....107..634T}%
  \BibitemOpen
  \bibfield  {author} {\bibinfo {author} {\bibfnamefont {S.}~\bibnamefont
  {{Tremaine}}}, \bibinfo {author} {\bibfnamefont {D.~O.}\ \bibnamefont
  {{Richstone}}}, \bibinfo {author} {\bibfnamefont {Y.-I.}\ \bibnamefont
  {{Byun}}}, \bibinfo {author} {\bibfnamefont {A.}~\bibnamefont {{Dressler}}},
  \bibinfo {author} {\bibfnamefont {S.~M.}\ \bibnamefont {{Faber}}}, \bibinfo
  {author} {\bibfnamefont {C.}~\bibnamefont {{Grillmair}}}, \bibinfo {author}
  {\bibfnamefont {J.}~\bibnamefont {{Kormendy}}}, \ and\ \bibinfo {author}
  {\bibfnamefont {T.~R.}\ \bibnamefont {{Lauer}}},\ }\bibfield  {title}
  {\enquote {\bibinfo {title} {{A family of models for spherical stellar
  systems}},}\ }\href {\doibase 10.1086/116883} {\bibfield  {journal} {\bibinfo
   {journal} {Astron. J.}\ }\textbf {\bibinfo {volume} {107}},\ \bibinfo
  {pages} {634--644} (\bibinfo {year} {1994})}\BibitemShut {NoStop}%
\bibitem [{\citenamefont {Dehnen}\ \emph {et~al.}(2006)\citenamefont {Dehnen},
  \citenamefont {McLaughlin},\ and\ \citenamefont {Sachania}}]{Dehnen:2006cm}%
  \BibitemOpen
  \bibfield  {author} {\bibinfo {author} {\bibfnamefont {Walter}\ \bibnamefont
  {Dehnen}}, \bibinfo {author} {\bibfnamefont {Dean}\ \bibnamefont
  {McLaughlin}}, \ and\ \bibinfo {author} {\bibfnamefont {Jalpesh}\
  \bibnamefont {Sachania}},\ }\bibfield  {title} {\enquote {\bibinfo {title}
  {{The velocity dispersion and mass profile of the milky way}},}\ }\href
  {\doibase 10.1111/j.1365-2966.2006.10404.x} {\bibfield  {journal} {\bibinfo
  {journal} {Mon. Not. Roy. Astron. Soc.}\ }\textbf {\bibinfo {volume} {369}},\
  \bibinfo {pages} {1688--1692} (\bibinfo {year} {2006})},\ \Eprint
  {http://arxiv.org/abs/astro-ph/0603825} {arXiv:astro-ph/0603825 [astro-ph]}
  \BibitemShut {NoStop}%
%%CITATION = ASTRO-PH/0603825;%%
\bibitem [{\citenamefont {Navarro}\ \emph {et~al.}(1996)\citenamefont
  {Navarro}, \citenamefont {Frenk},\ and\ \citenamefont
  {White}}]{Navarro:1995iw}%
  \BibitemOpen
  \bibfield  {author} {\bibinfo {author} {\bibfnamefont {Julio~F.}\
  \bibnamefont {Navarro}}, \bibinfo {author} {\bibfnamefont {Carlos~S.}\
  \bibnamefont {Frenk}}, \ and\ \bibinfo {author} {\bibfnamefont {Simon D.~M.}\
  \bibnamefont {White}},\ }\bibfield  {title} {\enquote {\bibinfo {title} {{The
  Structure of cold dark matter halos}},}\ }\href {\doibase 10.1086/177173}
  {\bibfield  {journal} {\bibinfo  {journal} {Astrophys. J.}\ }\textbf
  {\bibinfo {volume} {462}},\ \bibinfo {pages} {563--575} (\bibinfo {year}
  {1996})},\ \Eprint {http://arxiv.org/abs/astro-ph/9508025} {astro-ph/9508025}
  \BibitemShut {NoStop}%
%%CITATION = ASTRO-PH/9508025;%%
\bibitem [{\citenamefont {Navarro}\ \emph {et~al.}(1997)\citenamefont
  {Navarro}, \citenamefont {Frenk},\ and\ \citenamefont
  {White}}]{Navarro:1996gj}%
  \BibitemOpen
  \bibfield  {author} {\bibinfo {author} {\bibfnamefont {Julio~F.}\
  \bibnamefont {Navarro}}, \bibinfo {author} {\bibfnamefont {Carlos~S.}\
  \bibnamefont {Frenk}}, \ and\ \bibinfo {author} {\bibfnamefont {Simon D.~M.}\
  \bibnamefont {White}},\ }\bibfield  {title} {\enquote {\bibinfo {title} {{A
  Universal density profile from hierarchical clustering}},}\ }\href {\doibase
  10.1086/304888} {\bibfield  {journal} {\bibinfo  {journal} {Astrophys. J.}\
  }\textbf {\bibinfo {volume} {490}},\ \bibinfo {pages} {493--508} (\bibinfo
  {year} {1997})},\ \Eprint {http://arxiv.org/abs/astro-ph/9611107}
  {arXiv:astro-ph/9611107 [astro-ph]} \BibitemShut {NoStop}%
%%CITATION = ASTRO-PH/9611107;%%
\bibitem [{\citenamefont {Burkert}(1996)}]{Burkert:1995yz}%
  \BibitemOpen
  \bibfield  {author} {\bibinfo {author} {\bibfnamefont {A.}~\bibnamefont
  {Burkert}},\ }\bibfield  {title} {\enquote {\bibinfo {title} {{The Structure
  of dark matter halos in dwarf galaxies}},}\ }\bibfield  {booktitle} {\emph
  {\bibinfo {booktitle} {{IAU Symposium 171: New Light on Galaxy Evolution
  Heidelberg, Germany, June 26-30, 1995}}},\ }\href {\doibase 10.1086/309560}
  {\bibfield  {journal} {\bibinfo  {journal} {IAU Symp.}\ }\textbf {\bibinfo
  {volume} {171}},\ \bibinfo {pages} {175} (\bibinfo {year} {1996})},\ \bibinfo
  {note} {[Astrophys. J.447,L25(1995)]},\ \Eprint
  {http://arxiv.org/abs/astro-ph/9504041} {arXiv:astro-ph/9504041 [astro-ph]}
  \BibitemShut {NoStop}%
%%CITATION = ASTRO-PH/9504041;%%
\bibitem [{\citenamefont {Nesti}\ and\ \citenamefont
  {Salucci}(2013)}]{Nesti:2013uwa}%
  \BibitemOpen
  \bibfield  {author} {\bibinfo {author} {\bibfnamefont {Fabrizio}\
  \bibnamefont {Nesti}}\ and\ \bibinfo {author} {\bibfnamefont {Paolo}\
  \bibnamefont {Salucci}},\ }\bibfield  {title} {\enquote {\bibinfo {title}
  {{The Dark Matter halo of the Milky Way, AD 2013}},}\ }\href {\doibase
  10.1088/1475-7516/2013/07/016} {\bibfield  {journal} {\bibinfo  {journal}
  {JCAP}\ }\textbf {\bibinfo {volume} {1307}},\ \bibinfo {pages} {016}
  (\bibinfo {year} {2013})},\ \Eprint {http://arxiv.org/abs/1304.5127}
  {arXiv:1304.5127 [astro-ph.GA]} \BibitemShut {NoStop}%
%%CITATION = ARXIV:1304.5127;%%
\bibitem [{\citenamefont {Merritt}\ \emph {et~al.}(2007)\citenamefont
  {Merritt}, \citenamefont {Harfst},\ and\ \citenamefont
  {Bertone}}]{Merritt:2006mt}%
  \BibitemOpen
  \bibfield  {author} {\bibinfo {author} {\bibfnamefont {David}\ \bibnamefont
  {Merritt}}, \bibinfo {author} {\bibfnamefont {Stefan}\ \bibnamefont
  {Harfst}}, \ and\ \bibinfo {author} {\bibfnamefont {Gianfranco}\ \bibnamefont
  {Bertone}},\ }\bibfield  {title} {\enquote {\bibinfo {title} {{Collisionally
  Regenerated Dark Matter Structures in Galactic Nuclei}},}\ }\href {\doibase
  10.1103/PhysRevD.75.043517} {\bibfield  {journal} {\bibinfo  {journal} {Phys.
  Rev.}\ }\textbf {\bibinfo {volume} {D75}},\ \bibinfo {pages} {043517}
  (\bibinfo {year} {2007})},\ \Eprint {http://arxiv.org/abs/astro-ph/0610425}
  {arXiv:astro-ph/0610425 [astro-ph]} \BibitemShut {NoStop}%
%%CITATION = ASTRO-PH/0610425;%%
\bibitem [{\citenamefont {Monaco}\ \emph {et~al.}(2005)\citenamefont {Monaco},
  \citenamefont {Bellazzini}, \citenamefont {Ferraro},\ and\ \citenamefont
  {Pancino}}]{Monaco:2004ke}%
  \BibitemOpen
  \bibfield  {author} {\bibinfo {author} {\bibfnamefont {Lorenzo}\ \bibnamefont
  {Monaco}}, \bibinfo {author} {\bibfnamefont {M.}~\bibnamefont {Bellazzini}},
  \bibinfo {author} {\bibfnamefont {F.~R.}\ \bibnamefont {Ferraro}}, \ and\
  \bibinfo {author} {\bibfnamefont {E.}~\bibnamefont {Pancino}},\ }\bibfield
  {title} {\enquote {\bibinfo {title} {{The Central density cusp of the
  Sagittarius dwarf spheroidal galaxy}},}\ }\href {\doibase
  10.1111/j.1365-2966.2004.08579.x} {\bibfield  {journal} {\bibinfo  {journal}
  {Mon. Not. Roy. Astron. Soc.}\ }\textbf {\bibinfo {volume} {356}},\ \bibinfo
  {pages} {1396--1402} (\bibinfo {year} {2005})},\ \Eprint
  {http://arxiv.org/abs/astro-ph/0411107} {arXiv:astro-ph/0411107 [astro-ph]}
  \BibitemShut {NoStop}%
%%CITATION = ASTRO-PH/0411107;%%
\bibitem [{\citenamefont {{Ibata}}\ \emph {et~al.}(2009)\citenamefont
  {{Ibata}}, \citenamefont {{Bellazzini}}, \citenamefont {{Chapman}},
  \citenamefont {{Dalessandro}}, \citenamefont {{Ferraro}}, \citenamefont
  {{Irwin}}, \citenamefont {{Lanzoni}}, \citenamefont {{Lewis}}, \citenamefont
  {{Mackey}}, \citenamefont {{Miocchi}},\ and\ \citenamefont
  {{Varghese}}}]{2009ApJ...699L.169I}%
  \BibitemOpen
  \bibfield  {author} {\bibinfo {author} {\bibfnamefont {R.}~\bibnamefont
  {{Ibata}}}, \bibinfo {author} {\bibfnamefont {M.}~\bibnamefont
  {{Bellazzini}}}, \bibinfo {author} {\bibfnamefont {S.~C.}\ \bibnamefont
  {{Chapman}}}, \bibinfo {author} {\bibfnamefont {E.}~\bibnamefont
  {{Dalessandro}}}, \bibinfo {author} {\bibfnamefont {F.}~\bibnamefont
  {{Ferraro}}}, \bibinfo {author} {\bibfnamefont {M.}~\bibnamefont {{Irwin}}},
  \bibinfo {author} {\bibfnamefont {B.}~\bibnamefont {{Lanzoni}}}, \bibinfo
  {author} {\bibfnamefont {G.~F.}\ \bibnamefont {{Lewis}}}, \bibinfo {author}
  {\bibfnamefont {A.~D.}\ \bibnamefont {{Mackey}}}, \bibinfo {author}
  {\bibfnamefont {P.}~\bibnamefont {{Miocchi}}}, \ and\ \bibinfo {author}
  {\bibfnamefont {A.}~\bibnamefont {{Varghese}}},\ }\bibfield  {title}
  {\enquote {\bibinfo {title} {{Density and Kinematic Cusps in M54 at the Heart
  of the Sagittarius Dwarf Galaxy: Evidence for A 10$^{4}$ M $_{sun}$ Black
  Hole?}}}\ }\href {\doibase 10.1088/0004-637X/699/2/L169} {\bibfield
  {journal} {\bibinfo  {journal} {ApJL}\ }\textbf {\bibinfo {volume} {699}},\
  \bibinfo {pages} {L169--L173} (\bibinfo {year} {2009})},\ \Eprint
  {http://arxiv.org/abs/0906.4894} {arXiv:0906.4894 [astro-ph.GA]} \BibitemShut
  {NoStop}%
\bibitem [{\citenamefont {Kunder}\ and\ \citenamefont
  {Chaboyer}(2009)}]{Kunder:2009nb}%
  \BibitemOpen
  \bibfield  {author} {\bibinfo {author} {\bibfnamefont {Andrea}\ \bibnamefont
  {Kunder}}\ and\ \bibinfo {author} {\bibfnamefont {Brian}\ \bibnamefont
  {Chaboyer}},\ }\bibfield  {title} {\enquote {\bibinfo {title} {{Distance to
  the Sagittarius Dwarf Galaxy using MACHO Project RR Lyrae stars}},}\ }\href
  {\doibase 10.1088/0004-6256/137/5/4478} {\bibfield  {journal} {\bibinfo
  {journal} {Astron. J.}\ }\textbf {\bibinfo {volume} {137}},\ \bibinfo {pages}
  {4478} (\bibinfo {year} {2009})},\ \Eprint {http://arxiv.org/abs/0903.3040}
  {arXiv:0903.3040 [astro-ph.SR]} \BibitemShut {NoStop}%
%%CITATION = ARXIV:0903.3040;%%
\bibitem [{\citenamefont {Ivanova}\ \emph {et~al.}(2008)\citenamefont
  {Ivanova}, \citenamefont {Heinke}, \citenamefont {Rasio}, \citenamefont
  {Belczynski},\ and\ \citenamefont {Fregeau}}]{Ivanova:2007bu}%
  \BibitemOpen
  \bibfield  {author} {\bibinfo {author} {\bibfnamefont {N.}~\bibnamefont
  {Ivanova}}, \bibinfo {author} {\bibfnamefont {C.}~\bibnamefont {Heinke}},
  \bibinfo {author} {\bibfnamefont {F.~A.}\ \bibnamefont {Rasio}}, \bibinfo
  {author} {\bibfnamefont {K.}~\bibnamefont {Belczynski}}, \ and\ \bibinfo
  {author} {\bibfnamefont {J.}~\bibnamefont {Fregeau}},\ }\bibfield  {title}
  {\enquote {\bibinfo {title} {{Formation and evolution of compact binaries in
  globular clusters: II. Binaries with neutron stars}},}\ }\href {\doibase
  10.1111/j.1365-2966.2008.13064.x} {\bibfield  {journal} {\bibinfo  {journal}
  {Mon. Not. Roy. Astron. Soc.}\ }\textbf {\bibinfo {volume} {386}},\ \bibinfo
  {pages} {553--576} (\bibinfo {year} {2008})},\ \Eprint
  {http://arxiv.org/abs/0706.4096} {arXiv:0706.4096 [astro-ph]} \BibitemShut
  {NoStop}%
%%CITATION = ARXIV:0706.4096;%%
\bibitem [{\citenamefont {{Brown}}\ \emph {et~al.}(1999)\citenamefont
  {{Brown}}, \citenamefont {{Wallerstein}},\ and\ \citenamefont
  {{Gonzalez}}}]{1999AJ....118.1245B}%
  \BibitemOpen
  \bibfield  {author} {\bibinfo {author} {\bibfnamefont {J.~A.}\ \bibnamefont
  {{Brown}}}, \bibinfo {author} {\bibfnamefont {G.}~\bibnamefont
  {{Wallerstein}}}, \ and\ \bibinfo {author} {\bibfnamefont {G.}~\bibnamefont
  {{Gonzalez}}},\ }\bibfield  {title} {\enquote {\bibinfo {title} {{Elemental
  Abundances in Five Stars in M54, A Globular Cluster Associated with the
  Sagittarius Galaxy}},}\ }\href {\doibase 10.1086/300996} {\bibfield
  {journal} {\bibinfo  {journal} {Astron. J.}\ }\textbf {\bibinfo {volume}
  {118}},\ \bibinfo {pages} {1245--1251} (\bibinfo {year} {1999})}\BibitemShut
  {NoStop}%
\bibitem [{\citenamefont {Kuranov}\ and\ \citenamefont
  {Postnov}(2006)}]{Kuranov:2006kw}%
  \BibitemOpen
  \bibfield  {author} {\bibinfo {author} {\bibfnamefont {A.~G.}\ \bibnamefont
  {Kuranov}}\ and\ \bibinfo {author} {\bibfnamefont {K.~A.}\ \bibnamefont
  {Postnov}},\ }\bibfield  {title} {\enquote {\bibinfo {title} {{Neutron stars
  in globular clusters: formation and observational manifestations}},}\ }\href
  {\doibase 10.1134/S106377370606003X} {\bibfield  {journal} {\bibinfo
  {journal} {Astron. Lett.}\ }\textbf {\bibinfo {volume} {32}},\ \bibinfo
  {pages} {393} (\bibinfo {year} {2006})},\ \Eprint
  {http://arxiv.org/abs/astro-ph/0605115} {arXiv:astro-ph/0605115 [astro-ph]}
  \BibitemShut {NoStop}%
%%CITATION = ASTRO-PH/0605115;%%
\bibitem [{\citenamefont {Aharonian}(2008)}]{Aharonian:2007km}%
  \BibitemOpen
  \bibfield  {author} {\bibinfo {author} {\bibfnamefont {F.}~\bibnamefont
  {Aharonian}} (\bibinfo {collaboration} {H.E.S.S.}),\ }\bibfield  {title}
  {\enquote {\bibinfo {title} {{Observations of the Sagittarius Dwarf galaxy by
  the H.E.S.S. experiment and search for a Dark Matter signal}},}\ }\href
  {\doibase 10.1016/j.astropartphys.2007.11.007,
  10.1016/j.astropartphys.2010.01.007} {\bibfield  {journal} {\bibinfo
  {journal} {Astropart. Phys.}\ }\textbf {\bibinfo {volume} {29}},\ \bibinfo
  {pages} {55--62} (\bibinfo {year} {2008})},\ \bibinfo {note} {[Erratum:
  Astropart. Phys.33,274(2010)]},\ \Eprint {http://arxiv.org/abs/0711.2369}
  {arXiv:0711.2369 [astro-ph]} \BibitemShut {NoStop}%
%%CITATION = ARXIV:0711.2369;%%
\bibitem [{\citenamefont {Read}\ \emph {et~al.}(2016)\citenamefont {Read},
  \citenamefont {Agertz},\ and\ \citenamefont {Collins}}]{Read:2015sta}%
  \BibitemOpen
  \bibfield  {author} {\bibinfo {author} {\bibfnamefont {J.~I.}\ \bibnamefont
  {Read}}, \bibinfo {author} {\bibfnamefont {O.}~\bibnamefont {Agertz}}, \ and\
  \bibinfo {author} {\bibfnamefont {M.~L.~M.}\ \bibnamefont {Collins}},\
  }\bibfield  {title} {\enquote {\bibinfo {title} {{Dark matter cores all the
  way down}},}\ }\href {\doibase 10.1093/mnras/stw713} {\bibfield  {journal}
  {\bibinfo  {journal} {Mon. Not. Roy. Astron. Soc.}\ }\textbf {\bibinfo
  {volume} {459}},\ \bibinfo {pages} {2573--2590} (\bibinfo {year} {2016})},\
  \Eprint {http://arxiv.org/abs/1508.04143} {arXiv:1508.04143 [astro-ph.GA]}
  \BibitemShut {NoStop}%
%%CITATION = ARXIV:1508.04143;%%
\bibitem [{\citenamefont {Evans}\ \emph {et~al.}(2004)\citenamefont {Evans},
  \citenamefont {Ferrer},\ and\ \citenamefont {Sarkar}}]{PhysRevD.69.123501}%
  \BibitemOpen
  \bibfield  {author} {\bibinfo {author} {\bibfnamefont {N.~W.}\ \bibnamefont
  {Evans}}, \bibinfo {author} {\bibfnamefont {F.}~\bibnamefont {Ferrer}}, \
  and\ \bibinfo {author} {\bibfnamefont {S.}~\bibnamefont {Sarkar}},\
  }\bibfield  {title} {\enquote {\bibinfo {title} {A travel guide to the dark
  matter annihilation signal},}\ }\href {\doibase 10.1103/PhysRevD.69.123501}
  {\bibfield  {journal} {\bibinfo  {journal} {Phys. Rev. D}\ }\textbf {\bibinfo
  {volume} {69}},\ \bibinfo {pages} {123501} (\bibinfo {year}
  {2004})}\BibitemShut {NoStop}%
\bibitem [{\citenamefont {{Tamm}}\ \emph {et~al.}(2012)\citenamefont {{Tamm}},
  \citenamefont {{Tempel}}, \citenamefont {{Tenjes}}, \citenamefont
  {{Tihhonova}},\ and\ \citenamefont {{Tuvikene}}}]{2012A&A...546A...4T}%
  \BibitemOpen
  \bibfield  {author} {\bibinfo {author} {\bibfnamefont {A.}~\bibnamefont
  {{Tamm}}}, \bibinfo {author} {\bibfnamefont {E.}~\bibnamefont {{Tempel}}},
  \bibinfo {author} {\bibfnamefont {P.}~\bibnamefont {{Tenjes}}}, \bibinfo
  {author} {\bibfnamefont {O.}~\bibnamefont {{Tihhonova}}}, \ and\ \bibinfo
  {author} {\bibfnamefont {T.}~\bibnamefont {{Tuvikene}}},\ }\bibfield  {title}
  {\enquote {\bibinfo {title} {{Stellar mass map and dark matter distribution
  in M 31}},}\ }\href {\doibase 10.1051/0004-6361/201220065} {\bibfield
  {journal} {\bibinfo  {journal} {A\&A}\ }\textbf {\bibinfo {volume} {546}},\
  \bibinfo {eid} {A4} (\bibinfo {year} {2012})},\ \Eprint
  {http://arxiv.org/abs/1208.5712} {arXiv:1208.5712} \BibitemShut {NoStop}%
\bibitem [{\citenamefont {{Licquia}}\ and\ \citenamefont
  {{Newman}}(2015)}]{2015ApJ...806...96L}%
  \BibitemOpen
  \bibfield  {author} {\bibinfo {author} {\bibfnamefont {T.~C.}\ \bibnamefont
  {{Licquia}}}\ and\ \bibinfo {author} {\bibfnamefont {J.~A.}\ \bibnamefont
  {{Newman}}},\ }\bibfield  {title} {\enquote {\bibinfo {title} {{Improved
  Estimates of the Milky Way's Stellar Mass and Star Formation Rate from
  Hierarchical Bayesian Meta-Analysis}},}\ }\href {\doibase
  10.1088/0004-637X/806/1/96} {\bibfield  {journal} {\bibinfo  {journal} {ApJ}\
  }\textbf {\bibinfo {volume} {806}},\ \bibinfo {eid} {96} (\bibinfo {year}
  {2015})},\ \Eprint {http://arxiv.org/abs/1407.1078} {arXiv:1407.1078}
  \BibitemShut {NoStop}%
\bibitem [{\citenamefont {Banerjee}\ and\ \citenamefont
  {Jog}(2008)}]{Banerjee:2008kt}%
  \BibitemOpen
  \bibfield  {author} {\bibinfo {author} {\bibfnamefont {Arunima}\ \bibnamefont
  {Banerjee}}\ and\ \bibinfo {author} {\bibfnamefont {Chanda~J.}\ \bibnamefont
  {Jog}},\ }\bibfield  {title} {\enquote {\bibinfo {title} {{The Flattened Dark
  Matter Halo of M31 as Deduced from the Observed HI Scale Heights}},}\ }\href
  {\doibase 10.1086/591223} {\bibfield  {journal} {\bibinfo  {journal}
  {Astrophys. J.}\ }\textbf {\bibinfo {volume} {685}},\ \bibinfo {pages} {254}
  (\bibinfo {year} {2008})},\ \Eprint {http://arxiv.org/abs/0806.3610}
  {arXiv:0806.3610 [astro-ph]} \BibitemShut {NoStop}%
%%CITATION = ARXIV:0806.3610;%%
\bibitem [{\citenamefont {{Collins}}\ \emph {et~al.}(2011)\citenamefont
  {{Collins}}, \citenamefont {{Chapman}}, \citenamefont {{Ibata}},
  \citenamefont {{Irwin}}, \citenamefont {{Rich}}, \citenamefont {{Ferguson}},
  \citenamefont {{Lewis}}, \citenamefont {{Tanvir}},\ and\ \citenamefont
  {{Koch}}}]{2011MNRAS.413.1548C}%
  \BibitemOpen
  \bibfield  {author} {\bibinfo {author} {\bibfnamefont {M.~L.~M.}\
  \bibnamefont {{Collins}}}, \bibinfo {author} {\bibfnamefont {S.~C.}\
  \bibnamefont {{Chapman}}}, \bibinfo {author} {\bibfnamefont {R.~A.}\
  \bibnamefont {{Ibata}}}, \bibinfo {author} {\bibfnamefont {M.~J.}\
  \bibnamefont {{Irwin}}}, \bibinfo {author} {\bibfnamefont {R.~M.}\
  \bibnamefont {{Rich}}}, \bibinfo {author} {\bibfnamefont {A.~M.~N.}\
  \bibnamefont {{Ferguson}}}, \bibinfo {author} {\bibfnamefont {G.~F.}\
  \bibnamefont {{Lewis}}}, \bibinfo {author} {\bibfnamefont {N.}~\bibnamefont
  {{Tanvir}}}, \ and\ \bibinfo {author} {\bibfnamefont {A.}~\bibnamefont
  {{Koch}}},\ }\bibfield  {title} {\enquote {\bibinfo {title} {{The kinematic
  identification of a thick stellar disc in M31}},}\ }\href {\doibase
  10.1111/j.1365-2966.2011.18238.x} {\bibfield  {journal} {\bibinfo  {journal}
  {MNRAS}\ }\textbf {\bibinfo {volume} {413}},\ \bibinfo {pages} {1548--1568}
  (\bibinfo {year} {2011})},\ \Eprint {http://arxiv.org/abs/1010.5276}
  {arXiv:1010.5276} \BibitemShut {NoStop}%
\bibitem [{\citenamefont {Bender}\ \emph {et~al.}(2005)\citenamefont {Bender}
  \emph {et~al.}}]{Bender:2005rq}%
  \BibitemOpen
  \bibfield  {author} {\bibinfo {author} {\bibfnamefont {Ralf}\ \bibnamefont
  {Bender}} \emph {et~al.},\ }\bibfield  {title} {\enquote {\bibinfo {title}
  {{Hst stis spectroscopy of the triple nucleus of m31: two nested disks in
  keplerian rotation around a supermassive black hole}},}\ }\href {\doibase
  10.1086/432434} {\bibfield  {journal} {\bibinfo  {journal} {Astrophys. J.}\
  }\textbf {\bibinfo {volume} {631}},\ \bibinfo {pages} {280} (\bibinfo {year}
  {2005})},\ \Eprint {http://arxiv.org/abs/astro-ph/0509839}
  {arXiv:astro-ph/0509839 [astro-ph]} \BibitemShut {NoStop}%
%%CITATION = ASTRO-PH/0509839;%%
\bibitem [{\citenamefont {{Tremaine}}(1995)}]{1995AJ....110..628T}%
  \BibitemOpen
  \bibfield  {author} {\bibinfo {author} {\bibfnamefont {S.}~\bibnamefont
  {{Tremaine}}},\ }\bibfield  {title} {\enquote {\bibinfo {title} {{An
  Eccentric-Disk Model for the Nucleus of M31}},}\ }\href {\doibase
  10.1086/117548} {\bibfield  {journal} {\bibinfo  {journal} {AJ}\ }\textbf
  {\bibinfo {volume} {110}},\ \bibinfo {pages} {628} (\bibinfo {year}
  {1995})},\ \Eprint {http://arxiv.org/abs/astro-ph/9502065} {astro-ph/9502065}
  \BibitemShut {NoStop}%
\bibitem [{\citenamefont {Cowan}\ \emph {et~al.}(2011)\citenamefont {Cowan},
  \citenamefont {Cranmer}, \citenamefont {Gross},\ and\ \citenamefont
  {Vitells}}]{Cowan:2010js}%
  \BibitemOpen
  \bibfield  {author} {\bibinfo {author} {\bibfnamefont {Glen}\ \bibnamefont
  {Cowan}}, \bibinfo {author} {\bibfnamefont {Kyle}\ \bibnamefont {Cranmer}},
  \bibinfo {author} {\bibfnamefont {Eilam}\ \bibnamefont {Gross}}, \ and\
  \bibinfo {author} {\bibfnamefont {Ofer}\ \bibnamefont {Vitells}},\ }\bibfield
   {title} {\enquote {\bibinfo {title} {{Asymptotic formulae for
  likelihood-based tests of new physics}},}\ }\href {\doibase
  10.1140/epjc/s10052-011-1554-0, 10.1140/epjc/s10052-013-2501-z} {\bibfield
  {journal} {\bibinfo  {journal} {Eur. Phys. J.}\ }\textbf {\bibinfo {volume}
  {C71}},\ \bibinfo {pages} {1554} (\bibinfo {year} {2011})},\ \bibinfo {note}
  {[Erratum: Eur. Phys. J.C73,2501(2013)]},\ \Eprint
  {http://arxiv.org/abs/1007.1727} {arXiv:1007.1727 [physics.data-an]}
  \BibitemShut {NoStop}%
%%CITATION = ARXIV:1007.1727;%%
\bibitem [{\citenamefont {{Haslam}}\ \emph {et~al.}(1981)\citenamefont
  {{Haslam}}, \citenamefont {{Klein}}, \citenamefont {{Salter}}, \citenamefont
  {{Stoffel}}, \citenamefont {{Wilson}}, \citenamefont {{Cleary}},
  \citenamefont {{Cooke}},\ and\ \citenamefont
  {{Thomasson}}}]{1981A&A...100..209H}%
  \BibitemOpen
  \bibfield  {author} {\bibinfo {author} {\bibfnamefont {C.~G.~T.}\
  \bibnamefont {{Haslam}}}, \bibinfo {author} {\bibfnamefont {U.}~\bibnamefont
  {{Klein}}}, \bibinfo {author} {\bibfnamefont {C.~J.}\ \bibnamefont
  {{Salter}}}, \bibinfo {author} {\bibfnamefont {H.}~\bibnamefont {{Stoffel}}},
  \bibinfo {author} {\bibfnamefont {W.~E.}\ \bibnamefont {{Wilson}}}, \bibinfo
  {author} {\bibfnamefont {M.~N.}\ \bibnamefont {{Cleary}}}, \bibinfo {author}
  {\bibfnamefont {D.~J.}\ \bibnamefont {{Cooke}}}, \ and\ \bibinfo {author}
  {\bibfnamefont {P.}~\bibnamefont {{Thomasson}}},\ }\bibfield  {title}
  {\enquote {\bibinfo {title} {{A 408 MHz all-sky continuum survey. I -
  Observations at southern declinations and for the North Polar region}},}\
  }\href@noop {} {\bibfield  {journal} {\bibinfo  {journal} {A\&A}\ }\textbf
  {\bibinfo {volume} {100}},\ \bibinfo {pages} {209--219} (\bibinfo {year}
  {1981})}\BibitemShut {NoStop}%
\bibitem [{\citenamefont {{Haslam}}\ \emph {et~al.}(1982)\citenamefont
  {{Haslam}}, \citenamefont {{Salter}}, \citenamefont {{Stoffel}},\ and\
  \citenamefont {{Wilson}}}]{1982A&AS...47....1H}%
  \BibitemOpen
  \bibfield  {author} {\bibinfo {author} {\bibfnamefont {C.~G.~T.}\
  \bibnamefont {{Haslam}}}, \bibinfo {author} {\bibfnamefont {C.~J.}\
  \bibnamefont {{Salter}}}, \bibinfo {author} {\bibfnamefont {H.}~\bibnamefont
  {{Stoffel}}}, \ and\ \bibinfo {author} {\bibfnamefont {W.~E.}\ \bibnamefont
  {{Wilson}}},\ }\bibfield  {title} {\enquote {\bibinfo {title} {{A 408 MHz
  all-sky continuum survey. II - The atlas of contour maps}},}\ }\href@noop {}
  {\bibfield  {journal} {\bibinfo  {journal} {A\&AS}\ }\textbf {\bibinfo
  {volume} {47}},\ \bibinfo {pages} {1} (\bibinfo {year} {1982})}\BibitemShut
  {NoStop}%
\bibitem [{\citenamefont {{Remazeilles}}\ \emph {et~al.}(2015)\citenamefont
  {{Remazeilles}}, \citenamefont {{Dickinson}}, \citenamefont {{Banday}},
  \citenamefont {{Bigot-Sazy}},\ and\ \citenamefont
  {{Ghosh}}}]{2015MNRAS.451.4311R}%
  \BibitemOpen
  \bibfield  {author} {\bibinfo {author} {\bibfnamefont {M.}~\bibnamefont
  {{Remazeilles}}}, \bibinfo {author} {\bibfnamefont {C.}~\bibnamefont
  {{Dickinson}}}, \bibinfo {author} {\bibfnamefont {A.~J.}\ \bibnamefont
  {{Banday}}}, \bibinfo {author} {\bibfnamefont {M.-A.}\ \bibnamefont
  {{Bigot-Sazy}}}, \ and\ \bibinfo {author} {\bibfnamefont {T.}~\bibnamefont
  {{Ghosh}}},\ }\bibfield  {title} {\enquote {\bibinfo {title} {{An improved
  source-subtracted and destriped 408-MHz all-sky map}},}\ }\href {\doibase
  10.1093/mnras/stv1274} {\bibfield  {journal} {\bibinfo  {journal} {MNRAS}\
  }\textbf {\bibinfo {volume} {451}},\ \bibinfo {pages} {4311--4327} (\bibinfo
  {year} {2015})},\ \Eprint {http://arxiv.org/abs/1411.3628} {arXiv:1411.3628
  [astro-ph.IM]} \BibitemShut {NoStop}%
\bibitem [{\citenamefont {Platania}\ \emph {et~al.}(1998)\citenamefont
  {Platania}, \citenamefont {Bensadoun}, \citenamefont {Bersanelli},
  \citenamefont {De~Amici}, \citenamefont {Kogut}, \citenamefont {Levin},
  \citenamefont {Maino},\ and\ \citenamefont {Smoot}}]{Platania:1997zn}%
  \BibitemOpen
  \bibfield  {author} {\bibinfo {author} {\bibfnamefont {P.}~\bibnamefont
  {Platania}}, \bibinfo {author} {\bibfnamefont {M.}~\bibnamefont {Bensadoun}},
  \bibinfo {author} {\bibfnamefont {M.}~\bibnamefont {Bersanelli}}, \bibinfo
  {author} {\bibfnamefont {G.}~\bibnamefont {De~Amici}}, \bibinfo {author}
  {\bibfnamefont {A.}~\bibnamefont {Kogut}}, \bibinfo {author} {\bibfnamefont
  {S.}~\bibnamefont {Levin}}, \bibinfo {author} {\bibfnamefont
  {D.}~\bibnamefont {Maino}}, \ and\ \bibinfo {author} {\bibfnamefont
  {George~F.}\ \bibnamefont {Smoot}},\ }\bibfield  {title} {\enquote {\bibinfo
  {title} {{A determination of the spectral index of galactic synchrotron
  emission in the 1-10 GHz range}},}\ }\href {\doibase 10.1086/306175}
  {\bibfield  {journal} {\bibinfo  {journal} {Astrophys. J.}\ }\textbf
  {\bibinfo {volume} {505}},\ \bibinfo {pages} {473} (\bibinfo {year}
  {1998})},\ \Eprint {http://arxiv.org/abs/astro-ph/9707252}
  {arXiv:astro-ph/9707252 [astro-ph]} \BibitemShut {NoStop}%
%%CITATION = ASTRO-PH/9707252;%%
\bibitem [{\citenamefont {{Fixsen}}\ \emph {et~al.}(2011)\citenamefont
  {{Fixsen}}, \citenamefont {{Kogut}}, \citenamefont {{Levin}}, \citenamefont
  {{Limon}}, \citenamefont {{Lubin}}, \citenamefont {{Mirel}}, \citenamefont
  {{Seiffert}}, \citenamefont {{Singal}}, \citenamefont {{Wollack}},
  \citenamefont {{Villela}},\ and\ \citenamefont
  {{Wuensche}}}]{2011ApJ...734....5F}%
  \BibitemOpen
  \bibfield  {author} {\bibinfo {author} {\bibfnamefont {D.~J.}\ \bibnamefont
  {{Fixsen}}}, \bibinfo {author} {\bibfnamefont {A.}~\bibnamefont {{Kogut}}},
  \bibinfo {author} {\bibfnamefont {S.}~\bibnamefont {{Levin}}}, \bibinfo
  {author} {\bibfnamefont {M.}~\bibnamefont {{Limon}}}, \bibinfo {author}
  {\bibfnamefont {P.}~\bibnamefont {{Lubin}}}, \bibinfo {author} {\bibfnamefont
  {P.}~\bibnamefont {{Mirel}}}, \bibinfo {author} {\bibfnamefont
  {M.}~\bibnamefont {{Seiffert}}}, \bibinfo {author} {\bibfnamefont
  {J.}~\bibnamefont {{Singal}}}, \bibinfo {author} {\bibfnamefont
  {E.}~\bibnamefont {{Wollack}}}, \bibinfo {author} {\bibfnamefont
  {T.}~\bibnamefont {{Villela}}}, \ and\ \bibinfo {author} {\bibfnamefont
  {C.~A.}\ \bibnamefont {{Wuensche}}},\ }\bibfield  {title} {\enquote {\bibinfo
  {title} {{ARCADE 2 Measurement of the Absolute Sky Brightness at 3-90
  GHz}},}\ }\href {\doibase 10.1088/0004-637X/734/1/5} {\bibfield  {journal}
  {\bibinfo  {journal} {Astrophys. J.}\ }\textbf {\bibinfo {volume} {734}},\
  \bibinfo {eid} {5} (\bibinfo {year} {2011})},\ \Eprint
  {http://arxiv.org/abs/0901.0555} {arXiv:0901.0555} \BibitemShut {NoStop}%
\bibitem [{\citenamefont {Macquart}\ and\ \citenamefont
  {Kanekar}(2015)}]{Macquart:2015jfa}%
  \BibitemOpen
  \bibfield  {author} {\bibinfo {author} {\bibfnamefont {Jean-Pierre}\
  \bibnamefont {Macquart}}\ and\ \bibinfo {author} {\bibfnamefont {Nissim}\
  \bibnamefont {Kanekar}},\ }\bibfield  {title} {\enquote {\bibinfo {title}
  {{On Detecting Millisecond Pulsars at the Galactic Center}},}\ }\href
  {\doibase 10.1088/0004-637X/805/2/172} {\bibfield  {journal} {\bibinfo
  {journal} {Astrophys. J.}\ }\textbf {\bibinfo {volume} {805}},\ \bibinfo
  {pages} {172} (\bibinfo {year} {2015})},\ \Eprint
  {http://arxiv.org/abs/1504.02492} {arXiv:1504.02492 [astro-ph.HE]}
  \BibitemShut {NoStop}%
%%CITATION = ARXIV:1504.02492;%%
\bibitem [{\citenamefont {Rajwade}\ \emph {et~al.}(2017)\citenamefont
  {Rajwade}, \citenamefont {Lorimer},\ and\ \citenamefont
  {Anderson}}]{Rajwade:2016cto}%
  \BibitemOpen
  \bibfield  {author} {\bibinfo {author} {\bibfnamefont {Kaustubh}\
  \bibnamefont {Rajwade}}, \bibinfo {author} {\bibfnamefont {Duncan}\
  \bibnamefont {Lorimer}}, \ and\ \bibinfo {author} {\bibfnamefont {Loren}\
  \bibnamefont {Anderson}},\ }\bibfield  {title} {\enquote {\bibinfo {title}
  {{Detecting pulsars in the Galactic centre}},}\ }\href {\doibase
  10.1093/mnras/stx1661} {\bibfield  {journal} {\bibinfo  {journal} {Mon. Not.
  Roy. Astron. Soc.}\ }\textbf {\bibinfo {volume} {471}},\ \bibinfo {pages}
  {730--739} (\bibinfo {year} {2017})},\ \Eprint
  {http://arxiv.org/abs/1611.06977} {arXiv:1611.06977 [astro-ph.HE]}
  \BibitemShut {NoStop}%
%%CITATION = ARXIV:1611.06977;%%
\bibitem [{\citenamefont {Arik}\ \emph {et~al.}(2014)\citenamefont {Arik} \emph
  {et~al.}}]{Arik:2013nya}%
  \BibitemOpen
  \bibfield  {author} {\bibinfo {author} {\bibfnamefont {M.}~\bibnamefont
  {Arik}} \emph {et~al.} (\bibinfo {collaboration} {CAST}),\ }\bibfield
  {title} {\enquote {\bibinfo {title} {{Search for Solar Axions by the CERN
  Axion Solar Telescope with $^3$He Buffer Gas: Closing the Hot Dark Matter
  Gap}},}\ }\href {\doibase 10.1103/PhysRevLett.112.091302} {\bibfield
  {journal} {\bibinfo  {journal} {Phys. Rev. Lett.}\ }\textbf {\bibinfo
  {volume} {112}},\ \bibinfo {pages} {091302} (\bibinfo {year} {2014})},\
  \Eprint {http://arxiv.org/abs/1307.1985} {arXiv:1307.1985 [hep-ex]}
  \BibitemShut {NoStop}%
%%CITATION = ARXIV:1307.1985;%%
\bibitem [{\citenamefont {Arik}\ \emph {et~al.}(2015)\citenamefont {Arik} \emph
  {et~al.}}]{Arik:2015cjv}%
  \BibitemOpen
  \bibfield  {author} {\bibinfo {author} {\bibfnamefont {M.}~\bibnamefont
  {Arik}} \emph {et~al.} (\bibinfo {collaboration} {CAST}),\ }\bibfield
  {title} {\enquote {\bibinfo {title} {{New solar axion search using the CERN
  Axion Solar Telescope with $^4$He filling}},}\ }\href {\doibase
  10.1103/PhysRevD.92.021101} {\bibfield  {journal} {\bibinfo  {journal} {Phys.
  Rev.}\ }\textbf {\bibinfo {volume} {D92}},\ \bibinfo {pages} {021101}
  (\bibinfo {year} {2015})},\ \Eprint {http://arxiv.org/abs/1503.00610}
  {arXiv:1503.00610 [hep-ex]} \BibitemShut {NoStop}%
%%CITATION = ARXIV:1503.00610;%%
\bibitem [{\citenamefont {Anastassopoulos}\ \emph {et~al.}(2017)\citenamefont
  {Anastassopoulos} \emph {et~al.}}]{Anastassopoulos:2017ftl}%
  \BibitemOpen
  \bibfield  {author} {\bibinfo {author} {\bibfnamefont {V.}~\bibnamefont
  {Anastassopoulos}} \emph {et~al.} (\bibinfo {collaboration} {CAST}),\
  }\bibfield  {title} {\enquote {\bibinfo {title} {{New CAST Limit on the
  Axion-Photon Interaction}},}\ }\href {\doibase 10.1038/nphys4109} {\bibfield
  {journal} {\bibinfo  {journal} {Nature Phys.}\ }\textbf {\bibinfo {volume}
  {13}},\ \bibinfo {pages} {584--590} (\bibinfo {year} {2017})},\ \Eprint
  {http://arxiv.org/abs/1705.02290} {arXiv:1705.02290 [hep-ex]} \BibitemShut
  {NoStop}%
%%CITATION = ARXIV:1705.02290;%%
\end{thebibliography}%

\end{document}